\documentclass[11pt]{article}
\usepackage{courier}
\usepackage[margin=1in]{geometry}
\usepackage{amsmath,amssymb,amsthm}
\usepackage{enumitem}
\usepackage{booktabs,longtable,array}
\usepackage{float}
\usepackage{xcolor}
\usepackage{graphicx}
\usepackage{adjustbox}
\usepackage[most]{tcolorbox}
\usepackage{hhline}
\tcbset{atlasbox/.style={enhanced, breakable, colback=black!4, colframe=black!45, boxrule=0.4pt, arc=1pt,
  left=5pt, right=5pt, top=3pt, bottom=3pt, before skip=4pt, after skip=4pt, fonttitle=\scriptsize\scshape,
  coltitle=black!70, attach boxed title to top left={yshift=-2mm, xshift=3mm}, boxed title style={colback=white, boxrule=0pt, size=fbox}}}
\usepackage{tikz}
\usetikzlibrary{arrows.meta,positioning}
\usepackage[colorlinks=true,linkcolor=green!40!black,urlcolor=green!40!black]{hyperref}
\newcommand{\imp}{\Rightarrow} %
\newcommand{\nt}{\neg} %
\newcommand{\N}{\mathbb{N}}
\newcommand{\Z}{\mathbb{Z}}
\newcommand{\Q}{\mathbb{Q}}
\newcommand{\R}{\mathbb{R}}

\newcommand{\bx}[1]{\textsc{#1}} %

\definecolor{newblue}{RGB}{0,70,200}
\definecolor{galbeige}{RGB}{243,243,243} %
\definecolor{modelblue}{RGB}{176,0,32} %

\DeclareRobustCommand{\etc}{{\it\frenchspacing etc.}}
\DeclareRobustCommand{\ie}{{\it\frenchspacing i.e.}}
\DeclareRobustCommand{\eg}{{\it\frenchspacing e.g.}}

\def\beq#1{\begin{equation}\label{#1}}
\def\eeq{\end{equation}}
\def\beqa#1{\begin{eqnarray}\label{#1}}
\def\eeqa{\end{eqnarray}}

  \def\fig#1{Figure~\ref{#1}}
        
  \def\tab#1{Table~\ref{#1}}
  \def\Sec#1{Section~\ref{#1}}

\title{Defining the mathematical family tree\\with a minimal programming language}
\author{Max Tegmark}
\date{Last updated: November 23, 2026}

\begin{document}
\maketitle

\begin{abstract}\emergencystretch=3em
I introduce MLISP, a Turing-complete language with nine single-character commands
 \mbox{\texttt{a b c d f 0 i j k}}, and use it to define 88 mathematical structures and a family tree of 65 theories that they model, with three aims:\\
{\bf Pedagogy:} It presents the mathematical structures in a unified way, with standardized presentation tableaus that define them and axiomatized theories that describe them.\\
{\bf Connections:} It highlights how they are related in the family tree via added symbols, axioms, {\etc}
While the full tree is obviously infinite, its lower parts naturally separate into six main branches:
the relational branch, the algebraic branch, the logic branch, the number branch, the set branch and the continuum branch.\\
{\bf Foundations:} The definitions, key theorems and models of each entry are verified in {\it Lean}. 
Mathematical structures are defined as equivalence classes of presentations, which in a constructivist spirit are defined in a simple, explicit and self-contained way in standardized presentation tableaus with minimal human baggage. For finite structures, the generating functions are defined by lookup tables.
For countably infinite ones, they are instead defined computationally using MLISP, since more complex languages such as {\it Lean} are massive overkill for rigorously defining standard arithmetic operations (for example, the program \texttt{jibjbfid} defines subtraction).\\
An interactive version is available at \url{https://tegmark.org/mathtree}. The Lean code is public, also verifies the correctness of MLISP and includes 43\% of the theorems from Mathlib 2026-10-11.
\end{abstract}

\section{Introduction}\label{IntroSec}

After millennia of progress, mathematics is currently developing faster than ever because of artificial intelligence,
raising new opportunities and challenges \cite{MarathonStatement}.
For example, a recent declaration from leading mathematicians \cite{FieldsMedalistStatement} emphasizes the importance of not merely proving formal results, but also making them human-understandable.
In this spirit, the goal of this paper is pedagogical, defining and presenting a unified human-readable 
family tree of basic mathematical theories and structures.
This tree is obviously highly incomplete (the full tree is infinite) and should be viewed as work in progress;
it emphasizes mainly theories and structures that play a foundational role in
university courses, mathematical Wikipedia pages and {\it Lean}'s Mathlib \cite{Mathlib2020}.

\clearpage
\begin{figure}[htbp]
    \centering
    \def\treescale{1}\input{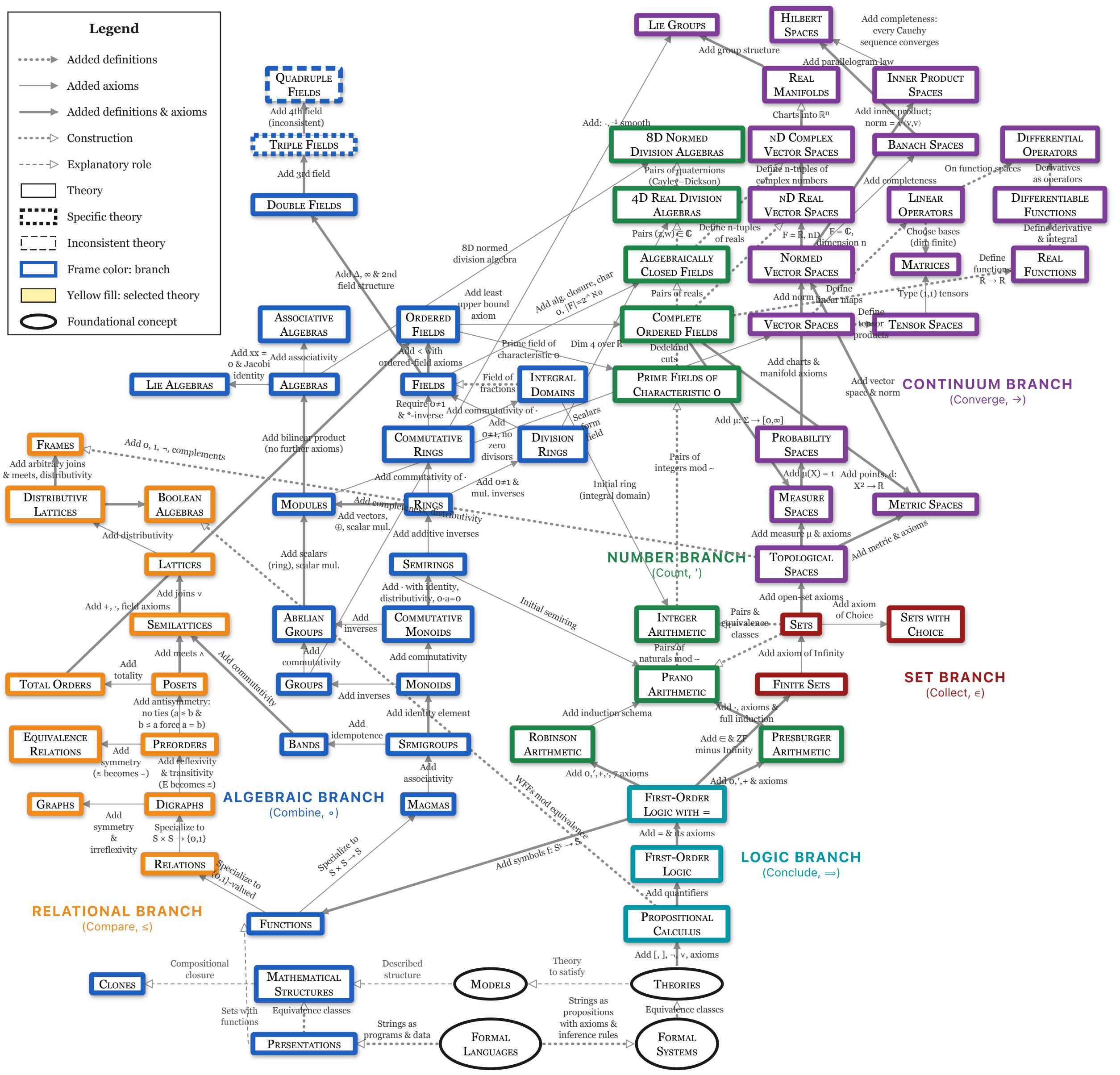} %
    \caption{Family tree of mathematical theories.
    If you are reading this on a computer, you can click any of the boxes for details.
    You will find a fully interactive version at \url{https://tegmark.org/mathtree}.
    Boxes denote mathematical theories (equivalence classes of formal systems). 
    Ovals, and the boxes \bx{Presentations}, \bx{Mathematical Structures} and \bx{Clones}, denote foundational concepts that are not themselves theories.
    }
    \label{TheoryTreeFig}
\end{figure}
\clearpage

\subsection{Mathematical theories}

Figure~\ref{TheoryTreeFig} shows our family tree\footnote{Technically, it is not a tree but a rooted directed acyclic graph --- but the same can be true of conventional family trees.} of basic mathematical theories, defined below as equivalence classes of formal systems that 
imply each other's axioms (after invertible translation).
Each box is described in the online theory atlas at \url{https://tegmark.org/mathtree}, and various mathematical structures that
are models of the theory are described in the online model atlas there.
This tree graphically extends the 1997 family tree from \cite{Tegmark9704009}, and the arrows agree with Mathlib's ``extends" constructions.

The arrows and lines connecting the boxes represent various types of relationships between theories, such as 
{\it specialization} (added axioms, so same symbols, more theorems), 
{\it definitional extension} (added symbols and definitions, so no new theorems about the old symbols) and
{\it construction} (the child's universe is built from the parent's via, \eg, pairs, quotients or cuts; models are not inherited along these arrows).
We refer to a mathematical theory as a {\it child} of another if it is a specialization, definitional extension or both
(these parent-child relations have a gray rather than white arrowhead in the figure).
Every model of the child is a model of the parent: forget the new symbols and the parent's axioms still hold.

While the full tree is infinite, its lower parts are seen to naturally separate into six main branches:
\begin{itemize}
\item {\bf Relational branch}, centered around binary Boolean functions $S\times S\to\{0,1\}$ interpreted as comparisons
\item {\bf Algebraic branch}, centered around binary operations $S\times S\to S$ interpreted as combining two elements
\item {\bf Logic branch}, centered around deduction rules for strings interpreted as well-formed formulas and theorems
\item {\bf Number branch}, centered around entities with arithmetic operations
\item {\bf Set branch}, centered around collecting elements into sets and subsets
\item {\bf Continuum branch}, centered around continuous spaces where limits can be taken
\end{itemize}
They can be alliteratively summarized as the ``compare'', ``combine'', ``conclude'', ``count'', ``collect'' and ``converge'' branches or, more succinctly, by their foundational symbols $\le$, $\circ$, $\implies$, $'$, $\in$ and $\to$, respectively.
We have grouped theories into branches according to what their typical models are, not by the language of their axioms: 
for example, we color \bx{Vector Spaces} purple (the Continuum branch) even though their axioms are algebraic.

\clearpage
\subsection{Mathematical structures and terminology}
\label{TerminologySubsec}

We must not conflate the map with the territory, \ie, the description with that which it describes.
Just as physics theories are distinct from the physical systems that they (approximately) describe,
mathematical theories must not be conflated with the mathematical structures that they (exactly) describe:
\begin{center}
\begin{tabular}{|l|l|}
\hline
Description			&Described\\
\hline
Map					&Territory\\
Physics theories		&Physical systems\\
Mathematical theories	&Mathematical structures\\
\hline
\end{tabular}
\end{center}

For both theories and structures, we also wish to distinguish between their mathematical content and arbitrary notational choices used to specify them.
We will therefore use the terminology of \tab{TheoryStructureTable}, as illustrated by the following examples:
\begin{itemize}
\item the sphere $S^2$ is a single mathematical structure regardless of which of the equivalent coordinate system atlases in \fig{DescriptionEquivalenceFig} is used to define it as a manifold;
\item the 4-element cyclic group $C_4$ is a single mathematical structure regardless of which of the equivalent group multiplication tables (Cayley tables) from \fig{DescriptionEquivalenceFig} is used to specify it;
\item Propositional Calculus is a single mathematical theory regardless of which of the equivalent formal systems is used to define it: the usual four axioms for NOT and OR or Nicod's single axiom for the Sheffer stroke (NAND).
\end{itemize}

\begin{figure}[H]
   \begin{center}
   \begin{center}
\setlength{\fboxsep}{0pt}%
\fbox{\begin{minipage}{0.9\textwidth}
\vspace*{10pt}
{\centering \textbf{The group $C_4$}\\[8pt]
\begin{tabular}[t]{@{}l@{}}
\begin{tikzpicture}[x=0.42cm,y=0.42cm,baseline=(current bounding box.north),every node/.style={font=\sffamily\bfseries,inner sep=0pt}]
\node at (0.5,-0.5) {$0$};
\node at (1.5,-0.5) {$1$};
\node at (2.5,-0.5) {$2$};
\node at (3.5,-0.5) {$3$};
\node at (0.5,-1.5) {$1$};
\node at (1.5,-1.5) {$2$};
\node at (2.5,-1.5) {$3$};
\node at (3.5,-1.5) {$0$};
\node at (0.5,-2.5) {$2$};
\node at (1.5,-2.5) {$3$};
\node at (2.5,-2.5) {$0$};
\node at (3.5,-2.5) {$1$};
\node at (0.5,-3.5) {$3$};
\node at (1.5,-3.5) {$0$};
\node at (2.5,-3.5) {$1$};
\node at (3.5,-3.5) {$2$};
\draw[black!25,line width=0.3pt] (1,0)--(1,-4);
\draw[black!25,line width=0.3pt] (0,-1)--(4,-1);
\draw[black!25,line width=0.3pt] (2,0)--(2,-4);
\draw[black!25,line width=0.3pt] (0,-2)--(4,-2);
\draw[black!25,line width=0.3pt] (3,0)--(3,-4);
\draw[black!25,line width=0.3pt] (0,-3)--(4,-3);
\draw[black,line width=0.9pt] (0,0) rectangle (4,-4);
\end{tikzpicture}\end{tabular}\hspace{1.1cm}%
\begin{tabular}[t]{@{}l@{}}
\begin{tikzpicture}[x=0.42cm,y=0.42cm,baseline=(current bounding box.north),every node/.style={font=\sffamily\bfseries,inner sep=0pt}]
\node at (0.5,-0.5) {$0$};
\node at (1.5,-0.5) {$1$};
\node at (2.5,-0.5) {$2$};
\node at (3.5,-0.5) {$3$};
\node at (0.5,-1.5) {$1$};
\node at (1.5,-1.5) {$3$};
\node at (2.5,-1.5) {$0$};
\node at (3.5,-1.5) {$2$};
\node at (0.5,-2.5) {$2$};
\node at (1.5,-2.5) {$0$};
\node at (2.5,-2.5) {$3$};
\node at (3.5,-2.5) {$1$};
\node at (0.5,-3.5) {$3$};
\node at (1.5,-3.5) {$2$};
\node at (2.5,-3.5) {$1$};
\node at (3.5,-3.5) {$0$};
\draw[black!25,line width=0.3pt] (1,0)--(1,-4);
\draw[black!25,line width=0.3pt] (0,-1)--(4,-1);
\draw[black!25,line width=0.3pt] (2,0)--(2,-4);
\draw[black!25,line width=0.3pt] (0,-2)--(4,-2);
\draw[black!25,line width=0.3pt] (3,0)--(3,-4);
\draw[black!25,line width=0.3pt] (0,-3)--(4,-3);
\draw[black,line width=0.9pt] (0,0) rectangle (4,-4);
\end{tikzpicture}\end{tabular}\hspace{1.1cm}%
\begin{tabular}[t]{@{}l@{}}
\begin{tikzpicture}[x=0.42cm,y=0.42cm,baseline=(current bounding box.north),every node/.style={font=\sffamily\bfseries,inner sep=0pt}]
\node at (0.5,-0.5) {$0$};
\node at (1.5,-0.5) {$1$};
\node at (2.5,-0.5) {$2$};
\node at (3.5,-0.5) {$3$};
\node at (0.5,-1.5) {$1$};
\node at (1.5,-1.5) {$0$};
\node at (2.5,-1.5) {$3$};
\node at (3.5,-1.5) {$2$};
\node at (0.5,-2.5) {$2$};
\node at (1.5,-2.5) {$3$};
\node at (2.5,-2.5) {$1$};
\node at (3.5,-2.5) {$0$};
\node at (0.5,-3.5) {$3$};
\node at (1.5,-3.5) {$2$};
\node at (2.5,-3.5) {$0$};
\node at (3.5,-3.5) {$1$};
\draw[black!25,line width=0.3pt] (1,0)--(1,-4);
\draw[black!25,line width=0.3pt] (0,-1)--(4,-1);
\draw[black!25,line width=0.3pt] (2,0)--(2,-4);
\draw[black!25,line width=0.3pt] (0,-2)--(4,-2);
\draw[black!25,line width=0.3pt] (3,0)--(3,-4);
\draw[black!25,line width=0.3pt] (0,-3)--(4,-3);
\draw[black,line width=0.9pt] (0,0) rectangle (4,-4);
\end{tikzpicture}\end{tabular}\par}
\medskip\hrule\medskip
{\centering \textbf{The 2-sphere $S^2$}\\[6pt]
\begin{tabular}[t]{@{}c@{}}
\begin{tikzpicture}[scale=0.55]
\useasboundingbox (-1.7,-1.4) rectangle (1.7,1.4);
\draw[line width=0.8pt] (0,0) circle (1);
\draw[black!55,dashed] (1,0) arc (0:180:1 and 0.32);
\draw[black!55] (-1,0) arc (180:360:1 and 0.32);
\fill (0,1) circle (1.6pt); \node[above,font=\scriptsize] at (0,1) {$N$};
\fill (0,-1) circle (1.6pt); \node[below,font=\scriptsize] at (0,-1) {$S$};
\end{tikzpicture}\\[2pt]
$\varphi_N(x,y,z)=\bigl(\tfrac{x}{1-z},\,\tfrac{y}{1-z}\bigr)$\\[5pt]
$\varphi_S(x,y,z)=\bigl(\tfrac{x}{1+z},\,\tfrac{y}{1+z}\bigr)$
\end{tabular}\hspace{2.2cm}%
\begin{tabular}[t]{@{}c@{}}
\begin{tikzpicture}[scale=0.55]
\useasboundingbox (-1.7,-1.4) rectangle (1.7,1.4);
\draw[line width=0.8pt] (0,0) circle (1);
\draw[black!55,dashed] (0,1) arc (90:270:0.32 and 1);
\draw[black!55] (0,-1) arc (270:450:0.32 and 1);
\fill (1,0) circle (1.6pt); \node[right,font=\scriptsize] at (1,0) {$E$};
\fill (-1,0) circle (1.6pt); \node[left,font=\scriptsize] at (-1,0) {$W$};
\end{tikzpicture}\\[2pt]
$\varphi_E(x,y,z)=\bigl(\tfrac{y}{1-x},\,\tfrac{z}{1-x}\bigr)$\\[5pt]
$\varphi_W(x,y,z)=\bigl(\tfrac{y}{1+x},\,\tfrac{z}{1+x}\bigr)$
\end{tabular}\par\medskip}
\vspace*{10pt}
\end{minipage}}
\end{center}
   \end{center}
   \caption{Equivalent specifications of the cyclic group of 4 elements (top) and the manifold $S^2$ (the sphere, bottom). We consider equivalent specifications as
   defining the same mathematical structure (here $C_4$ and $S^2$).
  The second and third group multiplication tables are related to the first by relabeling elements $2\leftrightarrow 3$ and $1\leftrightarrow 2$, respectively.
  Each chart $\varphi_i$ is defined on the whole unit sphere except at its projection point ($z=1$, $z=-1$, $x=1$ and $x=-1$, respectively).
   }
  \label{DescriptionEquivalenceFig}
\end{figure}

\begin{table}[H]
\begin{center}
\begin{footnotesize}
\begin{tabular}{|l|l|l|l|}
\hline
					&{\bf Mathematical theory}	&{\bf Mathematical structure}	&{\bf Closed mathematical }\\
					&						&						&{\bf structure (clone)}\\
\hhline{|=|=|=|=|}

{\bf Defined by}			&Formal system			&Presentation				&Presentation\\\hline
{\bf Equivalence} 		&Prove each other's axioms	&Relabeling				&Generate each other's\\ 
{\bf definition}			&(after invertible translation)	&						&generator functions\\\hline
{\bf Ingredients}			&Strings viewed			&Strings viewed			&Strings viewed\\
					&as propositions			&as programs				&as programs\\\hline
{\bf Starting point}		&Axioms					&Functions				&Functions\\\hline
{\bf Expansion method}	&Theorem deduction			&None					&Function composition\\
\hline
\end{tabular}
\end{footnotesize}
\end{center}
\caption{Our terminology for mathematical theories and structures. A translation is invertible if translating a formula there and back gives a formula provably equivalent to the original.
}
\label{TheoryStructureTable}
\end{table}

Specifically, we use the standard definition of a {\it formal system} in terms of a {\it specific} choice of symbols, axioms and deduction rules, and define a mathematical theory as an equivalence class of formal systems that prove each other's axioms after invertible translation.
This equivalence class determines the deductive closure, the (normally infinite) set of theorems deducible from any formal system in the class;
we let ``mathematical theory'' denote this class, while model theorists often use that word to denote that deductive closure.

Analogously, we define a {\it presentation} as a {\it specific} choice of functions on sets, and define a mathematical structure as an equivalence class of presentations under relabeling (two isomorphic presentations are deemed equivalent). 
This is a deliberate departure from the common model-theory nomenclature ``structure'' for what we call a presentation.
Our definition is the same as the one Carnap used in 1936 \cite{CarnapBachmann1936}, as discussed by Schiemer \cite[pp.~388--390]{Schiemer2020}.
It is more in line with structuralist mathematical views \cite{Bourbaki1950,Hilbert1918,MacLane1986,Corry1996,Gowers2008}
and with everyday usage by working mathematicians. 
For example, almost no mathematicians would say that the three Cayley tables in \fig{DescriptionEquivalenceFig} specify different groups, or that
the two shown function pairs specify different manifolds (which are indeed defined as equivalence classes of compatible atlases). Most mathematicians would probably also 
say that the 2-sphere (or the group $C_4$) counts as a mathematical structure.
According to our definitions, each group (counted in OEIS sequence A000001 \cite{OEIS}) is {\it one} mathematical structure.

\subsection{Presentations and models}

A presentation is said to be a {\it model} of a formal system if there exists an interpretation of each symbol of the formal system in terms of the sets and functions of the presentation such that every theorem holds in the presentation under the interpretation. 
In turn, a mathematical structure is said to be a {\it model} of a mathematical theory if a presentation in the equivalence class of the former is a model of a formal system in the equivalence class of the latter. 
All the model relations shown in the family tree are verified in Lean.
We will elaborate on the notion of a {\it closed mathematical structure} in \Sec{StructureSec}.

For {\it formal systems}, a very mature toolkit exists in the form of specialized computer languages such as 
Lean \cite{Lean4}, Rocq \cite{Rocq} and Isabelle \cite{Isabelle}. 
For {\it presentations}, less standardization has occurred, 
so we make a major effort in this direction.
Section~\ref{PresentationSec} presents our system for defining finite and countably infinite presentations as computations.
In \Sec{MLISPSec}, in the spirit of constructivist mathematics \cite{Bishop1967}, 
we define for this purpose a Turing-complete programming language {\it MLISP} that is dramatically simpler than traditional programming or proof languages, yet is able to define all functions in our presentations in simple halting 
programs that are often fewer than 10 characters long. A program implementing and running MLISP can be written in merely
195 characters --- in MLISP. The advantage here is that one can prove things 
about presentations without having to make assumptions about how traditional programming or proof languages implement their functions. For small finite sets, MLISP functions are also shown as human-friendly lookup tables.

The rest of this paper is organized as follows.
We describe how we define presentations and computations in \Sec{PresentationSec} and mathematical structures in \Sec{StructureSec}. 
\Sec{StructureSec} also relates this to physics, \Sec{ConclusionsSec} summarizes our conclusions and Appendix~\ref{CountingAppendix} describes how we counted closed mathematical structures.
An interactive version of the family tree of mathematical theories and structures is available at \url{https://tegmark.org/mathtree}.  
All MLISP and Lean code is at \url{https://github.com/mtegmark/mlisp9}.

\section{Computationally defining presentations}\label{PresentationSec}

In this section, we clarify how we define the left base of the tree (presentations, mathematical structures), 
since there is much less convergence on that in the mathematics literature than there is for the right part 
(formal systems, theories). 
We take a constructivist computational approach, since more complex languages such as {\it Lean} are massive overkill for this task.
Instead we use Lean to verify MLISP.
Reducing the trusted computing base not only makes foundations more human-understandable, 
but also reduces vulnerability to bugs in programming languages that are used to define presentations.
Numerous bugs have been found in various C compilers, and a bug was recently found in Lean itself that enabled proving false
propositions~\cite{LeanBug}.

\subsection{Presentations}\label{str:pres}
\noindent\textbf{Definition:} A presentation is a collection of sets and functions: sets $S_1,\dots,S_n$ together with functions $f_a\colon S_{i_1}\times\dots\times S_{i_k}\to S_j$ between them.
We refer to  $\{0,1\}$-valued functions in the presentation as {\it relations} or {\it Boolean functions}.
In other words,  a presentation is a concrete object, sets with functions defined by lookup tables or computations, before any axioms are imposed and before equivalent descriptions of the same thing are identified.  

\subsection{Finite presentations}

For {\it finite presentations} (when all sets are finite), we simply need to specify each set  followed by lookup tables defining each of its functions.
For example, we could specify the presentation of the Klein 4-group like this:
\begin{center}
\begin{minipage}{0.9\linewidth}
\textbf{1 set:}\quad $S_1=\{e,a,b,c\}$

\medskip
\textbf{3 functions:}
\begin{eqnarray*}
\circ&:& S_1\times S_1\to S_1 \\
{}^{-1}&:& S_1\to S_1 \\
e & \in& S_1
\end{eqnarray*}
\textbf{Lookup tables:}
\begin{center}\renewcommand{\arraystretch}{1.15}
\begin{tabular}{|c|cccc|}\hline $g\circ h$ & $e$ & $a$ & $b$ & $c$\\\hline $e$ & $e$ & $a$ & $b$ & $c$\\ $a$ & $a$ & $e$ & $c$ & $b$\\ $b$ & $b$ & $c$ & $e$ & $a$\\ $c$ & $c$ & $b$ & $a$ & $e$\\\hline\end{tabular}\qquad \begin{tabular}{|c|c|}\hline & $g^{-1}$\\\hline $e$ & $e$\\ $a$ & $a$\\ $b$ & $b$\\ $c$ & $c$\\\hline\end{tabular}\qquad \begin{tabular}{|c|}\hline $e$\\\hline $e$\\\hline\end{tabular}
\end{center}
\end{minipage}
\end{center}
However, we are keen to standardize the notation, simplify away as many arbitrary notational choices as possible and draw focus to the actual mathematical properties of the presentation.
We therefore relabel the elements of all finite sets as $\{0,1,\dots\}$, drop inferable labels and 
specify each presentation by standardized human-friendly {\it presentation tableaus} as in these examples:

\begin{center}
\noindent\setlength{\fboxsep}{6pt}\colorbox{galbeige}{\begin{minipage}{\dimexpr\textwidth-2\fboxsep-2\fboxrule\relax}\raggedright
\begin{tabular}[t]{@{}l@{}}\small Klein 4-group:\\[-6pt]\input{8FINITE_FIG/v4}\end{tabular}\hspace*{\fill}\begin{tabular}[t]{@{}l@{}}\small Total order of $\{0,\dots,5\}$:\\[-6pt]\begin{tikzpicture}[x=0.42cm,y=0.42cm,baseline=(current bounding box.north),every node/.style={font=\sffamily\bfseries,inner sep=0pt}]
\node[anchor=base west,font=\small\normalfont] at (0,1.45) {$6$\ \ \textcolor{red}{$2$}};
\node[text=red] at (0.80,-0.50) {$1$};
\node[text=red] at (1.80,-0.50) {$1$};
\node[text=red] at (2.80,-0.50) {$1$};
\node[text=red] at (3.80,-0.50) {$1$};
\node[text=red] at (4.80,-0.50) {$1$};
\node[text=red] at (5.80,-0.50) {$1$};
\node[text=red] at (0.80,-1.50) {$0$};
\node[text=red] at (1.80,-1.50) {$1$};
\node[text=red] at (2.80,-1.50) {$1$};
\node[text=red] at (3.80,-1.50) {$1$};
\node[text=red] at (4.80,-1.50) {$1$};
\node[text=red] at (5.80,-1.50) {$1$};
\node[text=red] at (0.80,-2.50) {$0$};
\node[text=red] at (1.80,-2.50) {$0$};
\node[text=red] at (2.80,-2.50) {$1$};
\node[text=red] at (3.80,-2.50) {$1$};
\node[text=red] at (4.80,-2.50) {$1$};
\node[text=red] at (5.80,-2.50) {$1$};
\node[text=red] at (0.80,-3.50) {$0$};
\node[text=red] at (1.80,-3.50) {$0$};
\node[text=red] at (2.80,-3.50) {$0$};
\node[text=red] at (3.80,-3.50) {$1$};
\node[text=red] at (4.80,-3.50) {$1$};
\node[text=red] at (5.80,-3.50) {$1$};
\node[text=red] at (0.80,-4.50) {$0$};
\node[text=red] at (1.80,-4.50) {$0$};
\node[text=red] at (2.80,-4.50) {$0$};
\node[text=red] at (3.80,-4.50) {$0$};
\node[text=red] at (4.80,-4.50) {$1$};
\node[text=red] at (5.80,-4.50) {$1$};
\node[text=red] at (0.80,-5.50) {$0$};
\node[text=red] at (1.80,-5.50) {$0$};
\node[text=red] at (2.80,-5.50) {$0$};
\node[text=red] at (3.80,-5.50) {$0$};
\node[text=red] at (4.80,-5.50) {$0$};
\node[text=red] at (5.80,-5.50) {$1$};
\draw[black!25,line width=0.3pt] (1.30,0)--(1.30,-6.00);
\draw[black!25,line width=0.3pt] (2.30,0)--(2.30,-6.00);
\draw[black!25,line width=0.3pt] (3.30,0)--(3.30,-6.00);
\draw[black!25,line width=0.3pt] (4.30,0)--(4.30,-6.00);
\draw[black!25,line width=0.3pt] (5.30,0)--(5.30,-6.00);
\draw[black!25,line width=0.3pt] (0.30,-1.00)--(6.30,-1.00);
\draw[black!25,line width=0.3pt] (0.30,-2.00)--(6.30,-2.00);
\draw[black!25,line width=0.3pt] (0.30,-3.00)--(6.30,-3.00);
\draw[black!25,line width=0.3pt] (0.30,-4.00)--(6.30,-4.00);
\draw[black!25,line width=0.3pt] (0.30,-5.00)--(6.30,-5.00);
\draw[black,line width=0.9pt] (0.30,0) rectangle (6.30,-6.00);
\node[anchor=south,font=\small\normalfont] at (3.30,0.3) {$\le$};
\draw[black!60,line width=0.5pt] (-0.3,1.05)--(6.60,1.05);
\draw[black,line width=1pt] (-0.3,-6.40) rectangle (6.60,2.2);
\end{tikzpicture}\end{tabular}\hspace*{\fill}\begin{tabular}[t]{@{}l@{}}\small Transformation monoid $T_2$:\\[-6pt]\input{8FINITE_FIG/t2}\end{tabular}\par\medskip
\begin{tabular}[t]{@{}l@{}}\small Frame of open sets of 3-point space:\\[-6pt]\input{8FINITE_FIG/fr5}\end{tabular}\hspace*{\fill}\begin{tabular}[t]{@{}l@{}}\small Rock--paper--\\[-1pt]\small scissors magma:\\[-6pt]\input{8FINITE_FIG/rps}\end{tabular}\hspace*{\fill}\begin{tabular}[t]{@{}l@{}}\small Diamond lattice $M_3$:\\[-6pt]\input{8FINITE_FIG/m3}\end{tabular}\par\medskip
\begin{tabular}[t]{@{}l@{}}\small Ring $\Z/4$:\\[2pt]\input{8FINITE_FIG/z4}\end{tabular}\hspace*{\fill}\begin{tabular}[t]{@{}l@{}}\small Double field $D_4$:\\[-6pt]\input{8FINITE_FIG/df4}\end{tabular}\par
\end{minipage}}   %
\end{center}

\bigskip
\noindent
The cardinality of each set is written in a separate color in the set specification area, above the horizontal line.
Note that the arity of each function can be inferred from the shape of its lookup table, and the identities of the input and output sets
can be inferred from the colors of the lookup table boxes and the digits therein. The operation labels (\eg\ $e$, $\circ$) are redundant, not part of the presentation itself, 
but we can optionally leave them in to make the tableaus more human-readable.

\subsection{Countably infinite presentations}

For presentations where at least one set is countably infinite, we could try to visualize them the same way, for example, by enumerating 
integers in the order  $0,1,-1,2,-2,\dots$ and  rationals in a signed extension of the Calkin--Wilf enumeration \cite{CalkinWilf2000}, $0,1,-1,\tfrac12,-\tfrac12,\dots$ (zero first, then each positive rational in Calkin--Wilf order, immediately followed by its negative):
 
\medskip
\begin{center}
\noindent\adjustbox{max width=\textwidth, max totalheight=0.8\textheight}{\setlength{\fboxsep}{6pt}\colorbox{galbeige}{\begin{minipage}{\dimexpr\textwidth-2\fboxsep-2\fboxrule\relax}\raggedright
\begin{tabular}[t]{@{}l@{}}\small Natural numbers $\N$ as Peano Arithmetic:\\[-6pt]\begin{tikzpicture}[x=0.42cm,y=0.42cm,baseline=(current bounding box.north),every node/.style={font=\sffamily\bfseries,inner sep=0pt}]
\node[anchor=base west,font=\small\normalfont] at (0,1.45) {\textcolor{black}{$\aleph_0$}};
\node[text=black] at (0.5,-0.5) {$0$};
\node[text=black] at (1.5,-0.5) {$1$};
\node[text=black] at (2.5,-0.5) {$2$};
\node[text=black] at (3.5,-0.5) {$3$};
\node[text=black] at (4.5,-0.5) {$4$};
\node[text=black] at (0.5,-1.5) {$1$};
\node[text=black] at (1.5,-1.5) {$2$};
\node[text=black] at (2.5,-1.5) {$3$};
\node[text=black] at (3.5,-1.5) {$4$};
\node[text=black] at (4.5,-1.5) {$5$};
\node[text=black] at (0.5,-2.5) {$2$};
\node[text=black] at (1.5,-2.5) {$3$};
\node[text=black] at (2.5,-2.5) {$4$};
\node[text=black] at (3.5,-2.5) {$5$};
\node[text=black] at (4.5,-2.5) {$6$};
\node[text=black] at (0.5,-3.5) {$3$};
\node[text=black] at (1.5,-3.5) {$4$};
\node[text=black] at (2.5,-3.5) {$5$};
\node[text=black] at (3.5,-3.5) {$6$};
\node[text=black] at (4.5,-3.5) {$7$};
\node[text=black] at (0.5,-4.5) {$4$};
\node[text=black] at (1.5,-4.5) {$5$};
\node[text=black] at (2.5,-4.5) {$6$};
\node[text=black] at (3.5,-4.5) {$7$};
\node[text=black] at (4.5,-4.5) {$8$};
\fill[black] (5.28,-0.5) circle (0.75pt);
\fill[black] (5.5,-0.5) circle (0.75pt);
\fill[black] (5.72,-0.5) circle (0.75pt);
\fill[black] (5.28,-1.5) circle (0.75pt);
\fill[black] (5.5,-1.5) circle (0.75pt);
\fill[black] (5.72,-1.5) circle (0.75pt);
\fill[black] (5.28,-2.5) circle (0.75pt);
\fill[black] (5.5,-2.5) circle (0.75pt);
\fill[black] (5.72,-2.5) circle (0.75pt);
\fill[black] (5.28,-3.5) circle (0.75pt);
\fill[black] (5.5,-3.5) circle (0.75pt);
\fill[black] (5.72,-3.5) circle (0.75pt);
\fill[black] (5.28,-4.5) circle (0.75pt);
\fill[black] (5.5,-4.5) circle (0.75pt);
\fill[black] (5.72,-4.5) circle (0.75pt);
\fill[black] (0.5,-5.28) circle (0.75pt);
\fill[black] (0.5,-5.5) circle (0.75pt);
\fill[black] (0.5,-5.72) circle (0.75pt);
\fill[black] (1.5,-5.28) circle (0.75pt);
\fill[black] (1.5,-5.5) circle (0.75pt);
\fill[black] (1.5,-5.72) circle (0.75pt);
\fill[black] (2.5,-5.28) circle (0.75pt);
\fill[black] (2.5,-5.5) circle (0.75pt);
\fill[black] (2.5,-5.72) circle (0.75pt);
\fill[black] (3.5,-5.28) circle (0.75pt);
\fill[black] (3.5,-5.5) circle (0.75pt);
\fill[black] (3.5,-5.72) circle (0.75pt);
\fill[black] (4.5,-5.28) circle (0.75pt);
\fill[black] (4.5,-5.5) circle (0.75pt);
\fill[black] (4.5,-5.72) circle (0.75pt);
\fill[black] (5.28,-5.72) circle (0.75pt);
\fill[black] (5.5,-5.5) circle (0.75pt);
\fill[black] (5.72,-5.28) circle (0.75pt);
\draw[black!25,line width=0.3pt] (1,0)--(1,-6);
\draw[black!25,line width=0.3pt] (2,0)--(2,-6);
\draw[black!25,line width=0.3pt] (3,0)--(3,-6);
\draw[black!25,line width=0.3pt] (4,0)--(4,-6);
\draw[black!25,line width=0.3pt] (5,0)--(5,-6);
\draw[black!25,line width=0.3pt] (0,-1)--(6,-1);
\draw[black!25,line width=0.3pt] (0,-2)--(6,-2);
\draw[black!25,line width=0.3pt] (0,-3)--(6,-3);
\draw[black!25,line width=0.3pt] (0,-4)--(6,-4);
\draw[black!25,line width=0.3pt] (0,-5)--(6,-5);
\draw[black,line width=0.9pt] (0,0) rectangle (6,-6);
\node[anchor=south,font=\small\normalfont] at (3,0.3) {$+$};
\node[text=black] at (7.05,-0.5) {$0$};
\node[text=black] at (8.45,-0.5) {$0$};
\node[text=black] at (9.85,-0.5) {$0$};
\node[text=black] at (11.25,-0.5) {$0$};
\node[text=black] at (12.65,-0.5) {$0$};
\node[text=black] at (7.05,-1.5) {$0$};
\node[text=black] at (8.45,-1.5) {$1$};
\node[text=black] at (9.85,-1.5) {$2$};
\node[text=black] at (11.25,-1.5) {$3$};
\node[text=black] at (12.65,-1.5) {$4$};
\node[text=black] at (7.05,-2.5) {$0$};
\node[text=black] at (8.45,-2.5) {$2$};
\node[text=black] at (9.85,-2.5) {$4$};
\node[text=black] at (11.25,-2.5) {$6$};
\node[text=black] at (12.65,-2.5) {$8$};
\node[text=black] at (7.05,-3.5) {$0$};
\node[text=black] at (8.45,-3.5) {$3$};
\node[text=black] at (9.85,-3.5) {$6$};
\node[text=black] at (11.25,-3.5) {$9$};
\node[text=black] at (12.65,-3.5) {$12$};
\node[text=black] at (7.05,-4.5) {$0$};
\node[text=black] at (8.45,-4.5) {$4$};
\node[text=black] at (9.85,-4.5) {$8$};
\node[text=black] at (11.25,-4.5) {$12$};
\node[text=black] at (12.65,-4.5) {$16$};
\fill[black] (13.83,-0.5) circle (0.75pt);
\fill[black] (14.05,-0.5) circle (0.75pt);
\fill[black] (14.27,-0.5) circle (0.75pt);
\fill[black] (13.83,-1.5) circle (0.75pt);
\fill[black] (14.05,-1.5) circle (0.75pt);
\fill[black] (14.27,-1.5) circle (0.75pt);
\fill[black] (13.83,-2.5) circle (0.75pt);
\fill[black] (14.05,-2.5) circle (0.75pt);
\fill[black] (14.27,-2.5) circle (0.75pt);
\fill[black] (13.83,-3.5) circle (0.75pt);
\fill[black] (14.05,-3.5) circle (0.75pt);
\fill[black] (14.27,-3.5) circle (0.75pt);
\fill[black] (13.83,-4.5) circle (0.75pt);
\fill[black] (14.05,-4.5) circle (0.75pt);
\fill[black] (14.27,-4.5) circle (0.75pt);
\fill[black] (7.05,-5.28) circle (0.75pt);
\fill[black] (7.05,-5.5) circle (0.75pt);
\fill[black] (7.05,-5.72) circle (0.75pt);
\fill[black] (8.45,-5.28) circle (0.75pt);
\fill[black] (8.45,-5.5) circle (0.75pt);
\fill[black] (8.45,-5.72) circle (0.75pt);
\fill[black] (9.85,-5.28) circle (0.75pt);
\fill[black] (9.85,-5.5) circle (0.75pt);
\fill[black] (9.85,-5.72) circle (0.75pt);
\fill[black] (11.25,-5.28) circle (0.75pt);
\fill[black] (11.25,-5.5) circle (0.75pt);
\fill[black] (11.25,-5.72) circle (0.75pt);
\fill[black] (12.65,-5.28) circle (0.75pt);
\fill[black] (12.65,-5.5) circle (0.75pt);
\fill[black] (12.65,-5.72) circle (0.75pt);
\fill[black] (13.83,-5.72) circle (0.75pt);
\fill[black] (14.05,-5.5) circle (0.75pt);
\fill[black] (14.27,-5.28) circle (0.75pt);
\draw[black!25,line width=0.3pt] (7.75,0)--(7.75,-6);
\draw[black!25,line width=0.3pt] (9.15,0)--(9.15,-6);
\draw[black!25,line width=0.3pt] (10.55,0)--(10.55,-6);
\draw[black!25,line width=0.3pt] (11.95,0)--(11.95,-6);
\draw[black!25,line width=0.3pt] (13.35,0)--(13.35,-6);
\draw[black!25,line width=0.3pt] (6.35,-1)--(14.75,-1);
\draw[black!25,line width=0.3pt] (6.35,-2)--(14.75,-2);
\draw[black!25,line width=0.3pt] (6.35,-3)--(14.75,-3);
\draw[black!25,line width=0.3pt] (6.35,-4)--(14.75,-4);
\draw[black!25,line width=0.3pt] (6.35,-5)--(14.75,-5);
\draw[black,line width=0.9pt] (6.35,0) rectangle (14.75,-6);
\node[anchor=south,font=\small\normalfont] at (10.55,0.3) {$\cdot$};
\node[text=black] at (15.6,-0.5) {$1$};
\node[text=black] at (15.6,-1.5) {$2$};
\node[text=black] at (15.6,-2.5) {$3$};
\node[text=black] at (15.6,-3.5) {$4$};
\node[text=black] at (15.6,-4.5) {$5$};
\fill[black] (15.6,-5.28) circle (0.75pt);
\fill[black] (15.6,-5.5) circle (0.75pt);
\fill[black] (15.6,-5.72) circle (0.75pt);
\draw[black!25,line width=0.3pt] (15.1,-1)--(16.1,-1);
\draw[black!25,line width=0.3pt] (15.1,-2)--(16.1,-2);
\draw[black!25,line width=0.3pt] (15.1,-3)--(16.1,-3);
\draw[black!25,line width=0.3pt] (15.1,-4)--(16.1,-4);
\draw[black!25,line width=0.3pt] (15.1,-5)--(16.1,-5);
\draw[black,line width=0.9pt] (15.1,0) rectangle (16.1,-6);
\node[anchor=south,font=\small\normalfont] at (15.6,0.3) {$'$};
\node[text=black] at (16.95,-0.5) {$0$};
\draw[black,line width=0.9pt] (16.45,0) rectangle (17.45,-1);
\node[anchor=south,font=\small\normalfont] at (16.95,0.3) {$0$};
\draw[black!60,line width=0.5pt] (-0.3,1.05)--(17.75,1.05);
\draw[black,line width=1pt] (-0.3,-6.4) rectangle (17.75,2.2);
\end{tikzpicture}\end{tabular}\hspace*{\fill}\begin{tabular}[t]{@{}r@{}}\small Natural number total order $(\N,\le)$:\\[-6pt]\begin{tikzpicture}[x=0.42cm,y=0.42cm,baseline=(current bounding box.north),every node/.style={font=\sffamily\bfseries,inner sep=0pt}]
\node[anchor=base west,font=\small\normalfont] at (0,1.45) {\textcolor{black}{$\aleph_0$}\ \ \textcolor{red}{2}};
\node[text=red] at (0.5,-0.5) {$1$};
\node[text=red] at (1.5,-0.5) {$1$};
\node[text=red] at (2.5,-0.5) {$1$};
\node[text=red] at (3.5,-0.5) {$1$};
\node[text=red] at (4.5,-0.5) {$1$};
\node[text=red] at (0.5,-1.5) {$0$};
\node[text=red] at (1.5,-1.5) {$1$};
\node[text=red] at (2.5,-1.5) {$1$};
\node[text=red] at (3.5,-1.5) {$1$};
\node[text=red] at (4.5,-1.5) {$1$};
\node[text=red] at (0.5,-2.5) {$0$};
\node[text=red] at (1.5,-2.5) {$0$};
\node[text=red] at (2.5,-2.5) {$1$};
\node[text=red] at (3.5,-2.5) {$1$};
\node[text=red] at (4.5,-2.5) {$1$};
\node[text=red] at (0.5,-3.5) {$0$};
\node[text=red] at (1.5,-3.5) {$0$};
\node[text=red] at (2.5,-3.5) {$0$};
\node[text=red] at (3.5,-3.5) {$1$};
\node[text=red] at (4.5,-3.5) {$1$};
\node[text=red] at (0.5,-4.5) {$0$};
\node[text=red] at (1.5,-4.5) {$0$};
\node[text=red] at (2.5,-4.5) {$0$};
\node[text=red] at (3.5,-4.5) {$0$};
\node[text=red] at (4.5,-4.5) {$1$};
\fill[red] (5.28,-0.5) circle (0.75pt);
\fill[red] (5.5,-0.5) circle (0.75pt);
\fill[red] (5.72,-0.5) circle (0.75pt);
\fill[red] (5.28,-1.5) circle (0.75pt);
\fill[red] (5.5,-1.5) circle (0.75pt);
\fill[red] (5.72,-1.5) circle (0.75pt);
\fill[red] (5.28,-2.5) circle (0.75pt);
\fill[red] (5.5,-2.5) circle (0.75pt);
\fill[red] (5.72,-2.5) circle (0.75pt);
\fill[red] (5.28,-3.5) circle (0.75pt);
\fill[red] (5.5,-3.5) circle (0.75pt);
\fill[red] (5.72,-3.5) circle (0.75pt);
\fill[red] (5.28,-4.5) circle (0.75pt);
\fill[red] (5.5,-4.5) circle (0.75pt);
\fill[red] (5.72,-4.5) circle (0.75pt);
\fill[red] (0.5,-5.28) circle (0.75pt);
\fill[red] (0.5,-5.5) circle (0.75pt);
\fill[red] (0.5,-5.72) circle (0.75pt);
\fill[red] (1.5,-5.28) circle (0.75pt);
\fill[red] (1.5,-5.5) circle (0.75pt);
\fill[red] (1.5,-5.72) circle (0.75pt);
\fill[red] (2.5,-5.28) circle (0.75pt);
\fill[red] (2.5,-5.5) circle (0.75pt);
\fill[red] (2.5,-5.72) circle (0.75pt);
\fill[red] (3.5,-5.28) circle (0.75pt);
\fill[red] (3.5,-5.5) circle (0.75pt);
\fill[red] (3.5,-5.72) circle (0.75pt);
\fill[red] (4.5,-5.28) circle (0.75pt);
\fill[red] (4.5,-5.5) circle (0.75pt);
\fill[red] (4.5,-5.72) circle (0.75pt);
\fill[red] (5.28,-5.72) circle (0.75pt);
\fill[red] (5.5,-5.5) circle (0.75pt);
\fill[red] (5.72,-5.28) circle (0.75pt);
\draw[black!25,line width=0.3pt] (1,0)--(1,-6);
\draw[black!25,line width=0.3pt] (2,0)--(2,-6);
\draw[black!25,line width=0.3pt] (3,0)--(3,-6);
\draw[black!25,line width=0.3pt] (4,0)--(4,-6);
\draw[black!25,line width=0.3pt] (5,0)--(5,-6);
\draw[black!25,line width=0.3pt] (0,-1)--(6,-1);
\draw[black!25,line width=0.3pt] (0,-2)--(6,-2);
\draw[black!25,line width=0.3pt] (0,-3)--(6,-3);
\draw[black!25,line width=0.3pt] (0,-4)--(6,-4);
\draw[black!25,line width=0.3pt] (0,-5)--(6,-5);
\draw[black,line width=0.9pt] (0,0) rectangle (6,-6);
\node[anchor=south,font=\small\normalfont] at (3,0.3) {$\le$};
\draw[black!60,line width=0.5pt] (-0.3,1.05)--(6.3,1.05);
\draw[black,line width=1pt] (-0.3,-6.4) rectangle (6.3,2.2);
\end{tikzpicture}\end{tabular}\par\medskip
\begin{tabular}[t]{@{}l@{}}\small Basic integer arithmetic $(\Z,0,-,+,\cdot)$\\[-1pt]\small enumerated $0,1,-1,2,-2,\dots$:\\[-6pt]\input{11COUNTABLE_FIG/Z}\end{tabular}\hspace*{\fill}\begin{tabular}[t]{@{}r@{}}\small Integer total order $(\Z,\le)$:\\[-6pt]\begin{tikzpicture}[x=0.42cm,y=0.42cm,baseline=(current bounding box.north),every node/.style={font=\sffamily\bfseries,inner sep=0pt}]
\node[anchor=base west,font=\small\normalfont] at (0,1.45) {\textcolor{black}{$\aleph_0$}\ \ \textcolor{red}{2}};
\node[text=red,anchor=east] at (0.88,-0.5) {$1$};
\node[text=red,anchor=east] at (1.88,-0.5) {$1$};
\node[text=red,anchor=east] at (2.88,-0.5) {$0$};
\node[text=red,anchor=east] at (3.88,-0.5) {$1$};
\node[text=red,anchor=east] at (4.88,-0.5) {$0$};
\node[text=red,anchor=east] at (0.88,-1.5) {$0$};
\node[text=red,anchor=east] at (1.88,-1.5) {$1$};
\node[text=red,anchor=east] at (2.88,-1.5) {$0$};
\node[text=red,anchor=east] at (3.88,-1.5) {$1$};
\node[text=red,anchor=east] at (4.88,-1.5) {$0$};
\node[text=red,anchor=east] at (0.88,-2.5) {$1$};
\node[text=red,anchor=east] at (1.88,-2.5) {$1$};
\node[text=red,anchor=east] at (2.88,-2.5) {$1$};
\node[text=red,anchor=east] at (3.88,-2.5) {$1$};
\node[text=red,anchor=east] at (4.88,-2.5) {$0$};
\node[text=red,anchor=east] at (0.88,-3.5) {$0$};
\node[text=red,anchor=east] at (1.88,-3.5) {$0$};
\node[text=red,anchor=east] at (2.88,-3.5) {$0$};
\node[text=red,anchor=east] at (3.88,-3.5) {$1$};
\node[text=red,anchor=east] at (4.88,-3.5) {$0$};
\node[text=red,anchor=east] at (0.88,-4.5) {$1$};
\node[text=red,anchor=east] at (1.88,-4.5) {$1$};
\node[text=red,anchor=east] at (2.88,-4.5) {$1$};
\node[text=red,anchor=east] at (3.88,-4.5) {$1$};
\node[text=red,anchor=east] at (4.88,-4.5) {$1$};
\fill[red] (5.28,-0.5) circle (0.75pt);
\fill[red] (5.5,-0.5) circle (0.75pt);
\fill[red] (5.72,-0.5) circle (0.75pt);
\fill[red] (5.28,-1.5) circle (0.75pt);
\fill[red] (5.5,-1.5) circle (0.75pt);
\fill[red] (5.72,-1.5) circle (0.75pt);
\fill[red] (5.28,-2.5) circle (0.75pt);
\fill[red] (5.5,-2.5) circle (0.75pt);
\fill[red] (5.72,-2.5) circle (0.75pt);
\fill[red] (5.28,-3.5) circle (0.75pt);
\fill[red] (5.5,-3.5) circle (0.75pt);
\fill[red] (5.72,-3.5) circle (0.75pt);
\fill[red] (5.28,-4.5) circle (0.75pt);
\fill[red] (5.5,-4.5) circle (0.75pt);
\fill[red] (5.72,-4.5) circle (0.75pt);
\fill[red] (0.5,-5.28) circle (0.75pt);
\fill[red] (0.5,-5.5) circle (0.75pt);
\fill[red] (0.5,-5.72) circle (0.75pt);
\fill[red] (1.5,-5.28) circle (0.75pt);
\fill[red] (1.5,-5.5) circle (0.75pt);
\fill[red] (1.5,-5.72) circle (0.75pt);
\fill[red] (2.5,-5.28) circle (0.75pt);
\fill[red] (2.5,-5.5) circle (0.75pt);
\fill[red] (2.5,-5.72) circle (0.75pt);
\fill[red] (3.5,-5.28) circle (0.75pt);
\fill[red] (3.5,-5.5) circle (0.75pt);
\fill[red] (3.5,-5.72) circle (0.75pt);
\fill[red] (4.5,-5.28) circle (0.75pt);
\fill[red] (4.5,-5.5) circle (0.75pt);
\fill[red] (4.5,-5.72) circle (0.75pt);
\fill[red] (5.28,-5.72) circle (0.75pt);
\fill[red] (5.5,-5.5) circle (0.75pt);
\fill[red] (5.72,-5.28) circle (0.75pt);
\draw[black!25,line width=0.3pt] (1,0)--(1,-6);
\draw[black!25,line width=0.3pt] (2,0)--(2,-6);
\draw[black!25,line width=0.3pt] (3,0)--(3,-6);
\draw[black!25,line width=0.3pt] (4,0)--(4,-6);
\draw[black!25,line width=0.3pt] (5,0)--(5,-6);
\draw[black!25,line width=0.3pt] (0,-1)--(6,-1);
\draw[black!25,line width=0.3pt] (0,-2)--(6,-2);
\draw[black!25,line width=0.3pt] (0,-3)--(6,-3);
\draw[black!25,line width=0.3pt] (0,-4)--(6,-4);
\draw[black!25,line width=0.3pt] (0,-5)--(6,-5);
\draw[black,line width=0.9pt] (0,0) rectangle (6,-6);
\node[anchor=south,font=\small\normalfont] at (3,0.3) {$\le$};
\draw[black!60,line width=0.5pt] (-0.3,1.05)--(6.3,1.05);
\draw[black,line width=1pt] (-0.3,-6.4) rectangle (6.3,2.2);
\end{tikzpicture}\end{tabular}\par\medskip
\begin{tabular}[t]{@{}l@{}}\small Rational numbers $\Q$ as the prime field of characteristic 0\\[-1pt]\small enumerated $0,1,-1,\tfrac12,-\tfrac12,\dots$:\\[-6pt]\begin{tikzpicture}[x=0.42cm,y=0.42cm,baseline=(current bounding box.north),every node/.style={font=\sffamily\bfseries,inner sep=0pt}]
\node[anchor=base west,font=\small\normalfont] at (0,1.45) {\textcolor{black}{$\aleph_0$}};
\node[text=black] at (0.85,-0.675) {$0$};
\node[text=black] at (2.55,-0.675) {$1$};
\node[text=black] at (4.25,-0.675) {$-1$};
\node[text=black] at (5.95,-0.675) {$\tfrac{1}{2}$};
\node[text=black] at (7.65,-0.675) {$-\tfrac{1}{2}$};
\node[text=black] at (0.85,-2.025) {$1$};
\node[text=black] at (2.55,-2.025) {$2$};
\node[text=black] at (4.25,-2.025) {$0$};
\node[text=black] at (5.95,-2.025) {$\tfrac{3}{2}$};
\node[text=black] at (7.65,-2.025) {$\tfrac{1}{2}$};
\node[text=black] at (0.85,-3.375) {$-1$};
\node[text=black] at (2.55,-3.375) {$0$};
\node[text=black] at (4.25,-3.375) {$-2$};
\node[text=black] at (5.95,-3.375) {$-\tfrac{1}{2}$};
\node[text=black] at (7.65,-3.375) {$-\tfrac{3}{2}$};
\node[text=black] at (0.85,-4.725) {$\tfrac{1}{2}$};
\node[text=black] at (2.55,-4.725) {$\tfrac{3}{2}$};
\node[text=black] at (4.25,-4.725) {$-\tfrac{1}{2}$};
\node[text=black] at (5.95,-4.725) {$1$};
\node[text=black] at (7.65,-4.725) {$0$};
\node[text=black] at (0.85,-6.075) {$-\tfrac{1}{2}$};
\node[text=black] at (2.55,-6.075) {$\tfrac{1}{2}$};
\node[text=black] at (4.25,-6.075) {$-\tfrac{3}{2}$};
\node[text=black] at (5.95,-6.075) {$0$};
\node[text=black] at (7.65,-6.075) {$-1$};
\fill[black] (9.13,-0.675) circle (0.75pt);
\fill[black] (9.35,-0.675) circle (0.75pt);
\fill[black] (9.57,-0.675) circle (0.75pt);
\fill[black] (9.13,-2.025) circle (0.75pt);
\fill[black] (9.35,-2.025) circle (0.75pt);
\fill[black] (9.57,-2.025) circle (0.75pt);
\fill[black] (9.13,-3.375) circle (0.75pt);
\fill[black] (9.35,-3.375) circle (0.75pt);
\fill[black] (9.57,-3.375) circle (0.75pt);
\fill[black] (9.13,-4.725) circle (0.75pt);
\fill[black] (9.35,-4.725) circle (0.75pt);
\fill[black] (9.57,-4.725) circle (0.75pt);
\fill[black] (9.13,-6.075) circle (0.75pt);
\fill[black] (9.35,-6.075) circle (0.75pt);
\fill[black] (9.57,-6.075) circle (0.75pt);
\fill[black] (0.85,-7.205) circle (0.75pt);
\fill[black] (0.85,-7.425) circle (0.75pt);
\fill[black] (0.85,-7.645) circle (0.75pt);
\fill[black] (2.55,-7.205) circle (0.75pt);
\fill[black] (2.55,-7.425) circle (0.75pt);
\fill[black] (2.55,-7.645) circle (0.75pt);
\fill[black] (4.25,-7.205) circle (0.75pt);
\fill[black] (4.25,-7.425) circle (0.75pt);
\fill[black] (4.25,-7.645) circle (0.75pt);
\fill[black] (5.95,-7.205) circle (0.75pt);
\fill[black] (5.95,-7.425) circle (0.75pt);
\fill[black] (5.95,-7.645) circle (0.75pt);
\fill[black] (7.65,-7.205) circle (0.75pt);
\fill[black] (7.65,-7.425) circle (0.75pt);
\fill[black] (7.65,-7.645) circle (0.75pt);
\fill[black] (9.13,-7.645) circle (0.75pt);
\fill[black] (9.35,-7.425) circle (0.75pt);
\fill[black] (9.57,-7.205) circle (0.75pt);
\draw[black!25,line width=0.3pt] (1.7,0)--(1.7,-8.1);
\draw[black!25,line width=0.3pt] (3.4,0)--(3.4,-8.1);
\draw[black!25,line width=0.3pt] (5.1,0)--(5.1,-8.1);
\draw[black!25,line width=0.3pt] (6.8,0)--(6.8,-8.1);
\draw[black!25,line width=0.3pt] (8.5,0)--(8.5,-8.1);
\draw[black!25,line width=0.3pt] (0,-1.35)--(10.2,-1.35);
\draw[black!25,line width=0.3pt] (0,-2.7)--(10.2,-2.7);
\draw[black!25,line width=0.3pt] (0,-4.05)--(10.2,-4.05);
\draw[black!25,line width=0.3pt] (0,-5.4)--(10.2,-5.4);
\draw[black!25,line width=0.3pt] (0,-6.75)--(10.2,-6.75);
\draw[black,line width=0.9pt] (0,0) rectangle (10.2,-8.1);
\node[anchor=south,font=\small\normalfont] at (5.1,0.3) {$+$};
\node[text=black] at (11.4,-0.675) {$0$};
\node[text=black] at (13.1,-0.675) {$0$};
\node[text=black] at (14.8,-0.675) {$0$};
\node[text=black] at (16.5,-0.675) {$0$};
\node[text=black] at (18.2,-0.675) {$0$};
\node[text=black] at (11.4,-2.025) {$0$};
\node[text=black] at (13.1,-2.025) {$1$};
\node[text=black] at (14.8,-2.025) {$-1$};
\node[text=black] at (16.5,-2.025) {$\tfrac{1}{2}$};
\node[text=black] at (18.2,-2.025) {$-\tfrac{1}{2}$};
\node[text=black] at (11.4,-3.375) {$0$};
\node[text=black] at (13.1,-3.375) {$-1$};
\node[text=black] at (14.8,-3.375) {$1$};
\node[text=black] at (16.5,-3.375) {$-\tfrac{1}{2}$};
\node[text=black] at (18.2,-3.375) {$\tfrac{1}{2}$};
\node[text=black] at (11.4,-4.725) {$0$};
\node[text=black] at (13.1,-4.725) {$\tfrac{1}{2}$};
\node[text=black] at (14.8,-4.725) {$-\tfrac{1}{2}$};
\node[text=black] at (16.5,-4.725) {$\tfrac{1}{4}$};
\node[text=black] at (18.2,-4.725) {$-\tfrac{1}{4}$};
\node[text=black] at (11.4,-6.075) {$0$};
\node[text=black] at (13.1,-6.075) {$-\tfrac{1}{2}$};
\node[text=black] at (14.8,-6.075) {$\tfrac{1}{2}$};
\node[text=black] at (16.5,-6.075) {$-\tfrac{1}{4}$};
\node[text=black] at (18.2,-6.075) {$\tfrac{1}{4}$};
\fill[black] (19.68,-0.675) circle (0.75pt);
\fill[black] (19.9,-0.675) circle (0.75pt);
\fill[black] (20.12,-0.675) circle (0.75pt);
\fill[black] (19.68,-2.025) circle (0.75pt);
\fill[black] (19.9,-2.025) circle (0.75pt);
\fill[black] (20.12,-2.025) circle (0.75pt);
\fill[black] (19.68,-3.375) circle (0.75pt);
\fill[black] (19.9,-3.375) circle (0.75pt);
\fill[black] (20.12,-3.375) circle (0.75pt);
\fill[black] (19.68,-4.725) circle (0.75pt);
\fill[black] (19.9,-4.725) circle (0.75pt);
\fill[black] (20.12,-4.725) circle (0.75pt);
\fill[black] (19.68,-6.075) circle (0.75pt);
\fill[black] (19.9,-6.075) circle (0.75pt);
\fill[black] (20.12,-6.075) circle (0.75pt);
\fill[black] (11.4,-7.205) circle (0.75pt);
\fill[black] (11.4,-7.425) circle (0.75pt);
\fill[black] (11.4,-7.645) circle (0.75pt);
\fill[black] (13.1,-7.205) circle (0.75pt);
\fill[black] (13.1,-7.425) circle (0.75pt);
\fill[black] (13.1,-7.645) circle (0.75pt);
\fill[black] (14.8,-7.205) circle (0.75pt);
\fill[black] (14.8,-7.425) circle (0.75pt);
\fill[black] (14.8,-7.645) circle (0.75pt);
\fill[black] (16.5,-7.205) circle (0.75pt);
\fill[black] (16.5,-7.425) circle (0.75pt);
\fill[black] (16.5,-7.645) circle (0.75pt);
\fill[black] (18.2,-7.205) circle (0.75pt);
\fill[black] (18.2,-7.425) circle (0.75pt);
\fill[black] (18.2,-7.645) circle (0.75pt);
\fill[black] (19.68,-7.645) circle (0.75pt);
\fill[black] (19.9,-7.425) circle (0.75pt);
\fill[black] (20.12,-7.205) circle (0.75pt);
\draw[black!25,line width=0.3pt] (12.25,0)--(12.25,-8.1);
\draw[black!25,line width=0.3pt] (13.95,0)--(13.95,-8.1);
\draw[black!25,line width=0.3pt] (15.65,0)--(15.65,-8.1);
\draw[black!25,line width=0.3pt] (17.35,0)--(17.35,-8.1);
\draw[black!25,line width=0.3pt] (19.05,0)--(19.05,-8.1);
\draw[black!25,line width=0.3pt] (10.55,-1.35)--(20.75,-1.35);
\draw[black!25,line width=0.3pt] (10.55,-2.7)--(20.75,-2.7);
\draw[black!25,line width=0.3pt] (10.55,-4.05)--(20.75,-4.05);
\draw[black!25,line width=0.3pt] (10.55,-5.4)--(20.75,-5.4);
\draw[black!25,line width=0.3pt] (10.55,-6.75)--(20.75,-6.75);
\draw[black,line width=0.9pt] (10.55,0) rectangle (20.75,-8.1);
\node[anchor=south,font=\small\normalfont] at (15.65,0.3) {$\cdot$};
\node[text=black] at (21.95,-0.675) {$0$};
\node[text=black] at (21.95,-2.025) {$-1$};
\node[text=black] at (21.95,-3.375) {$1$};
\node[text=black] at (21.95,-4.725) {$-\tfrac{1}{2}$};
\node[text=black] at (21.95,-6.075) {$\tfrac{1}{2}$};
\fill[black] (21.95,-7.205) circle (0.75pt);
\fill[black] (21.95,-7.425) circle (0.75pt);
\fill[black] (21.95,-7.645) circle (0.75pt);
\draw[black!25,line width=0.3pt] (21.1,-1.35)--(22.8,-1.35);
\draw[black!25,line width=0.3pt] (21.1,-2.7)--(22.8,-2.7);
\draw[black!25,line width=0.3pt] (21.1,-4.05)--(22.8,-4.05);
\draw[black!25,line width=0.3pt] (21.1,-5.4)--(22.8,-5.4);
\draw[black!25,line width=0.3pt] (21.1,-6.75)--(22.8,-6.75);
\draw[black,line width=0.9pt] (21.1,0) rectangle (22.8,-8.1);
\node[anchor=south,font=\small\normalfont] at (21.95,0.3) {$-$};
\node[text=black] at (23.65,-0.675) {$0$};
\draw[black,line width=0.9pt] (23.15,0) rectangle (24.15,-1.35);
\node[anchor=south,font=\small\normalfont] at (23.65,0.3) {$0$};
\node[text=black] at (25,-0.675) {$1$};
\draw[black,line width=0.9pt] (24.5,0) rectangle (25.5,-1.35);
\node[anchor=south,font=\small\normalfont] at (25,0.3) {$1$};
\draw[black!60,line width=0.5pt] (-0.3,1.05)--(25.8,1.05);
\draw[black,line width=1pt] (-0.3,-8.5) rectangle (25.8,2.2);
\end{tikzpicture}\end{tabular}\hspace*{\fill}\begin{tabular}[t]{@{}r@{}}\small Rational number total order $(\Q,\le)$:\\[-6pt]\begin{tikzpicture}[x=0.42cm,y=0.42cm,baseline=(current bounding box.north),every node/.style={font=\sffamily\bfseries,inner sep=0pt}]
\node[anchor=base west,font=\small\normalfont] at (0,1.45) {\textcolor{black}{$\aleph_0$}\ \ \textcolor{red}{2}};
\node[text=red] at (0.5,-0.5) {$1$};
\node[text=red] at (1.5,-0.5) {$1$};
\node[text=red] at (2.5,-0.5) {$0$};
\node[text=red] at (3.5,-0.5) {$1$};
\node[text=red] at (4.5,-0.5) {$0$};
\node[text=red] at (0.5,-1.5) {$0$};
\node[text=red] at (1.5,-1.5) {$1$};
\node[text=red] at (2.5,-1.5) {$0$};
\node[text=red] at (3.5,-1.5) {$0$};
\node[text=red] at (4.5,-1.5) {$0$};
\node[text=red] at (0.5,-2.5) {$1$};
\node[text=red] at (1.5,-2.5) {$1$};
\node[text=red] at (2.5,-2.5) {$1$};
\node[text=red] at (3.5,-2.5) {$1$};
\node[text=red] at (4.5,-2.5) {$1$};
\node[text=red] at (0.5,-3.5) {$0$};
\node[text=red] at (1.5,-3.5) {$1$};
\node[text=red] at (2.5,-3.5) {$0$};
\node[text=red] at (3.5,-3.5) {$1$};
\node[text=red] at (4.5,-3.5) {$0$};
\node[text=red] at (0.5,-4.5) {$1$};
\node[text=red] at (1.5,-4.5) {$1$};
\node[text=red] at (2.5,-4.5) {$0$};
\node[text=red] at (3.5,-4.5) {$1$};
\node[text=red] at (4.5,-4.5) {$1$};
\fill[red] (5.28,-0.5) circle (0.75pt);
\fill[red] (5.5,-0.5) circle (0.75pt);
\fill[red] (5.72,-0.5) circle (0.75pt);
\fill[red] (5.28,-1.5) circle (0.75pt);
\fill[red] (5.5,-1.5) circle (0.75pt);
\fill[red] (5.72,-1.5) circle (0.75pt);
\fill[red] (5.28,-2.5) circle (0.75pt);
\fill[red] (5.5,-2.5) circle (0.75pt);
\fill[red] (5.72,-2.5) circle (0.75pt);
\fill[red] (5.28,-3.5) circle (0.75pt);
\fill[red] (5.5,-3.5) circle (0.75pt);
\fill[red] (5.72,-3.5) circle (0.75pt);
\fill[red] (5.28,-4.5) circle (0.75pt);
\fill[red] (5.5,-4.5) circle (0.75pt);
\fill[red] (5.72,-4.5) circle (0.75pt);
\fill[red] (0.5,-5.28) circle (0.75pt);
\fill[red] (0.5,-5.5) circle (0.75pt);
\fill[red] (0.5,-5.72) circle (0.75pt);
\fill[red] (1.5,-5.28) circle (0.75pt);
\fill[red] (1.5,-5.5) circle (0.75pt);
\fill[red] (1.5,-5.72) circle (0.75pt);
\fill[red] (2.5,-5.28) circle (0.75pt);
\fill[red] (2.5,-5.5) circle (0.75pt);
\fill[red] (2.5,-5.72) circle (0.75pt);
\fill[red] (3.5,-5.28) circle (0.75pt);
\fill[red] (3.5,-5.5) circle (0.75pt);
\fill[red] (3.5,-5.72) circle (0.75pt);
\fill[red] (4.5,-5.28) circle (0.75pt);
\fill[red] (4.5,-5.5) circle (0.75pt);
\fill[red] (4.5,-5.72) circle (0.75pt);
\fill[red] (5.28,-5.72) circle (0.75pt);
\fill[red] (5.5,-5.5) circle (0.75pt);
\fill[red] (5.72,-5.28) circle (0.75pt);
\draw[black!25,line width=0.3pt] (1,0)--(1,-6);
\draw[black!25,line width=0.3pt] (2,0)--(2,-6);
\draw[black!25,line width=0.3pt] (3,0)--(3,-6);
\draw[black!25,line width=0.3pt] (4,0)--(4,-6);
\draw[black!25,line width=0.3pt] (5,0)--(5,-6);
\draw[black!25,line width=0.3pt] (0,-1)--(6,-1);
\draw[black!25,line width=0.3pt] (0,-2)--(6,-2);
\draw[black!25,line width=0.3pt] (0,-3)--(6,-3);
\draw[black!25,line width=0.3pt] (0,-4)--(6,-4);
\draw[black!25,line width=0.3pt] (0,-5)--(6,-5);
\draw[black,line width=0.9pt] (0,0) rectangle (6,-6);
\node[anchor=south,font=\small\normalfont] at (3,0.3) {$\le$};
\draw[black!60,line width=0.5pt] (-0.3,1.05)--(6.3,1.05);
\draw[black,line width=1pt] (-0.3,-6.4) rectangle (6.3,2.2);
\end{tikzpicture}\end{tabular}\par\medskip
\begin{tabular}[t]{@{}l@{}}\small Integer arithmetic\\[-1pt]\small with successor $(\Z,0,{}',-,+,\cdot)$:\\[-6pt]\begin{tikzpicture}[x=0.42cm,y=0.42cm,baseline=(current bounding box.north),every node/.style={font=\sffamily\bfseries,inner sep=0pt}]
\node[anchor=base west,font=\small\normalfont] at (0,1.45) {\textcolor{black}{$\aleph_0$}};
\node[text=black,anchor=east] at (1.28,-0.5) {$0$};
\node[text=black,anchor=east] at (2.68,-0.5) {$1$};
\node[text=black,anchor=east] at (4.08,-0.5) {$-1$};
\node[text=black,anchor=east] at (5.48,-0.5) {$2$};
\node[text=black,anchor=east] at (6.88,-0.5) {$-2$};
\node[text=black,anchor=east] at (1.28,-1.5) {$1$};
\node[text=black,anchor=east] at (2.68,-1.5) {$2$};
\node[text=black,anchor=east] at (4.08,-1.5) {$0$};
\node[text=black,anchor=east] at (5.48,-1.5) {$3$};
\node[text=black,anchor=east] at (6.88,-1.5) {$-1$};
\node[text=black,anchor=east] at (1.28,-2.5) {$-1$};
\node[text=black,anchor=east] at (2.68,-2.5) {$0$};
\node[text=black,anchor=east] at (4.08,-2.5) {$-2$};
\node[text=black,anchor=east] at (5.48,-2.5) {$1$};
\node[text=black,anchor=east] at (6.88,-2.5) {$-3$};
\node[text=black,anchor=east] at (1.28,-3.5) {$2$};
\node[text=black,anchor=east] at (2.68,-3.5) {$3$};
\node[text=black,anchor=east] at (4.08,-3.5) {$1$};
\node[text=black,anchor=east] at (5.48,-3.5) {$4$};
\node[text=black,anchor=east] at (6.88,-3.5) {$0$};
\node[text=black,anchor=east] at (1.28,-4.5) {$-2$};
\node[text=black,anchor=east] at (2.68,-4.5) {$-1$};
\node[text=black,anchor=east] at (4.08,-4.5) {$-3$};
\node[text=black,anchor=east] at (5.48,-4.5) {$0$};
\node[text=black,anchor=east] at (6.88,-4.5) {$-4$};
\fill[black] (7.48,-0.5) circle (0.75pt);
\fill[black] (7.7,-0.5) circle (0.75pt);
\fill[black] (7.92,-0.5) circle (0.75pt);
\fill[black] (7.48,-1.5) circle (0.75pt);
\fill[black] (7.7,-1.5) circle (0.75pt);
\fill[black] (7.92,-1.5) circle (0.75pt);
\fill[black] (7.48,-2.5) circle (0.75pt);
\fill[black] (7.7,-2.5) circle (0.75pt);
\fill[black] (7.92,-2.5) circle (0.75pt);
\fill[black] (7.48,-3.5) circle (0.75pt);
\fill[black] (7.7,-3.5) circle (0.75pt);
\fill[black] (7.92,-3.5) circle (0.75pt);
\fill[black] (7.48,-4.5) circle (0.75pt);
\fill[black] (7.7,-4.5) circle (0.75pt);
\fill[black] (7.92,-4.5) circle (0.75pt);
\fill[black] (0.7,-5.28) circle (0.75pt);
\fill[black] (0.7,-5.5) circle (0.75pt);
\fill[black] (0.7,-5.72) circle (0.75pt);
\fill[black] (2.1,-5.28) circle (0.75pt);
\fill[black] (2.1,-5.5) circle (0.75pt);
\fill[black] (2.1,-5.72) circle (0.75pt);
\fill[black] (3.5,-5.28) circle (0.75pt);
\fill[black] (3.5,-5.5) circle (0.75pt);
\fill[black] (3.5,-5.72) circle (0.75pt);
\fill[black] (4.9,-5.28) circle (0.75pt);
\fill[black] (4.9,-5.5) circle (0.75pt);
\fill[black] (4.9,-5.72) circle (0.75pt);
\fill[black] (6.3,-5.28) circle (0.75pt);
\fill[black] (6.3,-5.5) circle (0.75pt);
\fill[black] (6.3,-5.72) circle (0.75pt);
\fill[black] (7.48,-5.72) circle (0.75pt);
\fill[black] (7.7,-5.5) circle (0.75pt);
\fill[black] (7.92,-5.28) circle (0.75pt);
\draw[black!25,line width=0.3pt] (1.4,0)--(1.4,-6);
\draw[black!25,line width=0.3pt] (2.8,0)--(2.8,-6);
\draw[black!25,line width=0.3pt] (4.2,0)--(4.2,-6);
\draw[black!25,line width=0.3pt] (5.6,0)--(5.6,-6);
\draw[black!25,line width=0.3pt] (7,0)--(7,-6);
\draw[black!25,line width=0.3pt] (0,-1)--(8.4,-1);
\draw[black!25,line width=0.3pt] (0,-2)--(8.4,-2);
\draw[black!25,line width=0.3pt] (0,-3)--(8.4,-3);
\draw[black!25,line width=0.3pt] (0,-4)--(8.4,-4);
\draw[black!25,line width=0.3pt] (0,-5)--(8.4,-5);
\draw[black,line width=0.9pt] (0,0) rectangle (8.4,-6);
\node[anchor=south,font=\small\normalfont] at (4.2,0.3) {$+$};
\node[text=black,anchor=east] at (10.03,-0.5) {$0$};
\node[text=black,anchor=east] at (11.43,-0.5) {$0$};
\node[text=black,anchor=east] at (12.83,-0.5) {$0$};
\node[text=black,anchor=east] at (14.23,-0.5) {$0$};
\node[text=black,anchor=east] at (15.63,-0.5) {$0$};
\node[text=black,anchor=east] at (10.03,-1.5) {$0$};
\node[text=black,anchor=east] at (11.43,-1.5) {$1$};
\node[text=black,anchor=east] at (12.83,-1.5) {$-1$};
\node[text=black,anchor=east] at (14.23,-1.5) {$2$};
\node[text=black,anchor=east] at (15.63,-1.5) {$-2$};
\node[text=black,anchor=east] at (10.03,-2.5) {$0$};
\node[text=black,anchor=east] at (11.43,-2.5) {$-1$};
\node[text=black,anchor=east] at (12.83,-2.5) {$1$};
\node[text=black,anchor=east] at (14.23,-2.5) {$-2$};
\node[text=black,anchor=east] at (15.63,-2.5) {$2$};
\node[text=black,anchor=east] at (10.03,-3.5) {$0$};
\node[text=black,anchor=east] at (11.43,-3.5) {$2$};
\node[text=black,anchor=east] at (12.83,-3.5) {$-2$};
\node[text=black,anchor=east] at (14.23,-3.5) {$4$};
\node[text=black,anchor=east] at (15.63,-3.5) {$-4$};
\node[text=black,anchor=east] at (10.03,-4.5) {$0$};
\node[text=black,anchor=east] at (11.43,-4.5) {$-2$};
\node[text=black,anchor=east] at (12.83,-4.5) {$2$};
\node[text=black,anchor=east] at (14.23,-4.5) {$-4$};
\node[text=black,anchor=east] at (15.63,-4.5) {$4$};
\fill[black] (16.23,-0.5) circle (0.75pt);
\fill[black] (16.45,-0.5) circle (0.75pt);
\fill[black] (16.67,-0.5) circle (0.75pt);
\fill[black] (16.23,-1.5) circle (0.75pt);
\fill[black] (16.45,-1.5) circle (0.75pt);
\fill[black] (16.67,-1.5) circle (0.75pt);
\fill[black] (16.23,-2.5) circle (0.75pt);
\fill[black] (16.45,-2.5) circle (0.75pt);
\fill[black] (16.67,-2.5) circle (0.75pt);
\fill[black] (16.23,-3.5) circle (0.75pt);
\fill[black] (16.45,-3.5) circle (0.75pt);
\fill[black] (16.67,-3.5) circle (0.75pt);
\fill[black] (16.23,-4.5) circle (0.75pt);
\fill[black] (16.45,-4.5) circle (0.75pt);
\fill[black] (16.67,-4.5) circle (0.75pt);
\fill[black] (9.45,-5.28) circle (0.75pt);
\fill[black] (9.45,-5.5) circle (0.75pt);
\fill[black] (9.45,-5.72) circle (0.75pt);
\fill[black] (10.85,-5.28) circle (0.75pt);
\fill[black] (10.85,-5.5) circle (0.75pt);
\fill[black] (10.85,-5.72) circle (0.75pt);
\fill[black] (12.25,-5.28) circle (0.75pt);
\fill[black] (12.25,-5.5) circle (0.75pt);
\fill[black] (12.25,-5.72) circle (0.75pt);
\fill[black] (13.65,-5.28) circle (0.75pt);
\fill[black] (13.65,-5.5) circle (0.75pt);
\fill[black] (13.65,-5.72) circle (0.75pt);
\fill[black] (15.05,-5.28) circle (0.75pt);
\fill[black] (15.05,-5.5) circle (0.75pt);
\fill[black] (15.05,-5.72) circle (0.75pt);
\fill[black] (16.23,-5.72) circle (0.75pt);
\fill[black] (16.45,-5.5) circle (0.75pt);
\fill[black] (16.67,-5.28) circle (0.75pt);
\draw[black!25,line width=0.3pt] (10.15,0)--(10.15,-6);
\draw[black!25,line width=0.3pt] (11.55,0)--(11.55,-6);
\draw[black!25,line width=0.3pt] (12.95,0)--(12.95,-6);
\draw[black!25,line width=0.3pt] (14.35,0)--(14.35,-6);
\draw[black!25,line width=0.3pt] (15.75,0)--(15.75,-6);
\draw[black!25,line width=0.3pt] (8.75,-1)--(17.15,-1);
\draw[black!25,line width=0.3pt] (8.75,-2)--(17.15,-2);
\draw[black!25,line width=0.3pt] (8.75,-3)--(17.15,-3);
\draw[black!25,line width=0.3pt] (8.75,-4)--(17.15,-4);
\draw[black!25,line width=0.3pt] (8.75,-5)--(17.15,-5);
\draw[black,line width=0.9pt] (8.75,0) rectangle (17.15,-6);
\node[anchor=south,font=\small\normalfont] at (12.95,0.3) {$\cdot$};
\node[text=black,anchor=east] at (18.78,-0.5) {$1$};
\node[text=black,anchor=east] at (18.78,-1.5) {$2$};
\node[text=black,anchor=east] at (18.78,-2.5) {$0$};
\node[text=black,anchor=east] at (18.78,-3.5) {$3$};
\node[text=black,anchor=east] at (18.78,-4.5) {$-1$};
\fill[black] (18.2,-5.28) circle (0.75pt);
\fill[black] (18.2,-5.5) circle (0.75pt);
\fill[black] (18.2,-5.72) circle (0.75pt);
\draw[black!25,line width=0.3pt] (17.5,-1)--(18.9,-1);
\draw[black!25,line width=0.3pt] (17.5,-2)--(18.9,-2);
\draw[black!25,line width=0.3pt] (17.5,-3)--(18.9,-3);
\draw[black!25,line width=0.3pt] (17.5,-4)--(18.9,-4);
\draw[black!25,line width=0.3pt] (17.5,-5)--(18.9,-5);
\draw[black,line width=0.9pt] (17.5,0) rectangle (18.9,-6);
\node[anchor=south,font=\small\normalfont] at (18.2,0.3) {$'$};
\node[text=black,anchor=east] at (20.53,-0.5) {$0$};
\node[text=black,anchor=east] at (20.53,-1.5) {$-1$};
\node[text=black,anchor=east] at (20.53,-2.5) {$1$};
\node[text=black,anchor=east] at (20.53,-3.5) {$-2$};
\node[text=black,anchor=east] at (20.53,-4.5) {$2$};
\fill[black] (19.95,-5.28) circle (0.75pt);
\fill[black] (19.95,-5.5) circle (0.75pt);
\fill[black] (19.95,-5.72) circle (0.75pt);
\draw[black!25,line width=0.3pt] (19.25,-1)--(20.65,-1);
\draw[black!25,line width=0.3pt] (19.25,-2)--(20.65,-2);
\draw[black!25,line width=0.3pt] (19.25,-3)--(20.65,-3);
\draw[black!25,line width=0.3pt] (19.25,-4)--(20.65,-4);
\draw[black!25,line width=0.3pt] (19.25,-5)--(20.65,-5);
\draw[black,line width=0.9pt] (19.25,0) rectangle (20.65,-6);
\node[anchor=south,font=\small\normalfont] at (19.95,0.3) {$-$};
\node[text=black] at (21.5,-0.5) {$0$};
\draw[black,line width=0.9pt] (21,0) rectangle (22,-1);
\node[anchor=south,font=\small\normalfont] at (21.5,0.3) {$0$};
\draw[black!60,line width=0.5pt] (-0.3,1.05)--(22.3,1.05);
\draw[black,line width=1pt] (-0.3,-6.4) rectangle (22.3,2.2);
\end{tikzpicture}\end{tabular}\hspace*{\fill}\begin{tabular}[t]{@{}r@{}}\small Magma of integers\\[-1pt]\small under subtraction $(\Z,-)$:\\[-6pt]\begin{tikzpicture}[x=0.42cm,y=0.42cm,baseline=(current bounding box.north),every node/.style={font=\sffamily\bfseries,inner sep=0pt}]
\node[anchor=base west,font=\small\normalfont] at (0,1.45) {\textcolor{black}{$\aleph_0$}};
\node[text=black,anchor=east] at (1.28,-0.5) {$0$};
\node[text=black,anchor=east] at (2.68,-0.5) {$-1$};
\node[text=black,anchor=east] at (4.08,-0.5) {$1$};
\node[text=black,anchor=east] at (5.48,-0.5) {$-2$};
\node[text=black,anchor=east] at (6.88,-0.5) {$2$};
\node[text=black,anchor=east] at (1.28,-1.5) {$1$};
\node[text=black,anchor=east] at (2.68,-1.5) {$0$};
\node[text=black,anchor=east] at (4.08,-1.5) {$2$};
\node[text=black,anchor=east] at (5.48,-1.5) {$-1$};
\node[text=black,anchor=east] at (6.88,-1.5) {$3$};
\node[text=black,anchor=east] at (1.28,-2.5) {$-1$};
\node[text=black,anchor=east] at (2.68,-2.5) {$-2$};
\node[text=black,anchor=east] at (4.08,-2.5) {$0$};
\node[text=black,anchor=east] at (5.48,-2.5) {$-3$};
\node[text=black,anchor=east] at (6.88,-2.5) {$1$};
\node[text=black,anchor=east] at (1.28,-3.5) {$2$};
\node[text=black,anchor=east] at (2.68,-3.5) {$1$};
\node[text=black,anchor=east] at (4.08,-3.5) {$3$};
\node[text=black,anchor=east] at (5.48,-3.5) {$0$};
\node[text=black,anchor=east] at (6.88,-3.5) {$4$};
\node[text=black,anchor=east] at (1.28,-4.5) {$-2$};
\node[text=black,anchor=east] at (2.68,-4.5) {$-3$};
\node[text=black,anchor=east] at (4.08,-4.5) {$-1$};
\node[text=black,anchor=east] at (5.48,-4.5) {$-4$};
\node[text=black,anchor=east] at (6.88,-4.5) {$0$};
\fill[black] (7.48,-0.5) circle (0.75pt);
\fill[black] (7.7,-0.5) circle (0.75pt);
\fill[black] (7.92,-0.5) circle (0.75pt);
\fill[black] (7.48,-1.5) circle (0.75pt);
\fill[black] (7.7,-1.5) circle (0.75pt);
\fill[black] (7.92,-1.5) circle (0.75pt);
\fill[black] (7.48,-2.5) circle (0.75pt);
\fill[black] (7.7,-2.5) circle (0.75pt);
\fill[black] (7.92,-2.5) circle (0.75pt);
\fill[black] (7.48,-3.5) circle (0.75pt);
\fill[black] (7.7,-3.5) circle (0.75pt);
\fill[black] (7.92,-3.5) circle (0.75pt);
\fill[black] (7.48,-4.5) circle (0.75pt);
\fill[black] (7.7,-4.5) circle (0.75pt);
\fill[black] (7.92,-4.5) circle (0.75pt);
\fill[black] (0.7,-5.28) circle (0.75pt);
\fill[black] (0.7,-5.5) circle (0.75pt);
\fill[black] (0.7,-5.72) circle (0.75pt);
\fill[black] (2.1,-5.28) circle (0.75pt);
\fill[black] (2.1,-5.5) circle (0.75pt);
\fill[black] (2.1,-5.72) circle (0.75pt);
\fill[black] (3.5,-5.28) circle (0.75pt);
\fill[black] (3.5,-5.5) circle (0.75pt);
\fill[black] (3.5,-5.72) circle (0.75pt);
\fill[black] (4.9,-5.28) circle (0.75pt);
\fill[black] (4.9,-5.5) circle (0.75pt);
\fill[black] (4.9,-5.72) circle (0.75pt);
\fill[black] (6.3,-5.28) circle (0.75pt);
\fill[black] (6.3,-5.5) circle (0.75pt);
\fill[black] (6.3,-5.72) circle (0.75pt);
\fill[black] (7.48,-5.72) circle (0.75pt);
\fill[black] (7.7,-5.5) circle (0.75pt);
\fill[black] (7.92,-5.28) circle (0.75pt);
\draw[black!25,line width=0.3pt] (1.4,0)--(1.4,-6);
\draw[black!25,line width=0.3pt] (2.8,0)--(2.8,-6);
\draw[black!25,line width=0.3pt] (4.2,0)--(4.2,-6);
\draw[black!25,line width=0.3pt] (5.6,0)--(5.6,-6);
\draw[black!25,line width=0.3pt] (7,0)--(7,-6);
\draw[black!25,line width=0.3pt] (0,-1)--(8.4,-1);
\draw[black!25,line width=0.3pt] (0,-2)--(8.4,-2);
\draw[black!25,line width=0.3pt] (0,-3)--(8.4,-3);
\draw[black!25,line width=0.3pt] (0,-4)--(8.4,-4);
\draw[black!25,line width=0.3pt] (0,-5)--(8.4,-5);
\draw[black,line width=0.9pt] (0,0) rectangle (8.4,-6);
\node[anchor=south,font=\small\normalfont] at (4.2,0.3) {$-$};
\draw[black!60,line width=0.5pt] (-0.3,1.05)--(8.7,1.05);
\draw[black,line width=1pt] (-0.3,-6.4) rectangle (8.7,2.2);
\end{tikzpicture}\end{tabular}\par\medskip
\begin{tabular}[t]{@{}l@{}}\small Congruence mod 2\\[-1pt]\small equivalence relation on $\N$:\\[-6pt]\begin{tikzpicture}[x=0.42cm,y=0.42cm,baseline=(current bounding box.north),every node/.style={font=\sffamily\bfseries,inner sep=0pt}]
\node[anchor=base west,font=\small\normalfont] at (0,1.45) {\textcolor{black}{$\aleph_0$}\ \ \textcolor{red}{2}};
\node[text=red] at (0.5,-0.5) {$1$};
\node[text=red] at (1.5,-0.5) {$0$};
\node[text=red] at (2.5,-0.5) {$1$};
\node[text=red] at (3.5,-0.5) {$0$};
\node[text=red] at (4.5,-0.5) {$1$};
\node[text=red] at (0.5,-1.5) {$0$};
\node[text=red] at (1.5,-1.5) {$1$};
\node[text=red] at (2.5,-1.5) {$0$};
\node[text=red] at (3.5,-1.5) {$1$};
\node[text=red] at (4.5,-1.5) {$0$};
\node[text=red] at (0.5,-2.5) {$1$};
\node[text=red] at (1.5,-2.5) {$0$};
\node[text=red] at (2.5,-2.5) {$1$};
\node[text=red] at (3.5,-2.5) {$0$};
\node[text=red] at (4.5,-2.5) {$1$};
\node[text=red] at (0.5,-3.5) {$0$};
\node[text=red] at (1.5,-3.5) {$1$};
\node[text=red] at (2.5,-3.5) {$0$};
\node[text=red] at (3.5,-3.5) {$1$};
\node[text=red] at (4.5,-3.5) {$0$};
\node[text=red] at (0.5,-4.5) {$1$};
\node[text=red] at (1.5,-4.5) {$0$};
\node[text=red] at (2.5,-4.5) {$1$};
\node[text=red] at (3.5,-4.5) {$0$};
\node[text=red] at (4.5,-4.5) {$1$};
\fill[red] (5.28,-0.5) circle (0.75pt);
\fill[red] (5.5,-0.5) circle (0.75pt);
\fill[red] (5.72,-0.5) circle (0.75pt);
\fill[red] (5.28,-1.5) circle (0.75pt);
\fill[red] (5.5,-1.5) circle (0.75pt);
\fill[red] (5.72,-1.5) circle (0.75pt);
\fill[red] (5.28,-2.5) circle (0.75pt);
\fill[red] (5.5,-2.5) circle (0.75pt);
\fill[red] (5.72,-2.5) circle (0.75pt);
\fill[red] (5.28,-3.5) circle (0.75pt);
\fill[red] (5.5,-3.5) circle (0.75pt);
\fill[red] (5.72,-3.5) circle (0.75pt);
\fill[red] (5.28,-4.5) circle (0.75pt);
\fill[red] (5.5,-4.5) circle (0.75pt);
\fill[red] (5.72,-4.5) circle (0.75pt);
\fill[red] (0.5,-5.28) circle (0.75pt);
\fill[red] (0.5,-5.5) circle (0.75pt);
\fill[red] (0.5,-5.72) circle (0.75pt);
\fill[red] (1.5,-5.28) circle (0.75pt);
\fill[red] (1.5,-5.5) circle (0.75pt);
\fill[red] (1.5,-5.72) circle (0.75pt);
\fill[red] (2.5,-5.28) circle (0.75pt);
\fill[red] (2.5,-5.5) circle (0.75pt);
\fill[red] (2.5,-5.72) circle (0.75pt);
\fill[red] (3.5,-5.28) circle (0.75pt);
\fill[red] (3.5,-5.5) circle (0.75pt);
\fill[red] (3.5,-5.72) circle (0.75pt);
\fill[red] (4.5,-5.28) circle (0.75pt);
\fill[red] (4.5,-5.5) circle (0.75pt);
\fill[red] (4.5,-5.72) circle (0.75pt);
\fill[red] (5.28,-5.72) circle (0.75pt);
\fill[red] (5.5,-5.5) circle (0.75pt);
\fill[red] (5.72,-5.28) circle (0.75pt);
\draw[black!25,line width=0.3pt] (1,0)--(1,-6);
\draw[black!25,line width=0.3pt] (2,0)--(2,-6);
\draw[black!25,line width=0.3pt] (3,0)--(3,-6);
\draw[black!25,line width=0.3pt] (4,0)--(4,-6);
\draw[black!25,line width=0.3pt] (5,0)--(5,-6);
\draw[black!25,line width=0.3pt] (0,-1)--(6,-1);
\draw[black!25,line width=0.3pt] (0,-2)--(6,-2);
\draw[black!25,line width=0.3pt] (0,-3)--(6,-3);
\draw[black!25,line width=0.3pt] (0,-4)--(6,-4);
\draw[black!25,line width=0.3pt] (0,-5)--(6,-5);
\draw[black,line width=0.9pt] (0,0) rectangle (6,-6);
\node[anchor=south,font=\small\normalfont] at (3,0.3) {$\sim$};
\draw[black!60,line width=0.5pt] (-0.3,1.05)--(6.3,1.05);
\draw[black,line width=1pt] (-0.3,-6.4) rectangle (6.3,2.2);
\end{tikzpicture}\end{tabular}\hspace*{\fill}\begin{tabular}[t]{@{}l@{}}\small Distributive lattice\\[-1pt]\small of divisibility $(\N^+,\mid)$:\\[-6pt]\begin{tikzpicture}[x=0.42cm,y=0.42cm,baseline=(current bounding box.north),every node/.style={font=\sffamily\bfseries,inner sep=0pt}]
\node[anchor=base west,font=\small\normalfont] at (0,1.45) {\textcolor{black}{$\aleph_0$}\ \ \textcolor{red}{2}};
\node[text=red] at (0.5,-0.5) {$1$};
\node[text=red] at (1.5,-0.5) {$1$};
\node[text=red] at (2.5,-0.5) {$1$};
\node[text=red] at (3.5,-0.5) {$1$};
\node[text=red] at (4.5,-0.5) {$1$};
\node[text=red] at (0.5,-1.5) {$0$};
\node[text=red] at (1.5,-1.5) {$1$};
\node[text=red] at (2.5,-1.5) {$0$};
\node[text=red] at (3.5,-1.5) {$1$};
\node[text=red] at (4.5,-1.5) {$0$};
\node[text=red] at (0.5,-2.5) {$0$};
\node[text=red] at (1.5,-2.5) {$0$};
\node[text=red] at (2.5,-2.5) {$1$};
\node[text=red] at (3.5,-2.5) {$0$};
\node[text=red] at (4.5,-2.5) {$0$};
\node[text=red] at (0.5,-3.5) {$0$};
\node[text=red] at (1.5,-3.5) {$0$};
\node[text=red] at (2.5,-3.5) {$0$};
\node[text=red] at (3.5,-3.5) {$1$};
\node[text=red] at (4.5,-3.5) {$0$};
\node[text=red] at (0.5,-4.5) {$0$};
\node[text=red] at (1.5,-4.5) {$0$};
\node[text=red] at (2.5,-4.5) {$0$};
\node[text=red] at (3.5,-4.5) {$0$};
\node[text=red] at (4.5,-4.5) {$1$};
\fill[red] (5.28,-0.5) circle (0.75pt);
\fill[red] (5.5,-0.5) circle (0.75pt);
\fill[red] (5.72,-0.5) circle (0.75pt);
\fill[red] (5.28,-1.5) circle (0.75pt);
\fill[red] (5.5,-1.5) circle (0.75pt);
\fill[red] (5.72,-1.5) circle (0.75pt);
\fill[red] (5.28,-2.5) circle (0.75pt);
\fill[red] (5.5,-2.5) circle (0.75pt);
\fill[red] (5.72,-2.5) circle (0.75pt);
\fill[red] (5.28,-3.5) circle (0.75pt);
\fill[red] (5.5,-3.5) circle (0.75pt);
\fill[red] (5.72,-3.5) circle (0.75pt);
\fill[red] (5.28,-4.5) circle (0.75pt);
\fill[red] (5.5,-4.5) circle (0.75pt);
\fill[red] (5.72,-4.5) circle (0.75pt);
\fill[red] (0.5,-5.28) circle (0.75pt);
\fill[red] (0.5,-5.5) circle (0.75pt);
\fill[red] (0.5,-5.72) circle (0.75pt);
\fill[red] (1.5,-5.28) circle (0.75pt);
\fill[red] (1.5,-5.5) circle (0.75pt);
\fill[red] (1.5,-5.72) circle (0.75pt);
\fill[red] (2.5,-5.28) circle (0.75pt);
\fill[red] (2.5,-5.5) circle (0.75pt);
\fill[red] (2.5,-5.72) circle (0.75pt);
\fill[red] (3.5,-5.28) circle (0.75pt);
\fill[red] (3.5,-5.5) circle (0.75pt);
\fill[red] (3.5,-5.72) circle (0.75pt);
\fill[red] (4.5,-5.28) circle (0.75pt);
\fill[red] (4.5,-5.5) circle (0.75pt);
\fill[red] (4.5,-5.72) circle (0.75pt);
\fill[red] (5.28,-5.72) circle (0.75pt);
\fill[red] (5.5,-5.5) circle (0.75pt);
\fill[red] (5.72,-5.28) circle (0.75pt);
\draw[black!25,line width=0.3pt] (1,0)--(1,-6);
\draw[black!25,line width=0.3pt] (2,0)--(2,-6);
\draw[black!25,line width=0.3pt] (3,0)--(3,-6);
\draw[black!25,line width=0.3pt] (4,0)--(4,-6);
\draw[black!25,line width=0.3pt] (5,0)--(5,-6);
\draw[black!25,line width=0.3pt] (0,-1)--(6,-1);
\draw[black!25,line width=0.3pt] (0,-2)--(6,-2);
\draw[black!25,line width=0.3pt] (0,-3)--(6,-3);
\draw[black!25,line width=0.3pt] (0,-4)--(6,-4);
\draw[black!25,line width=0.3pt] (0,-5)--(6,-5);
\draw[black,line width=0.9pt] (0,0) rectangle (6,-6);
\node[anchor=south,font=\small\normalfont] at (3,0.3) {$\mid$};
\draw[black!60,line width=0.5pt] (-0.3,1.05)--(6.3,1.05);
\draw[black,line width=1pt] (-0.3,-6.4) rectangle (6.3,2.2);
\end{tikzpicture}\end{tabular}\hspace*{\fill}\begin{tabular}[t]{@{}r@{}}\small Divisibility\\[-1pt]\small preorder on $\Z$:\\[-6pt]\begin{tikzpicture}[x=0.42cm,y=0.42cm,baseline=(current bounding box.north),every node/.style={font=\sffamily\bfseries,inner sep=0pt}]
\node[anchor=base west,font=\small\normalfont] at (0,1.45) {\textcolor{black}{$\aleph_0$}\ \ \textcolor{red}{2}};
\node[text=red,anchor=east] at (0.88,-0.5) {$1$};
\node[text=red,anchor=east] at (1.88,-0.5) {$0$};
\node[text=red,anchor=east] at (2.88,-0.5) {$0$};
\node[text=red,anchor=east] at (3.88,-0.5) {$0$};
\node[text=red,anchor=east] at (4.88,-0.5) {$0$};
\node[text=red,anchor=east] at (0.88,-1.5) {$1$};
\node[text=red,anchor=east] at (1.88,-1.5) {$1$};
\node[text=red,anchor=east] at (2.88,-1.5) {$1$};
\node[text=red,anchor=east] at (3.88,-1.5) {$1$};
\node[text=red,anchor=east] at (4.88,-1.5) {$1$};
\node[text=red,anchor=east] at (0.88,-2.5) {$1$};
\node[text=red,anchor=east] at (1.88,-2.5) {$1$};
\node[text=red,anchor=east] at (2.88,-2.5) {$1$};
\node[text=red,anchor=east] at (3.88,-2.5) {$1$};
\node[text=red,anchor=east] at (4.88,-2.5) {$1$};
\node[text=red,anchor=east] at (0.88,-3.5) {$1$};
\node[text=red,anchor=east] at (1.88,-3.5) {$0$};
\node[text=red,anchor=east] at (2.88,-3.5) {$0$};
\node[text=red,anchor=east] at (3.88,-3.5) {$1$};
\node[text=red,anchor=east] at (4.88,-3.5) {$1$};
\node[text=red,anchor=east] at (0.88,-4.5) {$1$};
\node[text=red,anchor=east] at (1.88,-4.5) {$0$};
\node[text=red,anchor=east] at (2.88,-4.5) {$0$};
\node[text=red,anchor=east] at (3.88,-4.5) {$1$};
\node[text=red,anchor=east] at (4.88,-4.5) {$1$};
\fill[red] (5.28,-0.5) circle (0.75pt);
\fill[red] (5.5,-0.5) circle (0.75pt);
\fill[red] (5.72,-0.5) circle (0.75pt);
\fill[red] (5.28,-1.5) circle (0.75pt);
\fill[red] (5.5,-1.5) circle (0.75pt);
\fill[red] (5.72,-1.5) circle (0.75pt);
\fill[red] (5.28,-2.5) circle (0.75pt);
\fill[red] (5.5,-2.5) circle (0.75pt);
\fill[red] (5.72,-2.5) circle (0.75pt);
\fill[red] (5.28,-3.5) circle (0.75pt);
\fill[red] (5.5,-3.5) circle (0.75pt);
\fill[red] (5.72,-3.5) circle (0.75pt);
\fill[red] (5.28,-4.5) circle (0.75pt);
\fill[red] (5.5,-4.5) circle (0.75pt);
\fill[red] (5.72,-4.5) circle (0.75pt);
\fill[red] (0.5,-5.28) circle (0.75pt);
\fill[red] (0.5,-5.5) circle (0.75pt);
\fill[red] (0.5,-5.72) circle (0.75pt);
\fill[red] (1.5,-5.28) circle (0.75pt);
\fill[red] (1.5,-5.5) circle (0.75pt);
\fill[red] (1.5,-5.72) circle (0.75pt);
\fill[red] (2.5,-5.28) circle (0.75pt);
\fill[red] (2.5,-5.5) circle (0.75pt);
\fill[red] (2.5,-5.72) circle (0.75pt);
\fill[red] (3.5,-5.28) circle (0.75pt);
\fill[red] (3.5,-5.5) circle (0.75pt);
\fill[red] (3.5,-5.72) circle (0.75pt);
\fill[red] (4.5,-5.28) circle (0.75pt);
\fill[red] (4.5,-5.5) circle (0.75pt);
\fill[red] (4.5,-5.72) circle (0.75pt);
\fill[red] (5.28,-5.72) circle (0.75pt);
\fill[red] (5.5,-5.5) circle (0.75pt);
\fill[red] (5.72,-5.28) circle (0.75pt);
\draw[black!25,line width=0.3pt] (1,0)--(1,-6);
\draw[black!25,line width=0.3pt] (2,0)--(2,-6);
\draw[black!25,line width=0.3pt] (3,0)--(3,-6);
\draw[black!25,line width=0.3pt] (4,0)--(4,-6);
\draw[black!25,line width=0.3pt] (5,0)--(5,-6);
\draw[black!25,line width=0.3pt] (0,-1)--(6,-1);
\draw[black!25,line width=0.3pt] (0,-2)--(6,-2);
\draw[black!25,line width=0.3pt] (0,-3)--(6,-3);
\draw[black!25,line width=0.3pt] (0,-4)--(6,-4);
\draw[black!25,line width=0.3pt] (0,-5)--(6,-5);
\draw[black,line width=0.9pt] (0,0) rectangle (6,-6);
\node[anchor=south,font=\small\normalfont] at (3,0.3) {$\lesssim$};
\draw[black!60,line width=0.5pt] (-0.3,1.05)--(6.3,1.05);
\draw[black,line width=1pt] (-0.3,-6.4) rectangle (6.3,2.2);
\end{tikzpicture}\end{tabular}\par
\end{minipage}}}
\end{center}

\bigskip\noindent
However, this is unacceptable, because it relies on unspecified human intuition to guess how to fill in the parts indicated by triple dots. Instead, we define them computationally, using standardized {\it computational} presentation tableaus as in these examples:

\begin{center}
\noindent\adjustbox{max width=\textwidth, max totalheight=0.86\textheight}{\setlength{\fboxsep}{6pt}\colorbox{galbeige}{\begin{tabular}[t]{@{}l@{}}{\small Natural numbers $\N$ as Peano Arithmetic:}\\[-1pt]\setlength{\fboxsep}{0pt}\fcolorbox{black}{white}{\begin{tabular}{@{\hspace{5pt}}r@{\,:\ }l@{\ =\ }l@{\hspace{5pt}}}
\multicolumn{3}{@{\hspace{5pt}}l@{\hspace{5pt}}}{\rule{0pt}{1.3em}\textcolor{black}{$|S_1|=\aleph_0$;\ \ {\fontfamily{pcr}\selectfont "0"}$\,\in S_1$;\ \ $i\in S_1\Rightarrow$ {\fontfamily{pcr}\selectfont "0ic"}$\,\in S_1$}}\\[2pt]\hline
\rule{0pt}{1.25em}$0$ & $S_1$ & {\fontfamily{pcr}\selectfont "0"}\\
$'$ & $S_1\to S_1$ & {\fontfamily{pcr}\selectfont "0ic"}\\
$+$ & $S_1\times S_1\to S_1$ & {\fontfamily{pcr}\selectfont "iib0jcfjd"}\\
$\cdot$ & $S_1\times S_1\to S_1$ & {\fontfamily{pcr}\selectfont "iibjfj+0d"}\\[3pt]
\end{tabular}}
\\[8pt]
{\small Basic integer arithmetic $(\Z,0,-,+,\cdot)$:}\\[-1pt]\setlength{\fboxsep}{0pt}\fcolorbox{black}{white}{\begin{tabular}{@{\hspace{5pt}}r@{\,:\ }l@{\ =\ }l@{\hspace{5pt}}}
\multicolumn{3}{@{\hspace{5pt}}l@{\hspace{5pt}}}{\rule{0pt}{1.3em}\textcolor{black!50}{{\fontfamily{pcr}\selectfont "0"}$\,\in \mathbb N$;\ \ $i\in \mathbb N\Rightarrow$ {\fontfamily{pcr}\selectfont "0ic"}$\,\in \mathbb N$}}\\
\multicolumn{3}{@{\hspace{5pt}}l@{\hspace{5pt}}}{\textcolor{black}{$|S_1|=\aleph_0$;\ \ $i\in \mathbb N\Rightarrow$ {\fontfamily{pcr}\selectfont "i0c"}$\,\in S_1$, {\fontfamily{pcr}\selectfont "0ic"}$\,\in S_1$}}\\[2pt]\hline
\rule{0pt}{1.25em}$0$ & $S_1$ & {\fontfamily{pcr}\selectfont "00c"}\\
\multicolumn{2}{@{\hspace{5pt}}r@{\ =\ }}{\textcolor{black!50}{\texttt{+}}} & \textcolor{black!50}{{\fontfamily{pcr}\selectfont "iib0jcfjd"}}\\
\multicolumn{2}{@{\hspace{5pt}}r@{\ =\ }}{\textcolor{black!50}{\texttt{A}}} & \textcolor{black!50}{{\fontfamily{pcr}\selectfont "iaja+ibjb+c"}}\\
\multicolumn{2}{@{\hspace{5pt}}r@{\ =\ }}{\textcolor{black!50}{\texttt{1}}} & \textcolor{black!50}{{\fontfamily{pcr}\selectfont "00c"}}\\
\multicolumn{2}{@{\hspace{5pt}}r@{\ =\ }}{\textcolor{black!50}{\texttt{$<$}}} & \textcolor{black!50}{{\fontfamily{pcr}\selectfont "jiibjbf1d0d"}}\\
\multicolumn{2}{@{\hspace{5pt}}r@{\ =\ }}{\textcolor{black!50}{\texttt{$-$}}} & \textcolor{black!50}{{\fontfamily{pcr}\selectfont "jibjbfid"}}\\
\multicolumn{2}{@{\hspace{5pt}}r@{\ =\ }}{\textcolor{black!50}{\texttt{R}}} & \textcolor{black!50}{{\fontfamily{pcr}\selectfont "iaib$<$0ibia$-$ciaib$-$0cd"}}\\
$+$ & $S_1\times S_1\to S_1$ & {\fontfamily{pcr}\selectfont "ijAR"}\\
\multicolumn{2}{@{\hspace{5pt}}r@{\ =\ }}{\textcolor{black!50}{\texttt{S}}} & \textcolor{black!50}{{\fontfamily{pcr}\selectfont "iajb+ibja+c"}}\\
$-$ & $S_1\times S_1\to S_1$ & {\fontfamily{pcr}\selectfont "ijSR"}\\
\multicolumn{2}{@{\hspace{5pt}}r@{\ =\ }}{\textcolor{black!50}{\texttt{$\ast$}}} & \textcolor{black!50}{{\fontfamily{pcr}\selectfont "iibjfj+0d"}}\\
\multicolumn{2}{@{\hspace{5pt}}r@{\ =\ }}{\textcolor{black!50}{\texttt{M}}} & \textcolor{black!50}{{\fontfamily{pcr}\selectfont "iaja$\ast$ibjb$\ast$+iajb$\ast$ibja$\ast$+c"}}\\
$\cdot$ & $S_1\times S_1\to S_1$ & {\fontfamily{pcr}\selectfont "ijMR"}\\[3pt]
\end{tabular}}
\\[8pt]
{\small Rational numbers $\Q$ as the prime field of characteristic 0:}\\[-1pt]\setlength{\fboxsep}{0pt}\fcolorbox{black}{white}{\begin{tabular}{@{\hspace{5pt}}r@{\,:\ }l@{\ =\ }l@{\hspace{5pt}}}
\multicolumn{3}{@{\hspace{5pt}}l@{\hspace{5pt}}}{\rule{0pt}{1.3em}\textcolor{black!50}{{\fontfamily{pcr}\selectfont "10c1c"}$\,\in \mathbb Q^+$;\ \ $i\in \mathbb Q^+\Rightarrow$ {\fontfamily{pcr}\selectfont "iaiaib0cAac"}$\,\in \mathbb Q^+$, {\fontfamily{pcr}\selectfont "iaib0cAibc"}$\,\in \mathbb Q^+$}}\\
\multicolumn{3}{@{\hspace{5pt}}l@{\hspace{5pt}}}{\textcolor{black}{$|S_1|=\aleph_0$;\ \ {\fontfamily{pcr}\selectfont "00c1c"}$\,\in S_1$;\ \ $i\in \mathbb Q^+\Rightarrow$ {\fontfamily{pcr}\selectfont "i"}$\,\in S_1$, {\fontfamily{pcr}\selectfont "i$\circleddash$"}$\,\in S_1$}}\\[2pt]\hline
\multicolumn{2}{@{\hspace{5pt}}r@{\ =\ }}{\rule{0pt}{1.25em}\textcolor{black!50}{\texttt{1}}} & \textcolor{black!50}{{\fontfamily{pcr}\selectfont "00c"}}\\
\multicolumn{2}{@{\hspace{5pt}}r@{\ =\ }}{\textcolor{black!50}{\texttt{+}}} & \textcolor{black!50}{{\fontfamily{pcr}\selectfont "iib0jcfjd"}}\\
\multicolumn{2}{@{\hspace{5pt}}r@{\ =\ }}{\textcolor{black!50}{\texttt{A}}} & \textcolor{black!50}{{\fontfamily{pcr}\selectfont "iaja+ibjb+c"}}\\
\multicolumn{2}{@{\hspace{5pt}}r@{\ =\ }}{\textcolor{black!50}{\texttt{N}}} & \textcolor{black!50}{{\fontfamily{pcr}\selectfont "ibiac"}}\\
\multicolumn{2}{@{\hspace{5pt}}r@{\ =\ }}{\textcolor{black!50}{\texttt{$\circleddash$}}} & \textcolor{black!50}{{\fontfamily{pcr}\selectfont "iaNibc"}}\\
$0$ & $S_1$ & {\fontfamily{pcr}\selectfont "00c1c"}\\
$1$ & $S_1$ & {\fontfamily{pcr}\selectfont "10c1c"}\\
$-$ & $S_1\to S_1$ & {\fontfamily{pcr}\selectfont "iaNibc"}\\
\multicolumn{2}{@{\hspace{5pt}}r@{\ =\ }}{\textcolor{black!50}{\texttt{$\ast$}}} & \textcolor{black!50}{{\fontfamily{pcr}\selectfont "iibjfj+0d"}}\\
\multicolumn{2}{@{\hspace{5pt}}r@{\ =\ }}{\textcolor{black!50}{\texttt{M}}} & \textcolor{black!50}{{\fontfamily{pcr}\selectfont "iaja$\ast$ibjb$\ast$+iajb$\ast$ibja$\ast$+c"}}\\
\multicolumn{2}{@{\hspace{5pt}}r@{\ =\ }}{\textcolor{black!50}{\texttt{$\odot$}}} & \textcolor{black!50}{{\fontfamily{pcr}\selectfont "ij0cM"}}\\
\multicolumn{2}{@{\hspace{5pt}}r@{\ =\ }}{\textcolor{black!50}{\texttt{$<$}}} & \textcolor{black!50}{{\fontfamily{pcr}\selectfont "jiibjbf1d0d"}}\\
\multicolumn{2}{@{\hspace{5pt}}r@{\ =\ }}{\textcolor{black!50}{\texttt{$-$}}} & \textcolor{black!50}{{\fontfamily{pcr}\selectfont "jibjbfid"}}\\
\multicolumn{2}{@{\hspace{5pt}}r@{\ =\ }}{\textcolor{black!50}{\texttt{R}}} & \textcolor{black!50}{{\fontfamily{pcr}\selectfont "iaib$<$0ibia$-$ciaib$-$0cd"}}\\
\multicolumn{2}{@{\hspace{5pt}}r@{\ =\ }}{\textcolor{black!50}{\texttt{Q}}} & \textcolor{black!50}{{\fontfamily{pcr}\selectfont "ij$<$00ij$-$jfcd"}}\\
\multicolumn{2}{@{\hspace{5pt}}r@{\ =\ }}{\textcolor{black!50}{\texttt{\%}}} & \textcolor{black!50}{{\fontfamily{pcr}\selectfont "ij$<$iij$-$jfd"}}\\
\multicolumn{2}{@{\hspace{5pt}}r@{\ =\ }}{\textcolor{black!50}{\texttt{G}}} & \textcolor{black!50}{{\fontfamily{pcr}\selectfont "jjij\%fid"}}\\
\multicolumn{2}{@{\hspace{5pt}}r@{\ =\ }}{\textcolor{black!50}{\texttt{g}}} & \textcolor{black!50}{{\fontfamily{pcr}\selectfont "iaaiab+ibG"}}\\
\multicolumn{2}{@{\hspace{5pt}}r@{\ =\ }}{\textcolor{black!50}{\texttt{Y}}} & \textcolor{black!50}{{\fontfamily{pcr}\selectfont "iaaigQiabigQcibigQc"}}\\
$+$ & $S_1\times S_1\to S_1$ & {\fontfamily{pcr}\selectfont "iajb$\odot$jaib$\odot$ARibjb$\ast$cY"}\\
$\cdot$ & $S_1\times S_1\to S_1$ & {\fontfamily{pcr}\selectfont "iajaMibjb$\ast$cY"}\\[3pt]
\end{tabular}}
\\[8pt]
{\small Integer arithmetic with successor $(\Z,0,{}',-,+,\cdot)$:}\\[-1pt]\setlength{\fboxsep}{0pt}\fcolorbox{black}{white}{\begin{tabular}{@{\hspace{5pt}}r@{\,:\ }l@{\ =\ }l@{\hspace{5pt}}}
\multicolumn{3}{@{\hspace{5pt}}l@{\hspace{5pt}}}{\rule{0pt}{1.3em}\textcolor{black!50}{{\fontfamily{pcr}\selectfont "0"}$\,\in \mathbb N$;\ \ $i\in \mathbb N\Rightarrow$ {\fontfamily{pcr}\selectfont "0ic"}$\,\in \mathbb N$}}\\
\multicolumn{3}{@{\hspace{5pt}}l@{\hspace{5pt}}}{\textcolor{black}{$|S_1|=\aleph_0$;\ \ $i\in \mathbb N\Rightarrow$ {\fontfamily{pcr}\selectfont "i0c"}$\,\in S_1$, {\fontfamily{pcr}\selectfont "0ic"}$\,\in S_1$}}\\[2pt]\hline
\rule{0pt}{1.25em}$0$ & $S_1$ & {\fontfamily{pcr}\selectfont "00c"}\\
\multicolumn{2}{@{\hspace{5pt}}r@{\ =\ }}{\textcolor{black!50}{\texttt{+}}} & \textcolor{black!50}{{\fontfamily{pcr}\selectfont "iib0jcfjd"}}\\
\multicolumn{2}{@{\hspace{5pt}}r@{\ =\ }}{\textcolor{black!50}{\texttt{A}}} & \textcolor{black!50}{{\fontfamily{pcr}\selectfont "iaja+ibjb+c"}}\\
\multicolumn{2}{@{\hspace{5pt}}r@{\ =\ }}{\textcolor{black!50}{\texttt{1}}} & \textcolor{black!50}{{\fontfamily{pcr}\selectfont "00c"}}\\
\multicolumn{2}{@{\hspace{5pt}}r@{\ =\ }}{\textcolor{black!50}{\texttt{$<$}}} & \textcolor{black!50}{{\fontfamily{pcr}\selectfont "jiibjbf1d0d"}}\\
\multicolumn{2}{@{\hspace{5pt}}r@{\ =\ }}{\textcolor{black!50}{\texttt{$-$}}} & \textcolor{black!50}{{\fontfamily{pcr}\selectfont "jibjbfid"}}\\
\multicolumn{2}{@{\hspace{5pt}}r@{\ =\ }}{\textcolor{black!50}{\texttt{R}}} & \textcolor{black!50}{{\fontfamily{pcr}\selectfont "iaib$<$0ibia$-$ciaib$-$0cd"}}\\
\multicolumn{2}{@{\hspace{5pt}}r@{\ =\ }}{\textcolor{black!50}{\texttt{$\boxplus$}}} & \textcolor{black!50}{{\fontfamily{pcr}\selectfont "ijAR"}}\\
$'$ & $S_1\to S_1$ & {\fontfamily{pcr}\selectfont "i00c0c$\boxplus$"}\\
$-$ & $S_1\to S_1$ & {\fontfamily{pcr}\selectfont "ibiac"}\\
$+$ & $S_1\times S_1\to S_1$ & {\fontfamily{pcr}\selectfont "ijAR"}\\
\multicolumn{2}{@{\hspace{5pt}}r@{\ =\ }}{\textcolor{black!50}{\texttt{$\ast$}}} & \textcolor{black!50}{{\fontfamily{pcr}\selectfont "iibjfj+0d"}}\\
\multicolumn{2}{@{\hspace{5pt}}r@{\ =\ }}{\textcolor{black!50}{\texttt{M}}} & \textcolor{black!50}{{\fontfamily{pcr}\selectfont "iaja$\ast$ibjb$\ast$+iajb$\ast$ibja$\ast$+c"}}\\
$\cdot$ & $S_1\times S_1\to S_1$ & {\fontfamily{pcr}\selectfont "ijMR"}\\[3pt]
\end{tabular}}
\end{tabular}\hspace{14pt}\begin{tabular}[t]{@{}l@{}}{\small Natural number total order $(\N,\le)$:}\\[-1pt]\setlength{\fboxsep}{0pt}\fcolorbox{black}{white}{\begin{tabular}{@{\hspace{5pt}}r@{\,:\ }l@{\ =\ }l@{\hspace{5pt}}}
\multicolumn{3}{@{\hspace{5pt}}l@{\hspace{5pt}}}{\rule{0pt}{1.3em}\textcolor{black}{$|S_1|=\aleph_0$;\ \ {\fontfamily{pcr}\selectfont "0"}$\,\in S_1$;\ \ $i\in S_1\Rightarrow$ {\fontfamily{pcr}\selectfont "0ic"}$\,\in S_1$}}\\
\multicolumn{3}{@{\hspace{5pt}}l@{\hspace{5pt}}}{\textcolor{red}{$|S_2|=2$;\ \ {\fontfamily{pcr}\selectfont "0"}, {\fontfamily{pcr}\selectfont "00c"}$\,\in S_2$}}\\[2pt]\hline
\multicolumn{2}{@{\hspace{5pt}}r@{\ =\ }}{\rule{0pt}{1.25em}\textcolor{black!50}{\texttt{1}}} & \textcolor{black!50}{{\fontfamily{pcr}\selectfont "00c"}}\\
\multicolumn{2}{@{\hspace{5pt}}r@{\ =\ }}{\textcolor{black!50}{\texttt{$<$}}} & \textcolor{black!50}{{\fontfamily{pcr}\selectfont "jiibjbf1d0d"}}\\
\multicolumn{2}{@{\hspace{5pt}}r@{\ =\ }}{\textcolor{black!50}{\texttt{$\lnot$}}} & \textcolor{black!50}{{\fontfamily{pcr}\selectfont "i01d"}}\\
$\le$ & $S_1\times S_1\to S_2$ & {\fontfamily{pcr}\selectfont "ji$<$$\lnot$"}\\[3pt]
\end{tabular}}
\\[8pt]
{\small Integer total order $(\Z,\le)$:}\\[-1pt]\setlength{\fboxsep}{0pt}\fcolorbox{black}{white}{\begin{tabular}{@{\hspace{5pt}}r@{\,:\ }l@{\ =\ }l@{\hspace{5pt}}}
\multicolumn{3}{@{\hspace{5pt}}l@{\hspace{5pt}}}{\rule{0pt}{1.3em}\textcolor{black!50}{{\fontfamily{pcr}\selectfont "0"}$\,\in \mathbb N$;\ \ $i\in \mathbb N\Rightarrow$ {\fontfamily{pcr}\selectfont "0ic"}$\,\in \mathbb N$}}\\
\multicolumn{3}{@{\hspace{5pt}}l@{\hspace{5pt}}}{\textcolor{black}{$|S_1|=\aleph_0$;\ \ $i\in \mathbb N\Rightarrow$ {\fontfamily{pcr}\selectfont "i0c"}$\,\in S_1$, {\fontfamily{pcr}\selectfont "0ic"}$\,\in S_1$}}\\
\multicolumn{3}{@{\hspace{5pt}}l@{\hspace{5pt}}}{\textcolor{red}{$|S_2|=2$;\ \ {\fontfamily{pcr}\selectfont "0"}, {\fontfamily{pcr}\selectfont "00c"}$\,\in S_2$}}\\[2pt]\hline
\multicolumn{2}{@{\hspace{5pt}}r@{\ =\ }}{\rule{0pt}{1.25em}\textcolor{black!50}{\texttt{+}}} & \textcolor{black!50}{{\fontfamily{pcr}\selectfont "iib0jcfjd"}}\\
\multicolumn{2}{@{\hspace{5pt}}r@{\ =\ }}{\textcolor{black!50}{\texttt{1}}} & \textcolor{black!50}{{\fontfamily{pcr}\selectfont "00c"}}\\
\multicolumn{2}{@{\hspace{5pt}}r@{\ =\ }}{\textcolor{black!50}{\texttt{$<$}}} & \textcolor{black!50}{{\fontfamily{pcr}\selectfont "jiibjbf1d0d"}}\\
\multicolumn{2}{@{\hspace{5pt}}r@{\ =\ }}{\textcolor{black!50}{\texttt{$\lnot$}}} & \textcolor{black!50}{{\fontfamily{pcr}\selectfont "i01d"}}\\
$\le$ & $S_1\times S_1\to S_2$ & {\fontfamily{pcr}\selectfont "jaib+iajb+$<$$\lnot$"}\\[3pt]
\end{tabular}}
\\[8pt]
{\small Rational number total order $(\Q,\le)$:}\\[-1pt]\setlength{\fboxsep}{0pt}\fcolorbox{black}{white}{\begin{tabular}{@{\hspace{5pt}}r@{\,:\ }l@{\ =\ }l@{\hspace{5pt}}}
\multicolumn{3}{@{\hspace{5pt}}l@{\hspace{5pt}}}{\rule{0pt}{1.3em}\textcolor{black!50}{{\fontfamily{pcr}\selectfont "10c1c"}$\,\in \mathbb Q^+$;\ \ $i\in \mathbb Q^+\Rightarrow$ {\fontfamily{pcr}\selectfont "iaiaib0cAac"}$\,\in \mathbb Q^+$, {\fontfamily{pcr}\selectfont "iaib0cAibc"}$\,\in \mathbb Q^+$}}\\
\multicolumn{3}{@{\hspace{5pt}}l@{\hspace{5pt}}}{\textcolor{black}{$|S_1|=\aleph_0$;\ \ {\fontfamily{pcr}\selectfont "00c1c"}$\,\in S_1$;\ \ $i\in \mathbb Q^+\Rightarrow$ {\fontfamily{pcr}\selectfont "i"}$\,\in S_1$, {\fontfamily{pcr}\selectfont "i$\circleddash$"}$\,\in S_1$}}\\
\multicolumn{3}{@{\hspace{5pt}}l@{\hspace{5pt}}}{\textcolor{red}{$|S_2|=2$;\ \ {\fontfamily{pcr}\selectfont "0"}, {\fontfamily{pcr}\selectfont "00c"}$\,\in S_2$}}\\[2pt]\hline
\multicolumn{2}{@{\hspace{5pt}}r@{\ =\ }}{\rule{0pt}{1.25em}\textcolor{black!50}{\texttt{1}}} & \textcolor{black!50}{{\fontfamily{pcr}\selectfont "00c"}}\\
\multicolumn{2}{@{\hspace{5pt}}r@{\ =\ }}{\textcolor{black!50}{\texttt{+}}} & \textcolor{black!50}{{\fontfamily{pcr}\selectfont "iib0jcfjd"}}\\
\multicolumn{2}{@{\hspace{5pt}}r@{\ =\ }}{\textcolor{black!50}{\texttt{A}}} & \textcolor{black!50}{{\fontfamily{pcr}\selectfont "iaja+ibjb+c"}}\\
\multicolumn{2}{@{\hspace{5pt}}r@{\ =\ }}{\textcolor{black!50}{\texttt{N}}} & \textcolor{black!50}{{\fontfamily{pcr}\selectfont "ibiac"}}\\
\multicolumn{2}{@{\hspace{5pt}}r@{\ =\ }}{\textcolor{black!50}{\texttt{$\circleddash$}}} & \textcolor{black!50}{{\fontfamily{pcr}\selectfont "iaNibc"}}\\
\multicolumn{2}{@{\hspace{5pt}}r@{\ =\ }}{\textcolor{black!50}{\texttt{$\ast$}}} & \textcolor{black!50}{{\fontfamily{pcr}\selectfont "iibjfj+0d"}}\\
\multicolumn{2}{@{\hspace{5pt}}r@{\ =\ }}{\textcolor{black!50}{\texttt{M}}} & \textcolor{black!50}{{\fontfamily{pcr}\selectfont "iaja$\ast$ibjb$\ast$+iajb$\ast$ibja$\ast$+c"}}\\
\multicolumn{2}{@{\hspace{5pt}}r@{\ =\ }}{\textcolor{black!50}{\texttt{$\odot$}}} & \textcolor{black!50}{{\fontfamily{pcr}\selectfont "ij0cM"}}\\
\multicolumn{2}{@{\hspace{5pt}}r@{\ =\ }}{\textcolor{black!50}{\texttt{$<$}}} & \textcolor{black!50}{{\fontfamily{pcr}\selectfont "jiibjbf1d0d"}}\\
\multicolumn{2}{@{\hspace{5pt}}r@{\ =\ }}{\textcolor{black!50}{\texttt{$\lnot$}}} & \textcolor{black!50}{{\fontfamily{pcr}\selectfont "i01d"}}\\
\multicolumn{2}{@{\hspace{5pt}}r@{\ =\ }}{\textcolor{black!50}{\texttt{L}}} & \textcolor{black!50}{{\fontfamily{pcr}\selectfont "jaib+iajb+$<$$\lnot$"}}\\
$\le$ & $S_1\times S_1\to S_2$ & {\fontfamily{pcr}\selectfont "iajb$\odot$jaib$\odot$L"}\\[3pt]
\end{tabular}}
\\[8pt]
{\small Magma of integers under subtraction $(\Z,-)$:}\\[-1pt]\setlength{\fboxsep}{0pt}\fcolorbox{black}{white}{\begin{tabular}{@{\hspace{5pt}}r@{\,:\ }l@{\ =\ }l@{\hspace{5pt}}}
\multicolumn{3}{@{\hspace{5pt}}l@{\hspace{5pt}}}{\rule{0pt}{1.3em}\textcolor{black!50}{{\fontfamily{pcr}\selectfont "0"}$\,\in \mathbb N$;\ \ $i\in \mathbb N\Rightarrow$ {\fontfamily{pcr}\selectfont "0ic"}$\,\in \mathbb N$}}\\
\multicolumn{3}{@{\hspace{5pt}}l@{\hspace{5pt}}}{\textcolor{black}{$|S_1|=\aleph_0$;\ \ $i\in \mathbb N\Rightarrow$ {\fontfamily{pcr}\selectfont "i0c"}$\,\in S_1$, {\fontfamily{pcr}\selectfont "0ic"}$\,\in S_1$}}\\[2pt]\hline
\multicolumn{2}{@{\hspace{5pt}}r@{\ =\ }}{\rule{0pt}{1.25em}\textcolor{black!50}{\texttt{+}}} & \textcolor{black!50}{{\fontfamily{pcr}\selectfont "iib0jcfjd"}}\\
\multicolumn{2}{@{\hspace{5pt}}r@{\ =\ }}{\textcolor{black!50}{\texttt{S}}} & \textcolor{black!50}{{\fontfamily{pcr}\selectfont "iajb+ibja+c"}}\\
\multicolumn{2}{@{\hspace{5pt}}r@{\ =\ }}{\textcolor{black!50}{\texttt{1}}} & \textcolor{black!50}{{\fontfamily{pcr}\selectfont "00c"}}\\
\multicolumn{2}{@{\hspace{5pt}}r@{\ =\ }}{\textcolor{black!50}{\texttt{$<$}}} & \textcolor{black!50}{{\fontfamily{pcr}\selectfont "jiibjbf1d0d"}}\\
\multicolumn{2}{@{\hspace{5pt}}r@{\ =\ }}{\textcolor{black!50}{\texttt{$-$}}} & \textcolor{black!50}{{\fontfamily{pcr}\selectfont "jibjbfid"}}\\
\multicolumn{2}{@{\hspace{5pt}}r@{\ =\ }}{\textcolor{black!50}{\texttt{R}}} & \textcolor{black!50}{{\fontfamily{pcr}\selectfont "iaib$<$0ibia$-$ciaib$-$0cd"}}\\
$-$ & $S_1\times S_1\to S_1$ & {\fontfamily{pcr}\selectfont "ijSR"}\\[3pt]
\end{tabular}}
\\[8pt]
{\small Congruence mod 2 equivalence relation on $\N$:}\\[-1pt]\setlength{\fboxsep}{0pt}\fcolorbox{black}{white}{\begin{tabular}{@{\hspace{5pt}}r@{\,:\ }l@{\ =\ }l@{\hspace{5pt}}}
\multicolumn{3}{@{\hspace{5pt}}l@{\hspace{5pt}}}{\rule{0pt}{1.3em}\textcolor{black}{$|S_1|=\aleph_0$;\ \ {\fontfamily{pcr}\selectfont "0"}$\,\in S_1$;\ \ $i\in S_1\Rightarrow$ {\fontfamily{pcr}\selectfont "0ic"}$\,\in S_1$}}\\
\multicolumn{3}{@{\hspace{5pt}}l@{\hspace{5pt}}}{\textcolor{red}{$|S_2|=2$;\ \ {\fontfamily{pcr}\selectfont "0"}, {\fontfamily{pcr}\selectfont "00c"}$\,\in S_2$}}\\[2pt]\hline
\multicolumn{2}{@{\hspace{5pt}}r@{\ =\ }}{\rule{0pt}{1.25em}\textcolor{black!50}{\texttt{1}}} & \textcolor{black!50}{{\fontfamily{pcr}\selectfont "00c"}}\\
\multicolumn{2}{@{\hspace{5pt}}r@{\ =\ }}{\textcolor{black!50}{\texttt{$\lnot$}}} & \textcolor{black!50}{{\fontfamily{pcr}\selectfont "i01d"}}\\
\multicolumn{2}{@{\hspace{5pt}}r@{\ =\ }}{\textcolor{black!50}{\texttt{E}}} & \textcolor{black!50}{{\fontfamily{pcr}\selectfont "iibf$\lnot$1d"}}\\
\multicolumn{2}{@{\hspace{5pt}}r@{\ =\ }}{\textcolor{black!50}{\texttt{e}}} & \textcolor{black!50}{{\fontfamily{pcr}\selectfont "ijiajafibjbf0d0dj01dd"}}\\
$\sim$ & $S_1\times S_1\to S_2$ & {\fontfamily{pcr}\selectfont "iEjEe"}\\[3pt]
\end{tabular}}
\\[8pt]
{\small Distributive lattice of divisibility $(\N^+,\mid)$:}\\[-1pt]\setlength{\fboxsep}{0pt}\fcolorbox{black}{white}{\begin{tabular}{@{\hspace{5pt}}r@{\,:\ }l@{\ =\ }l@{\hspace{5pt}}}
\multicolumn{3}{@{\hspace{5pt}}l@{\hspace{5pt}}}{\rule{0pt}{1.3em}\textcolor{black}{$|S_1|=\aleph_0$;\ \ {\fontfamily{pcr}\selectfont "00c"}$\,\in S_1$;\ \ $i\in S_1\Rightarrow$ {\fontfamily{pcr}\selectfont "0ic"}$\,\in S_1$}}\\
\multicolumn{3}{@{\hspace{5pt}}l@{\hspace{5pt}}}{\textcolor{red}{$|S_2|=2$;\ \ {\fontfamily{pcr}\selectfont "0"}, {\fontfamily{pcr}\selectfont "00c"}$\,\in S_2$}}\\[2pt]\hline
\multicolumn{2}{@{\hspace{5pt}}r@{\ =\ }}{\rule{0pt}{1.25em}\textcolor{black!50}{\texttt{1}}} & \textcolor{black!50}{{\fontfamily{pcr}\selectfont "00c"}}\\
\multicolumn{2}{@{\hspace{5pt}}r@{\ =\ }}{\textcolor{black!50}{\texttt{$<$}}} & \textcolor{black!50}{{\fontfamily{pcr}\selectfont "jiibjbf1d0d"}}\\
\multicolumn{2}{@{\hspace{5pt}}r@{\ =\ }}{\textcolor{black!50}{\texttt{$-$}}} & \textcolor{black!50}{{\fontfamily{pcr}\selectfont "jibjbfid"}}\\
\multicolumn{2}{@{\hspace{5pt}}r@{\ =\ }}{\textcolor{black!50}{\texttt{$\lnot$}}} & \textcolor{black!50}{{\fontfamily{pcr}\selectfont "i01d"}}\\
\multicolumn{2}{@{\hspace{5pt}}r@{\ =\ }}{\textcolor{black!50}{\texttt{\%}}} & \textcolor{black!50}{{\fontfamily{pcr}\selectfont "ij$<$iij$-$jfd"}}\\
\multicolumn{2}{@{\hspace{5pt}}r@{\ =\ }}{\textcolor{black!50}{\texttt{$\Delta$}}} & \textcolor{black!50}{{\fontfamily{pcr}\selectfont "ij\%$\lnot$"}}\\
$\mid$ & $S_1\times S_1\to S_2$ & {\fontfamily{pcr}\selectfont "ji$\Delta$"}\\[3pt]
\end{tabular}}
\\[8pt]
{\small Divisibility preorder on $\Z$:}\\[-1pt]\setlength{\fboxsep}{0pt}\fcolorbox{black}{white}{\begin{tabular}{@{\hspace{5pt}}r@{\,:\ }l@{\ =\ }l@{\hspace{5pt}}}
\multicolumn{3}{@{\hspace{5pt}}l@{\hspace{5pt}}}{\rule{0pt}{1.3em}\textcolor{black!50}{{\fontfamily{pcr}\selectfont "0"}$\,\in \mathbb N$;\ \ $i\in \mathbb N\Rightarrow$ {\fontfamily{pcr}\selectfont "0ic"}$\,\in \mathbb N$}}\\
\multicolumn{3}{@{\hspace{5pt}}l@{\hspace{5pt}}}{\textcolor{black}{$|S_1|=\aleph_0$;\ \ $i\in \mathbb N\Rightarrow$ {\fontfamily{pcr}\selectfont "i0c"}$\,\in S_1$, {\fontfamily{pcr}\selectfont "0ic"}$\,\in S_1$}}\\
\multicolumn{3}{@{\hspace{5pt}}l@{\hspace{5pt}}}{\textcolor{red}{$|S_2|=2$;\ \ {\fontfamily{pcr}\selectfont "0"}, {\fontfamily{pcr}\selectfont "00c"}$\,\in S_2$}}\\[2pt]\hline
\multicolumn{2}{@{\hspace{5pt}}r@{\ =\ }}{\rule{0pt}{1.25em}\textcolor{black!50}{\texttt{+}}} & \textcolor{black!50}{{\fontfamily{pcr}\selectfont "iib0jcfjd"}}\\
\multicolumn{2}{@{\hspace{5pt}}r@{\ =\ }}{\textcolor{black!50}{\texttt{1}}} & \textcolor{black!50}{{\fontfamily{pcr}\selectfont "00c"}}\\
\multicolumn{2}{@{\hspace{5pt}}r@{\ =\ }}{\textcolor{black!50}{\texttt{$<$}}} & \textcolor{black!50}{{\fontfamily{pcr}\selectfont "jiibjbf1d0d"}}\\
\multicolumn{2}{@{\hspace{5pt}}r@{\ =\ }}{\textcolor{black!50}{\texttt{$-$}}} & \textcolor{black!50}{{\fontfamily{pcr}\selectfont "jibjbfid"}}\\
\multicolumn{2}{@{\hspace{5pt}}r@{\ =\ }}{\textcolor{black!50}{\texttt{$\lnot$}}} & \textcolor{black!50}{{\fontfamily{pcr}\selectfont "i01d"}}\\
\multicolumn{2}{@{\hspace{5pt}}r@{\ =\ }}{\textcolor{black!50}{\texttt{\%}}} & \textcolor{black!50}{{\fontfamily{pcr}\selectfont "ij$<$iij$-$jfd"}}\\
\multicolumn{2}{@{\hspace{5pt}}r@{\ =\ }}{\textcolor{black!50}{\texttt{$\Delta$}}} & \textcolor{black!50}{{\fontfamily{pcr}\selectfont "ij\%$\lnot$"}}\\
$\lesssim$ & $S_1\times S_1\to S_2$ & {\fontfamily{pcr}\selectfont "iaib+jajb+iaib+$\Delta$jajb+01dd"}\\[3pt]
\end{tabular}}
\end{tabular}}}
\end{center}

Here the mysterious-looking strings such as ``\texttt{jibjbfid}'' are ultra-short computer programs that  fully define each function in a minimal LISP-like language  called MLISP that I created for this purpose and will describe below.
The additional definitions shown in gray are helper functions, not part of the set of
functions defining the presentation.

\subsection{Defining functions computationally with MLISP}\label{MLISPSec}

MLISP is a minimal yet Turing-complete programming language helpful for defining functions. 
A program is a string of single-character symbols, each acting in turn on a stack of binary trees, as when parsing reverse Polish notation. MLISP has only the 9 basic 1-symbol operations\footnote{The even more 
minimal (but still Turing-complete) language mLISP using only 7 symbols is described and implemented
at \url{https://tegmark.org/mathtree/mlisp7.html}; we stick with the 9-symbol version 
here because its programs tend to be slightly shorter and more human-readable. 
}
\texttt{a b c d f 0 i j k} defined as follows:
\begin{table}[H]\centering
\begin{tabular}{@{}cp{12.5cm}c@{}}\hline
\textbf{Operation} & \textbf{What it returns} & \textbf{Args}\\\hline
\texttt{a} & {\bf a}-part: the first part of the pair $i$ & 1\\
\texttt{b} & {\bf b}-part: the second part of the pair $i$& 1\\
\texttt{c} & {\bf c}ombine: the ordered pair $(i,j)$ & 2\\
\texttt{d} & {\bf d}ecide: $j$ if $i$ is a pair, otherwise $k$ (only the chosen branch is evaluated) & 3\\
\texttt{f} & {\bf f}unction being defined: for recursion; as many arguments as that function & \\
\texttt{0} & the atom $0$ & 0\\
\texttt{i} & the first input& 0\\
\texttt{j} & the second input& 0\\
\texttt{k} & the third input& 0\\
\hline\end{tabular}
\caption{The nine built-in MLISP commands}
\label{tab:MLISPdefTable}
\end{table}

\begin{figure}[H]
    \centering
    \includegraphics[width=\textwidth]{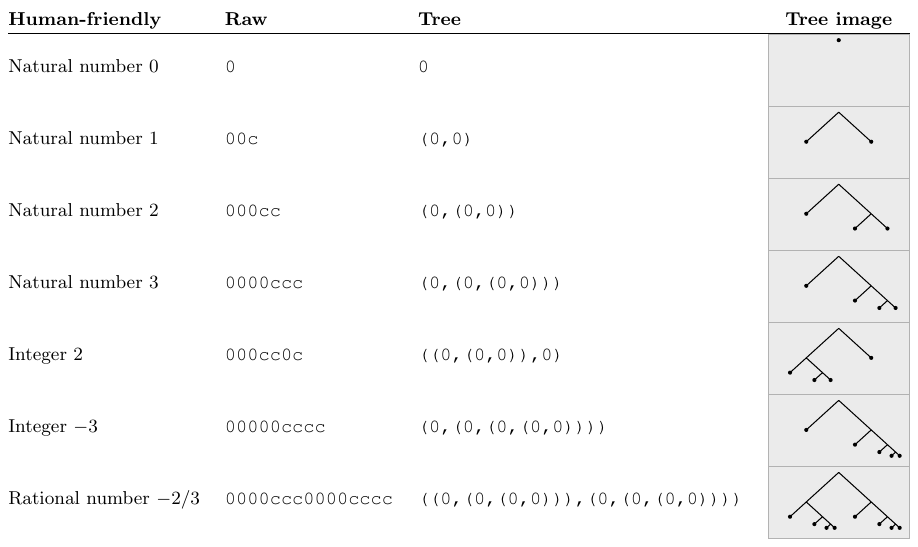}
    \caption{MLISP variables are MLISP programs using 0 and c in reverse Polish notation (``raw" column), and can be interpreted as, \eg,  binary trees, natural numbers, integers and rational numbers.}
    \label{TreeRepFig}
\end{figure}

The computational logic of the language is independent of how the binary trees are represented.
As illustrated in Figure~\ref{TreeRepFig}, it is intuitive to write them as nested lists such as \texttt{(0,(0,(0,0)))} and to draw them as actual trees.
In practice, under the slogan ``data is code", we instead represent them as the MLISP program generating them (the ``raw" column). So the example above corresponds to ``0000ccc'', which is simply the nested list in reverse Polish notation.
There are many convenient ways to interpret such variables \texttt{i}.
We can interpret them as bit strings if we replace ``c" by ``1'', making this example ``0000111''.
We can interpret them as nested lists of zeros, with \texttt{ia} interpreted as the first list element and \texttt{ib} as the rest.
For our purposes of defining presentations, we can further interpret a list of $n$ zeros as the natural number $n$.
We can also interpret a pair of natural numbers $(i,j)$ as an integer $i-j$ whose standard form
is $(i,0)=i$ and $(0,i)=-i$. We can interpret a pair $(i,j)$ of an integer $i$ and a positive natural number $j$ as a rational number $(i,j)=i/j$.
Alternatively, as described in the next section, we can represent natural numbers as binary strings to allow more efficient computation.
The possibilities are endless, but none of this is part of the definition of the MLISP language itself. At \url{https://tegmark.org/mathtree/mlisp.html}, you can
run MLISP programs in an interactive calculator and download reference implementations in Python, C and Lean.

Any undefined symbols are ignored, but you can define new programs: for example, \texttt{+iib0jcfjd\#} defines \texttt{+} to mean \texttt{iib0jcfjd}. The definitions in \tab{tab:MLISPprograms} let you do arithmetic with natural numbers $\N$ represented as nested pairs:
\texttt{0}, \texttt{1=(0,0)}, \texttt{2=(0,(0,0))}, \texttt{3=(0,(0,(0,0)))}.

\bigskip
\bigskip
\bigskip

{\small

\begin{table}[H]\centering\small
\setlength{\tabcolsep}{5.3pt}
\begin{tabular}{|l|l|l|l|l|l|}
\hline
		& 		& 		& \textbf{Input} & \textbf{Output} & \\
\textbf{Symbol} & \textbf{Interpretation} & \textbf{Args} & \textbf{types} & \textbf{type} & \textbf{Definition}\\
\hline
\texttt{1} & one, true & 0 &  & $\mathbb N$ & \texttt{00c}\\
\texttt{2} & two & 0 &  & $\mathbb N$ & \texttt{01c}\\
\texttt{e} & equal ((0,0) if i = j, else 0) & 2 & any, any & 0 or 1 & \texttt{ijiajafibjbf0d0dj01dd}\\
\texttt{$'$} & successor & 1 & $\mathbb N$ & $\mathbb N$ & \texttt{0ic}\\
\texttt{$-$} & minus (undefined for i $<$ j) & 2 & $\mathbb N$, $\mathbb N$ & $\mathbb N$ & \texttt{jibjbfid}\\
\texttt{+} & plus & 2 & $\mathbb N$, $\mathbb N$ & $\mathbb N$ & \texttt{iib0jcfjd}\\
\texttt{*} & times & 2 & $\mathbb N$, $\mathbb N$ & $\mathbb N$ & \texttt{iibjfj+0d}\\
\texttt{\^{}} & power & 2 & $\mathbb N$, $\mathbb N$ & $\mathbb N$ & \texttt{jijbfi*1d}\\
\texttt{!} & factorial & 1 & $\mathbb N$ & $\mathbb N$ & \texttt{iiibf*1d}\\
\texttt{$\neg$} & not & 1 & 0 or 1 & 0 or 1 & \texttt{i01d}\\
\texttt{\&} & and & 2 & 0 or 1, 0 or 1 & 0 or 1 & \texttt{ij10d0d}\\
\texttt{$\lor$} & or & 2 & 0 or 1, 0 or 1 & 0 or 1 & \texttt{i1j10dd}\\
\texttt{$<$} & less than (i$<$j) & 2 & $\mathbb N$, $\mathbb N$ & 0 or 1 & \texttt{jiibjbf1d0d}\\
\texttt{\%} & remainder (i mod j) & 2 & $\mathbb N$, $\mathbb N$ & $\mathbb N$ & \texttt{ij$<$iij$-$jfd}\\
\texttt{$\Delta$} & divisible (i by j) & 2 & $\mathbb N$, $\mathbb N$ & 0 or 1 & \texttt{ij\%$\neg$}\\
\texttt{P} & i indivisible by all of 2$\dots$j & 2 & $\mathbb N$, $\mathbb N$ & 0 or 1 & \texttt{jbij$\Delta$$\neg$ijbf\&1d}\\
\texttt{$\Pi$} & prime (i $\ge$ 2) & 1 & $\mathbb N$ & 0 or 1 & \texttt{iibP}\\
\hline\end{tabular}
\caption{MLISP programs defining arithmetic with natural numbers. The first symbol of each program is defined by the rest.}
\label{tab:MLISPprograms}
\end{table}

\clearpage

\medskip\noindent\textbf{Worked example: \texttt{0ic} (the successor $'$, one argument, with i = 3)}\par\nopagebreak\smallskip
\begin{longtable}{@{}clp{8cm}@{}}\hline
\textbf{Symbol} & \textbf{Stack afterwards} & \textbf{Comment}\\\hline\endhead
\texttt{0} & \texttt{0} & push the atom\\
\texttt{i} & \texttt{0,\ 3} & push the first input\\
\texttt{c} & \texttt{(0,3)} & combine the top two into the pair (0,3), which is the number 4\\
\hline\end{longtable}\addtocounter{table}{-1}

\medskip\noindent\textbf{Worked example: \texttt{jibjbfid} (minus, two arguments: i $-$ j, with i = 3, j = 1)}\par\nopagebreak\smallskip
\begin{longtable}{@{}clp{8cm}@{}}\hline
\textbf{Symbol} & \textbf{Stack afterwards} & \textbf{Comment}\\\hline\endhead
\texttt{j} & \texttt{1} & push the second input\\
\texttt{i} & \texttt{1,\ 3} & push the first input\\
\texttt{b} & \texttt{1,\ 2} & b-part of 3 = (0,2) is 2: the predecessor of i\\
\texttt{j} & \texttt{1,\ 2,\ 1} & \\
\texttt{b} & \texttt{1,\ 2,\ 0} & the predecessor of j\\
\texttt{f} & \texttt{1,\ f(2,0)} & the function itself applied to the top two values (a recursive call, not yet evaluated)\\
\texttt{i} & \texttt{1,\ f(2,0),\ 3} & \\
\texttt{d} & \texttt{d(1,\ f(2,0),\ 3)} & decide: the first of the three is 1, a pair, so the result is the second, f(2,0); with $j=0$, $i$ would have been returned instead\\
\hline\end{longtable}\addtocounter{table}{-1}
}

Note that MLISP programs, like all finite programs, can represent and compute with many exact real numbers (as opposed to finite approximations such as 64-bit floating-point numbers), but they cannot individually specify all real numbers: there are only countably many finite programs, whereas a generic real number requires infinitely many bits to specify.
Indeed, constructivist mathematicians have questioned whether real numbers exist mathematically in the same sense that integers or rational numbers do \cite{BishopBridges1985,Wildberger2005}
and it has been claimed that there is no evidence that real numbers 
are needed to describe our physical universe
\cite{TegmarkMathematicalUniversebook}.

\clearpage

\section{Defining mathematical structures}\label{StructureSec}

Equivalence class constructions are ubiquitous in mathematics, as illustrated in Table~\ref{EquivalenceTable}.
Generally, the goal is to focus on the mathematical properties by stripping away (via a quotient) irrelevant
details related to notation.
The essence of a rational number is its value, not whether we write it as ${1\over 2}$ or  ${3\over 6}$.
Analogously, the essence of a mathematical theory such as Propositional Calculus is which theorems it proves, 
not whether the NOT function is written with the symbol $\sim$ or $\neg$, 
or whether it is defined via the usual four axioms or via the equivalent single axiom of Nicod \cite{Nicod2017}.

\begin{table}[H]
\begin{center}
\setlength{\tabcolsep}{4.5pt}
\begin{tabular}{|l|l|l|l|}
\hline
{\bf Set} 					&{\bf Object} 			&{\bf Equivalence relation} &{\bf Normal form}\\
\hline
integers $\Z$ 				& $(a,b)\in\N\times\N$ 	& $a_1+b_2=a_2+b_1$ 	& smallest $b$\\\hline
rationals $\Q$ 				& $(a,b)\in\Z\times\N$, $b\ne 0$		& $a_1b_2=a_2b_1$	& smallest $b$\\\hline
reals $\R$ 				&Cauchy sequence $a_i\in\Q$ & $a_i-b_i\to 0$ in $\Q$	& --\\\hline
mathematical theories		&formal system			&prove each other's axioms&simplest?\\
						&					&after invertible translation&	            \\\hline
mathematical structures		&presentation			&isomorphism	&simplest?\\\hline
closed mathematical 		&presentation			&isomorphic closures	&simplest?\\
{\hskip1cm}structures (clones)	&					&					&\\\hline
\end{tabular}
\end{center}
\caption{Many important mathematical objects are defined as equivalence classes in a set of other objects,
that is, subsets whose members are all equivalent. It is often convenient to define
a mapping to a standardized ``normal form" serving as a representative of each equivalence class, for example
``1/2" for all rational numbers $(a,b)$ with $b=2a$.
\label{EquivalenceTable}}
\end{table}

In this same spirit, we define a {\it mathematical structure} as an equivalence class of presentations.
Now how exactly is the equivalence relation best defined --- how broad should it be? 
The formal and colloquial usage of the term ``mathematical structure'' among mathematicians is not always consistent, so let us now discuss and motivate 
the choice made in our atlas.

\clearpage
\subsection{Presentation isomorphism}

There is universal agreement that the labels used to refer to set elements and functions have no mathematical meaning.
Two presentations are therefore defined to be {\it isomorphic} if there is a one-to-one renaming of the elements of each set that turns every lookup table (or program output) of one presentation into the corresponding one of the other. 
In a human-friendly tableau, renaming set elements labeled $0,1,\dots$ permutes the digits inside the tables and reorders the rows and columns to match, as in the three isomorphic presentation tableaus defining the group $C_4$ in \fig{PresentationIsomorphismFig} (top panel).
Analogously, the bottom panel illustrates that the presentation of the Boolean functions (NOT, AND) is isomorphic to 
that for (NOT, OR), since they become identical if 0 and 1 are swapped (De Morgan's laws).

\begin{figure}[H]
\centering
\noindent\setlength{\fboxsep}{6pt}\colorbox{galbeige}{\begin{minipage}{\dimexpr\textwidth-2\fboxsep-2\fboxrule\relax}\raggedright
\begin{tabular}[t]{@{}l@{}}\small $C_4$, generator $1$:\\[-6pt]\begin{tikzpicture}[x=0.42cm,y=0.42cm,baseline=(current bounding box.north),every node/.style={font=\sffamily\bfseries,inner sep=0pt}]
\node[anchor=base west,font=\small\normalfont] at (0,1.45) {\textcolor{black}{4}};
\node[text=black] at (0.5,-0.5) {$0$};
\node[text=black] at (1.5,-0.5) {$1$};
\node[text=black] at (2.5,-0.5) {$2$};
\node[text=black] at (3.5,-0.5) {$3$};
\node[text=black] at (0.5,-1.5) {$1$};
\node[text=black] at (1.5,-1.5) {$2$};
\node[text=black] at (2.5,-1.5) {$3$};
\node[text=black] at (3.5,-1.5) {$0$};
\node[text=black] at (0.5,-2.5) {$2$};
\node[text=black] at (1.5,-2.5) {$3$};
\node[text=black] at (2.5,-2.5) {$0$};
\node[text=black] at (3.5,-2.5) {$1$};
\node[text=black] at (0.5,-3.5) {$3$};
\node[text=black] at (1.5,-3.5) {$0$};
\node[text=black] at (2.5,-3.5) {$1$};
\node[text=black] at (3.5,-3.5) {$2$};
\draw[black!25,line width=0.3pt] (1,0)--(1,-4);
\draw[black!25,line width=0.3pt] (2,0)--(2,-4);
\draw[black!25,line width=0.3pt] (3,0)--(3,-4);
\draw[black!25,line width=0.3pt] (0,-1)--(4,-1);
\draw[black!25,line width=0.3pt] (0,-2)--(4,-2);
\draw[black!25,line width=0.3pt] (0,-3)--(4,-3);
\draw[black,line width=0.9pt] (0,0) rectangle (4,-4);
\node[anchor=south,font=\small\normalfont] at (2,0.3) {$\circ$};
\node[text=black] at (4.85,-0.5) {$0$};
\node[text=black] at (4.85,-1.5) {$3$};
\node[text=black] at (4.85,-2.5) {$2$};
\node[text=black] at (4.85,-3.5) {$1$};
\draw[black!25,line width=0.3pt] (4.35,-1)--(5.35,-1);
\draw[black!25,line width=0.3pt] (4.35,-2)--(5.35,-2);
\draw[black!25,line width=0.3pt] (4.35,-3)--(5.35,-3);
\draw[black,line width=0.9pt] (4.35,0) rectangle (5.35,-4);
\node[anchor=south,font=\small\normalfont] at (4.85,0.3) {${}^{-1}$};
\node[text=black] at (6.2,-0.5) {$0$};
\draw[black,line width=0.9pt] (5.7,0) rectangle (6.7,-1);
\node[anchor=south,font=\small\normalfont] at (6.2,0.3) {$e$};
\draw[black!60,line width=0.5pt] (-0.3,1.05)--(7,1.05);
\draw[black,line width=1pt] (-0.3,-4.4) rectangle (7,2.2);
\end{tikzpicture}\end{tabular}\hspace*{\fill}\begin{tabular}[t]{@{}l@{}}\small $C_4$, elements $2\leftrightarrow3$:\\[-6pt]\begin{tikzpicture}[x=0.42cm,y=0.42cm,baseline=(current bounding box.north),every node/.style={font=\sffamily\bfseries,inner sep=0pt}]
\node[anchor=base west,font=\small\normalfont] at (0,1.45) {\textcolor{black}{4}};
\node[text=black] at (0.5,-0.5) {$0$};
\node[text=black] at (1.5,-0.5) {$1$};
\node[text=black] at (2.5,-0.5) {$2$};
\node[text=black] at (3.5,-0.5) {$3$};
\node[text=black] at (0.5,-1.5) {$1$};
\node[text=black] at (1.5,-1.5) {$3$};
\node[text=black] at (2.5,-1.5) {$0$};
\node[text=black] at (3.5,-1.5) {$2$};
\node[text=black] at (0.5,-2.5) {$2$};
\node[text=black] at (1.5,-2.5) {$0$};
\node[text=black] at (2.5,-2.5) {$3$};
\node[text=black] at (3.5,-2.5) {$1$};
\node[text=black] at (0.5,-3.5) {$3$};
\node[text=black] at (1.5,-3.5) {$2$};
\node[text=black] at (2.5,-3.5) {$1$};
\node[text=black] at (3.5,-3.5) {$0$};
\draw[black!25,line width=0.3pt] (1,0)--(1,-4);
\draw[black!25,line width=0.3pt] (2,0)--(2,-4);
\draw[black!25,line width=0.3pt] (3,0)--(3,-4);
\draw[black!25,line width=0.3pt] (0,-1)--(4,-1);
\draw[black!25,line width=0.3pt] (0,-2)--(4,-2);
\draw[black!25,line width=0.3pt] (0,-3)--(4,-3);
\draw[black,line width=0.9pt] (0,0) rectangle (4,-4);
\node[anchor=south,font=\small\normalfont] at (2,0.3) {$\circ$};
\node[text=black] at (4.85,-0.5) {$0$};
\node[text=black] at (4.85,-1.5) {$2$};
\node[text=black] at (4.85,-2.5) {$1$};
\node[text=black] at (4.85,-3.5) {$3$};
\draw[black!25,line width=0.3pt] (4.35,-1)--(5.35,-1);
\draw[black!25,line width=0.3pt] (4.35,-2)--(5.35,-2);
\draw[black!25,line width=0.3pt] (4.35,-3)--(5.35,-3);
\draw[black,line width=0.9pt] (4.35,0) rectangle (5.35,-4);
\node[anchor=south,font=\small\normalfont] at (4.85,0.3) {${}^{-1}$};
\node[text=black] at (6.2,-0.5) {$0$};
\draw[black,line width=0.9pt] (5.7,0) rectangle (6.7,-1);
\node[anchor=south,font=\small\normalfont] at (6.2,0.3) {$e$};
\draw[black!60,line width=0.5pt] (-0.3,1.05)--(7,1.05);
\draw[black,line width=1pt] (-0.3,-4.4) rectangle (7,2.2);
\end{tikzpicture}\end{tabular}\hspace*{\fill}\begin{tabular}[t]{@{}l@{}}\small $C_4$, elements $1\leftrightarrow2$:\\[-6pt]\begin{tikzpicture}[x=0.42cm,y=0.42cm,baseline=(current bounding box.north),every node/.style={font=\sffamily\bfseries,inner sep=0pt}]
\node[anchor=base west,font=\small\normalfont] at (0,1.45) {\textcolor{black}{4}};
\node[text=black] at (0.5,-0.5) {$0$};
\node[text=black] at (1.5,-0.5) {$1$};
\node[text=black] at (2.5,-0.5) {$2$};
\node[text=black] at (3.5,-0.5) {$3$};
\node[text=black] at (0.5,-1.5) {$1$};
\node[text=black] at (1.5,-1.5) {$0$};
\node[text=black] at (2.5,-1.5) {$3$};
\node[text=black] at (3.5,-1.5) {$2$};
\node[text=black] at (0.5,-2.5) {$2$};
\node[text=black] at (1.5,-2.5) {$3$};
\node[text=black] at (2.5,-2.5) {$1$};
\node[text=black] at (3.5,-2.5) {$0$};
\node[text=black] at (0.5,-3.5) {$3$};
\node[text=black] at (1.5,-3.5) {$2$};
\node[text=black] at (2.5,-3.5) {$0$};
\node[text=black] at (3.5,-3.5) {$1$};
\draw[black!25,line width=0.3pt] (1,0)--(1,-4);
\draw[black!25,line width=0.3pt] (2,0)--(2,-4);
\draw[black!25,line width=0.3pt] (3,0)--(3,-4);
\draw[black!25,line width=0.3pt] (0,-1)--(4,-1);
\draw[black!25,line width=0.3pt] (0,-2)--(4,-2);
\draw[black!25,line width=0.3pt] (0,-3)--(4,-3);
\draw[black,line width=0.9pt] (0,0) rectangle (4,-4);
\node[anchor=south,font=\small\normalfont] at (2,0.3) {$\circ$};
\node[text=black] at (4.85,-0.5) {$0$};
\node[text=black] at (4.85,-1.5) {$1$};
\node[text=black] at (4.85,-2.5) {$3$};
\node[text=black] at (4.85,-3.5) {$2$};
\draw[black!25,line width=0.3pt] (4.35,-1)--(5.35,-1);
\draw[black!25,line width=0.3pt] (4.35,-2)--(5.35,-2);
\draw[black!25,line width=0.3pt] (4.35,-3)--(5.35,-3);
\draw[black,line width=0.9pt] (4.35,0) rectangle (5.35,-4);
\node[anchor=south,font=\small\normalfont] at (4.85,0.3) {${}^{-1}$};
\node[text=black] at (6.2,-0.5) {$0$};
\draw[black,line width=0.9pt] (5.7,0) rectangle (6.7,-1);
\node[anchor=south,font=\small\normalfont] at (6.2,0.3) {$e$};
\draw[black!60,line width=0.5pt] (-0.3,1.05)--(7,1.05);
\draw[black,line width=1pt] (-0.3,-4.4) rectangle (7,2.2);
\end{tikzpicture}\end{tabular}\par\medskip
\begin{tabular}[t]{@{}l@{}}\small $\{0,1\}$ with \textsc{not}, \textsc{and}:\\[-6pt]\begin{tikzpicture}[x=0.42cm,y=0.42cm,baseline=(current bounding box.north),every node/.style={font=\sffamily\bfseries,inner sep=0pt}]
\node[anchor=base west,font=\small\normalfont] at (0,1.45) {\textcolor{black}{2}};
\node[text=black] at (0.5,-0.5) {$0$};
\node[text=black] at (1.5,-0.5) {$0$};
\node[text=black] at (0.5,-1.5) {$0$};
\node[text=black] at (1.5,-1.5) {$1$};
\draw[black!25,line width=0.3pt] (1,0)--(1,-2);
\draw[black!25,line width=0.3pt] (0,-1)--(2,-1);
\draw[black,line width=0.9pt] (0,0) rectangle (2,-2);
\node[anchor=south,font=\small\normalfont] at (1,0.3) {$\land$};
\node[text=black] at (2.85,-0.5) {$1$};
\node[text=black] at (2.85,-1.5) {$0$};
\draw[black!25,line width=0.3pt] (2.35,-1)--(3.35,-1);
\draw[black,line width=0.9pt] (2.35,0) rectangle (3.35,-2);
\node[anchor=south,font=\small\normalfont] at (2.85,0.3) {$\nt$};
\draw[black!60,line width=0.5pt] (-0.3,1.05)--(3.65,1.05);
\draw[black,line width=1pt] (-0.3,-2.4) rectangle (3.65,2.2);
\end{tikzpicture}\end{tabular}\hspace*{3em}\begin{tabular}[t]{@{}l@{}}\small $\{0,1\}$ with \textsc{not}, \textsc{or}:\\[-6pt]\begin{tikzpicture}[x=0.42cm,y=0.42cm,baseline=(current bounding box.north),every node/.style={font=\sffamily\bfseries,inner sep=0pt}]
\node[anchor=base west,font=\small\normalfont] at (0,1.45) {\textcolor{black}{2}};
\node[text=black] at (0.5,-0.5) {$0$};
\node[text=black] at (1.5,-0.5) {$1$};
\node[text=black] at (0.5,-1.5) {$1$};
\node[text=black] at (1.5,-1.5) {$1$};
\draw[black!25,line width=0.3pt] (1,0)--(1,-2);
\draw[black!25,line width=0.3pt] (0,-1)--(2,-1);
\draw[black,line width=0.9pt] (0,0) rectangle (2,-2);
\node[anchor=south,font=\small\normalfont] at (1,0.3) {$\lor$};
\node[text=black] at (2.85,-0.5) {$1$};
\node[text=black] at (2.85,-1.5) {$0$};
\draw[black!25,line width=0.3pt] (2.35,-1)--(3.35,-1);
\draw[black,line width=0.9pt] (2.35,0) rectangle (3.35,-2);
\node[anchor=south,font=\small\normalfont] at (2.85,0.3) {$\nt$};
\draw[black!60,line width=0.5pt] (-0.3,1.05)--(3.65,1.05);
\draw[black,line width=1pt] (-0.3,-2.4) rectangle (3.65,2.2);
\end{tikzpicture}\end{tabular}\par
\end{minipage}}

\caption{Equivalence by isomorphism. The top three presentations are isomorphic, differing in how their elements are labeled, and all represent the same 
mathematical structure: the cyclic group $C_4$.
Analogously, the bottom two presentations are also isomorphic.}
\label{PresentationIsomorphismFig}
\end{figure}

\clearpage

\fig{NPresentationIsomorphismFig} shows an analogous example of  isomorphism for a countably infinite presentation
of the natural numbers with the operations of Peano Arithmetic, representing $n$ either in unary or in binary 
(essentially the difference between representing 13 as a string of length $n$ like ``1111111111111'' versus a bit string of length of order $\log_2 n$ like ``1101").
They are isomorphic because there are unary MLISP functions 
$f$ = ``\texttt{iibf$'$0d}" and  its inverse function $g$ =  ``\texttt{iibfibf+ia+0d}"
that form a bijection between unary and binary number labels (strings) preserving all functions.
For example, letting $\cdot$ and $\times$ denote multiplication in the two presentations, 
we have $f(i\cdot j) = f(i)\times f(j)$.
The binary presentation is vastly more efficient computationally, requiring 
$\mathcal{O}\left((\log_2 n)^2\right)$ operations instead of $\mathcal{O}\left(n^2\right)$ to implement multiplication,
but because it is isomorphic, it defines the exact same mathematical structure. 

\begin{figure}[H]
\centering
\noindent\setlength{\fboxsep}{6pt}\colorbox{galbeige}{\begin{minipage}{\dimexpr\textwidth-2\fboxsep-2\fboxrule\relax}\raggedright
\begin{tabular}[t]{@{}l@{}}\small $(\mathbb N,0,{}',+,\cdot)$, elements as unary strings:\\[-1pt]\setlength{\fboxsep}{0pt}\fcolorbox{black}{white}{\begin{tabular}{@{\hspace{5pt}}r@{\,:\ }l@{\ =\ }l@{\hspace{5pt}}}
\multicolumn{3}{@{\hspace{5pt}}l@{\hspace{5pt}}}{\rule{0pt}{1.3em}\textcolor{black}{$|S_1|=\aleph_0$;\ \ {\fontfamily{pcr}\selectfont "0"}$\,\in S_1$;\ \ $i\in S_1\Rightarrow$ {\fontfamily{pcr}\selectfont "0ic"}$\,\in S_1$}}\\[2pt]\hline
\rule{0pt}{1.25em}$0$ & $S_1$ & {\fontfamily{pcr}\selectfont "0"}\\
$'$ & $S_1\to S_1$ & {\fontfamily{pcr}\selectfont "0ic"}\\
$+$ & $S_1\times S_1\to S_1$ & {\fontfamily{pcr}\selectfont "iib0jcfjd"}\\
$\cdot$ & $S_1\times S_1\to S_1$ & {\fontfamily{pcr}\selectfont "iibjfj+0d"}\\[3pt]
\end{tabular}}
\end{tabular}\par\medskip
\adjustbox{max width=\linewidth}{\begin{tabular}[t]{@{}l@{}}\small $(\mathbb N,0,{}',+,\cdot)$, elements as binary strings:\\[-1pt]\setlength{\fboxsep}{0pt}\fcolorbox{black}{white}{\begin{tabular}{@{\hspace{5pt}}r@{\,:\ }l@{\ =\ }l@{\hspace{5pt}}}
\multicolumn{3}{@{\hspace{5pt}}l@{\hspace{5pt}}}{\rule{0pt}{1.3em}$|S_1|=\aleph_0$;\ \ {\fontfamily{pcr}\selectfont "0"}$\,\in S_1$;\ \ $i\in S_1\Rightarrow$ {\fontfamily{pcr}\selectfont "1ic"}$\,\in S_1$;\ \ $i\in S_1,\ i\neq${\fontfamily{pcr}\selectfont "0"} $\Rightarrow$ {\fontfamily{pcr}\selectfont "0ic"}$\,\in S_1$}\\[2pt]\hline
\multicolumn{2}{@{\hspace{5pt}}r@{\ =\ }}{\rule{0pt}{1.25em}$\textcolor{black!50}{1}$} & \textcolor{black!50}{{\fontfamily{pcr}\selectfont "00c"}}\\
$0$ & $S_1$ & {\fontfamily{pcr}\selectfont "0"}\\
$'$ & $S_1\to S_1$ & {\fontfamily{pcr}\selectfont "iia0ibfc1ibcd10cd"}\\
$+$ & $S_1\times S_1\to S_1$ & {\fontfamily{pcr}\selectfont "ijiaja0ibjbf$'$c1ibjbfcdja1ibjbfc0ibjbfcddidjd"}\\
\multicolumn{2}{@{\hspace{5pt}}r@{\ =\ }}{$\textcolor{black!50}{w}$} & \textcolor{black!50}{{\fontfamily{pcr}\selectfont "i0icid"}}\\
$\cdot$ & $S_1\times S_1\to S_1$ & {\fontfamily{pcr}\selectfont "iiajibjfw+ibjfwd0d"}\\[3pt]
\end{tabular}}
\end{tabular}}\par
\end{minipage}}

\caption{Two computational presentations of the natural numbers $(\mathbb N,0,{}',+,\cdot)$, with the elements written as unary strings (top) and as binary strings (bottom).}
\label{NPresentationIsomorphismFig}
\end{figure}

\clearpage

\subsection{Compositional equivalence and clones}

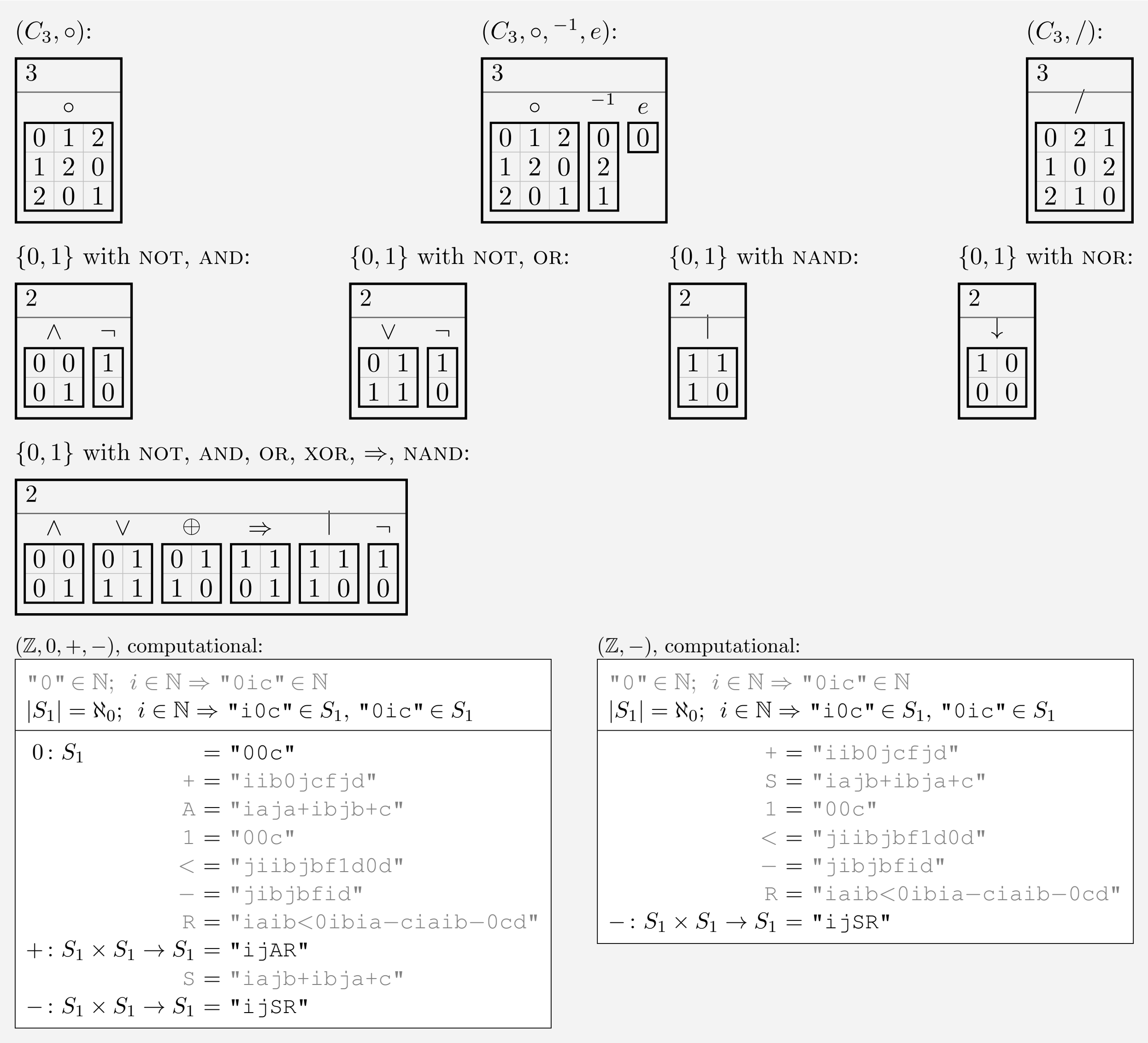
\begin{figure}[H]
\centering
\noindent\setlength{\fboxsep}{6pt}\colorbox{galbeige}{\begin{minipage}{\dimexpr\textwidth-2\fboxsep-2\fboxrule\relax}\raggedright
\begin{tabular}[t]{@{}l@{}}\small $(C_3,\circ)$:\\[-6pt]\begin{tikzpicture}[x=0.42cm,y=0.42cm,baseline=(current bounding box.north),every node/.style={font=\sffamily\bfseries,inner sep=0pt}]
\node[anchor=base west,font=\small\normalfont] at (0,1.45) {\textcolor{black}{3}};
\node[text=black] at (0.5,-0.5) {$0$};
\node[text=black] at (1.5,-0.5) {$1$};
\node[text=black] at (2.5,-0.5) {$2$};
\node[text=black] at (0.5,-1.5) {$1$};
\node[text=black] at (1.5,-1.5) {$2$};
\node[text=black] at (2.5,-1.5) {$0$};
\node[text=black] at (0.5,-2.5) {$2$};
\node[text=black] at (1.5,-2.5) {$0$};
\node[text=black] at (2.5,-2.5) {$1$};
\draw[black!25,line width=0.3pt] (1,0)--(1,-3);
\draw[black!25,line width=0.3pt] (2,0)--(2,-3);
\draw[black!25,line width=0.3pt] (0,-1)--(3,-1);
\draw[black!25,line width=0.3pt] (0,-2)--(3,-2);
\draw[black,line width=0.9pt] (0,0) rectangle (3,-3);
\node[anchor=south,font=\small\normalfont] at (1.5,0.3) {$\circ$};
\draw[black!60,line width=0.5pt] (-0.3,1.05)--(3.3,1.05);
\draw[black,line width=1pt] (-0.3,-3.4) rectangle (3.3,2.2);
\end{tikzpicture}\end{tabular}\hspace*{\fill}\begin{tabular}[t]{@{}l@{}}\small $(C_3,\circ,{}^{-1},e)$:\\[-6pt]\begin{tikzpicture}[x=0.42cm,y=0.42cm,baseline=(current bounding box.north),every node/.style={font=\sffamily\bfseries,inner sep=0pt}]
\node[anchor=base west,font=\small\normalfont] at (0,1.45) {\textcolor{black}{3}};
\node[text=black] at (0.5,-0.5) {$0$};
\node[text=black] at (1.5,-0.5) {$1$};
\node[text=black] at (2.5,-0.5) {$2$};
\node[text=black] at (0.5,-1.5) {$1$};
\node[text=black] at (1.5,-1.5) {$2$};
\node[text=black] at (2.5,-1.5) {$0$};
\node[text=black] at (0.5,-2.5) {$2$};
\node[text=black] at (1.5,-2.5) {$0$};
\node[text=black] at (2.5,-2.5) {$1$};
\draw[black!25,line width=0.3pt] (1,0)--(1,-3);
\draw[black!25,line width=0.3pt] (2,0)--(2,-3);
\draw[black!25,line width=0.3pt] (0,-1)--(3,-1);
\draw[black!25,line width=0.3pt] (0,-2)--(3,-2);
\draw[black,line width=0.9pt] (0,0) rectangle (3,-3);
\node[anchor=south,font=\small\normalfont] at (1.5,0.3) {$\circ$};
\node[text=black] at (3.85,-0.5) {$0$};
\node[text=black] at (3.85,-1.5) {$2$};
\node[text=black] at (3.85,-2.5) {$1$};
\draw[black!25,line width=0.3pt] (3.35,-1)--(4.35,-1);
\draw[black!25,line width=0.3pt] (3.35,-2)--(4.35,-2);
\draw[black,line width=0.9pt] (3.35,0) rectangle (4.35,-3);
\node[anchor=south,font=\small\normalfont] at (3.85,0.3) {${}^{-1}$};
\node[text=black] at (5.2,-0.5) {$0$};
\draw[black,line width=0.9pt] (4.7,0) rectangle (5.7,-1);
\node[anchor=south,font=\small\normalfont] at (5.2,0.3) {$e$};
\draw[black!60,line width=0.5pt] (-0.3,1.05)--(6,1.05);
\draw[black,line width=1pt] (-0.3,-3.4) rectangle (6,2.2);
\end{tikzpicture}\end{tabular}\hspace*{\fill}\begin{tabular}[t]{@{}l@{}}\small $(C_3,/)$:\\[-6pt]\begin{tikzpicture}[x=0.42cm,y=0.42cm,baseline=(current bounding box.north),every node/.style={font=\sffamily\bfseries,inner sep=0pt}]
\node[anchor=base west,font=\small\normalfont] at (0,1.45) {\textcolor{black}{3}};
\node[text=black] at (0.5,-0.5) {$0$};
\node[text=black] at (1.5,-0.5) {$2$};
\node[text=black] at (2.5,-0.5) {$1$};
\node[text=black] at (0.5,-1.5) {$1$};
\node[text=black] at (1.5,-1.5) {$0$};
\node[text=black] at (2.5,-1.5) {$2$};
\node[text=black] at (0.5,-2.5) {$2$};
\node[text=black] at (1.5,-2.5) {$1$};
\node[text=black] at (2.5,-2.5) {$0$};
\draw[black!25,line width=0.3pt] (1,0)--(1,-3);
\draw[black!25,line width=0.3pt] (2,0)--(2,-3);
\draw[black!25,line width=0.3pt] (0,-1)--(3,-1);
\draw[black!25,line width=0.3pt] (0,-2)--(3,-2);
\draw[black,line width=0.9pt] (0,0) rectangle (3,-3);
\node[anchor=south,font=\small\normalfont] at (1.5,0.3) {$/$};
\draw[black!60,line width=0.5pt] (-0.3,1.05)--(3.3,1.05);
\draw[black,line width=1pt] (-0.3,-3.4) rectangle (3.3,2.2);
\end{tikzpicture}\end{tabular}\par\medskip
\begin{tabular}[t]{@{}l@{}}\small $\{0,1\}$ with \textsc{not}, \textsc{and}:\\[-6pt]\begin{tikzpicture}[x=0.42cm,y=0.42cm,baseline=(current bounding box.north),every node/.style={font=\sffamily\bfseries,inner sep=0pt}]
\node[anchor=base west,font=\small\normalfont] at (0,1.45) {\textcolor{black}{2}};
\node[text=black] at (0.5,-0.5) {$0$};
\node[text=black] at (1.5,-0.5) {$0$};
\node[text=black] at (0.5,-1.5) {$0$};
\node[text=black] at (1.5,-1.5) {$1$};
\draw[black!25,line width=0.3pt] (1,0)--(1,-2);
\draw[black!25,line width=0.3pt] (0,-1)--(2,-1);
\draw[black,line width=0.9pt] (0,0) rectangle (2,-2);
\node[anchor=south,font=\small\normalfont] at (1,0.3) {$\land$};
\node[text=black] at (2.85,-0.5) {$1$};
\node[text=black] at (2.85,-1.5) {$0$};
\draw[black!25,line width=0.3pt] (2.35,-1)--(3.35,-1);
\draw[black,line width=0.9pt] (2.35,0) rectangle (3.35,-2);
\node[anchor=south,font=\small\normalfont] at (2.85,0.3) {$\nt$};
\draw[black!60,line width=0.5pt] (-0.3,1.05)--(3.65,1.05);
\draw[black,line width=1pt] (-0.3,-2.4) rectangle (3.65,2.2);
\end{tikzpicture}\end{tabular}\hspace*{\fill}\begin{tabular}[t]{@{}l@{}}\small $\{0,1\}$ with \textsc{not}, \textsc{or}:\\[-6pt]\begin{tikzpicture}[x=0.42cm,y=0.42cm,baseline=(current bounding box.north),every node/.style={font=\sffamily\bfseries,inner sep=0pt}]
\node[anchor=base west,font=\small\normalfont] at (0,1.45) {\textcolor{black}{2}};
\node[text=black] at (0.5,-0.5) {$0$};
\node[text=black] at (1.5,-0.5) {$1$};
\node[text=black] at (0.5,-1.5) {$1$};
\node[text=black] at (1.5,-1.5) {$1$};
\draw[black!25,line width=0.3pt] (1,0)--(1,-2);
\draw[black!25,line width=0.3pt] (0,-1)--(2,-1);
\draw[black,line width=0.9pt] (0,0) rectangle (2,-2);
\node[anchor=south,font=\small\normalfont] at (1,0.3) {$\lor$};
\node[text=black] at (2.85,-0.5) {$1$};
\node[text=black] at (2.85,-1.5) {$0$};
\draw[black!25,line width=0.3pt] (2.35,-1)--(3.35,-1);
\draw[black,line width=0.9pt] (2.35,0) rectangle (3.35,-2);
\node[anchor=south,font=\small\normalfont] at (2.85,0.3) {$\nt$};
\draw[black!60,line width=0.5pt] (-0.3,1.05)--(3.65,1.05);
\draw[black,line width=1pt] (-0.3,-2.4) rectangle (3.65,2.2);
\end{tikzpicture}\end{tabular}\hspace*{\fill}\begin{tabular}[t]{@{}l@{}}\small $\{0,1\}$ with \textsc{nand}:\\[-6pt]\begin{tikzpicture}[x=0.42cm,y=0.42cm,baseline=(current bounding box.north),every node/.style={font=\sffamily\bfseries,inner sep=0pt}]
\node[anchor=base west,font=\small\normalfont] at (0,1.45) {\textcolor{black}{2}};
\node[text=black] at (0.5,-0.5) {$1$};
\node[text=black] at (1.5,-0.5) {$1$};
\node[text=black] at (0.5,-1.5) {$1$};
\node[text=black] at (1.5,-1.5) {$0$};
\draw[black!25,line width=0.3pt] (1,0)--(1,-2);
\draw[black!25,line width=0.3pt] (0,-1)--(2,-1);
\draw[black,line width=0.9pt] (0,0) rectangle (2,-2);
\node[anchor=south,font=\small\normalfont] at (1,0.3) {$|$};
\draw[black!60,line width=0.5pt] (-0.3,1.05)--(2.3,1.05);
\draw[black,line width=1pt] (-0.3,-2.4) rectangle (2.3,2.2);
\end{tikzpicture}\end{tabular}\hspace*{\fill}\begin{tabular}[t]{@{}l@{}}\small $\{0,1\}$ with \textsc{nor}:\\[-6pt]\begin{tikzpicture}[x=0.42cm,y=0.42cm,baseline=(current bounding box.north),every node/.style={font=\sffamily\bfseries,inner sep=0pt}]
\node[anchor=base west,font=\small\normalfont] at (0,1.45) {\textcolor{black}{2}};
\node[text=black] at (0.5,-0.5) {$1$};
\node[text=black] at (1.5,-0.5) {$0$};
\node[text=black] at (0.5,-1.5) {$0$};
\node[text=black] at (1.5,-1.5) {$0$};
\draw[black!25,line width=0.3pt] (1,0)--(1,-2);
\draw[black!25,line width=0.3pt] (0,-1)--(2,-1);
\draw[black,line width=0.9pt] (0,0) rectangle (2,-2);
\node[anchor=south,font=\small\normalfont] at (1,0.3) {$\downarrow$};
\draw[black!60,line width=0.5pt] (-0.3,1.05)--(2.3,1.05);
\draw[black,line width=1pt] (-0.3,-2.4) rectangle (2.3,2.2);
\end{tikzpicture}\end{tabular}\par\medskip
\begin{tabular}[t]{@{}l@{}}\small $\{0,1\}$ with \textsc{not}, \textsc{and}, \textsc{or}, \textsc{xor}, $\Rightarrow$, \textsc{nand}:\\[-6pt]\begin{tikzpicture}[x=0.42cm,y=0.42cm,baseline=(current bounding box.north),every node/.style={font=\sffamily\bfseries,inner sep=0pt}]
\node[anchor=base west,font=\small\normalfont] at (0,1.45) {\textcolor{black}{2}};
\node[text=black] at (0.5,-0.5) {$0$};
\node[text=black] at (1.5,-0.5) {$0$};
\node[text=black] at (0.5,-1.5) {$0$};
\node[text=black] at (1.5,-1.5) {$1$};
\draw[black!25,line width=0.3pt] (1,0)--(1,-2);
\draw[black!25,line width=0.3pt] (0,-1)--(2,-1);
\draw[black,line width=0.9pt] (0,0) rectangle (2,-2);
\node[anchor=south,font=\small\normalfont] at (1,0.3) {$\land$};
\node[text=black] at (2.85,-0.5) {$0$};
\node[text=black] at (3.85,-0.5) {$1$};
\node[text=black] at (2.85,-1.5) {$1$};
\node[text=black] at (3.85,-1.5) {$1$};
\draw[black!25,line width=0.3pt] (3.35,0)--(3.35,-2);
\draw[black!25,line width=0.3pt] (2.35,-1)--(4.35,-1);
\draw[black,line width=0.9pt] (2.35,0) rectangle (4.35,-2);
\node[anchor=south,font=\small\normalfont] at (3.35,0.3) {$\lor$};
\node[text=black] at (5.2,-0.5) {$0$};
\node[text=black] at (6.2,-0.5) {$1$};
\node[text=black] at (5.2,-1.5) {$1$};
\node[text=black] at (6.2,-1.5) {$0$};
\draw[black!25,line width=0.3pt] (5.7,0)--(5.7,-2);
\draw[black!25,line width=0.3pt] (4.7,-1)--(6.7,-1);
\draw[black,line width=0.9pt] (4.7,0) rectangle (6.7,-2);
\node[anchor=south,font=\small\normalfont] at (5.7,0.3) {$\oplus$};
\node[text=black] at (7.55,-0.5) {$1$};
\node[text=black] at (8.55,-0.5) {$1$};
\node[text=black] at (7.55,-1.5) {$0$};
\node[text=black] at (8.55,-1.5) {$1$};
\draw[black!25,line width=0.3pt] (8.05,0)--(8.05,-2);
\draw[black!25,line width=0.3pt] (7.05,-1)--(9.05,-1);
\draw[black,line width=0.9pt] (7.05,0) rectangle (9.05,-2);
\node[anchor=south,font=\small\normalfont] at (8.05,0.3) {$\Rightarrow$};
\node[text=black] at (9.9,-0.5) {$1$};
\node[text=black] at (10.9,-0.5) {$1$};
\node[text=black] at (9.9,-1.5) {$1$};
\node[text=black] at (10.9,-1.5) {$0$};
\draw[black!25,line width=0.3pt] (10.4,0)--(10.4,-2);
\draw[black!25,line width=0.3pt] (9.4,-1)--(11.4,-1);
\draw[black,line width=0.9pt] (9.4,0) rectangle (11.4,-2);
\node[anchor=south,font=\small\normalfont] at (10.4,0.3) {$|$};
\node[text=black] at (12.25,-0.5) {$1$};
\node[text=black] at (12.25,-1.5) {$0$};
\draw[black!25,line width=0.3pt] (11.75,-1)--(12.75,-1);
\draw[black,line width=0.9pt] (11.75,0) rectangle (12.75,-2);
\node[anchor=south,font=\small\normalfont] at (12.25,0.3) {$\nt$};
\draw[black!60,line width=0.5pt] (-0.3,1.05)--(13.05,1.05);
\draw[black,line width=1pt] (-0.3,-2.4) rectangle (13.05,2.2);
\end{tikzpicture}\end{tabular}\par\medskip
\adjustbox{max width=\linewidth}{\begin{tabular}[t]{@{}l@{}}\small $(\mathbb Z,0,+,-)$, computational:\\[-1pt]\setlength{\fboxsep}{0pt}\fcolorbox{black}{white}{\begin{tabular}{@{\hspace{5pt}}r@{\,:\ }l@{\ =\ }l@{\hspace{5pt}}}
\multicolumn{3}{@{\hspace{5pt}}l@{\hspace{5pt}}}{\rule{0pt}{1.3em}\textcolor{black!50}{{\fontfamily{pcr}\selectfont "0"}$\,\in \mathbb N$;\ \ $i\in \mathbb N\Rightarrow$ {\fontfamily{pcr}\selectfont "0ic"}$\,\in \mathbb N$}}\\
\multicolumn{3}{@{\hspace{5pt}}l@{\hspace{5pt}}}{\textcolor{black}{$|S_1|=\aleph_0$;\ \ $i\in \mathbb N\Rightarrow$ {\fontfamily{pcr}\selectfont "i0c"}$\,\in S_1$, {\fontfamily{pcr}\selectfont "0ic"}$\,\in S_1$}}\\[2pt]\hline
\rule{0pt}{1.25em}$0$ & $S_1$ & {\fontfamily{pcr}\selectfont "00c"}\\
\multicolumn{2}{@{\hspace{5pt}}r@{\ =\ }}{\textcolor{black!50}{\texttt{+}}} & \textcolor{black!50}{{\fontfamily{pcr}\selectfont "iib0jcfjd"}}\\
\multicolumn{2}{@{\hspace{5pt}}r@{\ =\ }}{\textcolor{black!50}{\texttt{A}}} & \textcolor{black!50}{{\fontfamily{pcr}\selectfont "iaja+ibjb+c"}}\\
\multicolumn{2}{@{\hspace{5pt}}r@{\ =\ }}{\textcolor{black!50}{\texttt{1}}} & \textcolor{black!50}{{\fontfamily{pcr}\selectfont "00c"}}\\
\multicolumn{2}{@{\hspace{5pt}}r@{\ =\ }}{\textcolor{black!50}{\texttt{$<$}}} & \textcolor{black!50}{{\fontfamily{pcr}\selectfont "jiibjbf1d0d"}}\\
\multicolumn{2}{@{\hspace{5pt}}r@{\ =\ }}{\textcolor{black!50}{\texttt{$-$}}} & \textcolor{black!50}{{\fontfamily{pcr}\selectfont "jibjbfid"}}\\
\multicolumn{2}{@{\hspace{5pt}}r@{\ =\ }}{\textcolor{black!50}{\texttt{R}}} & \textcolor{black!50}{{\fontfamily{pcr}\selectfont "iaib$<$0ibia$-$ciaib$-$0cd"}}\\
$+$ & $S_1\times S_1\to S_1$ & {\fontfamily{pcr}\selectfont "ijAR"}\\
\multicolumn{2}{@{\hspace{5pt}}r@{\ =\ }}{\textcolor{black!50}{\texttt{S}}} & \textcolor{black!50}{{\fontfamily{pcr}\selectfont "iajb+ibja+c"}}\\
$-$ & $S_1\times S_1\to S_1$ & {\fontfamily{pcr}\selectfont "ijSR"}\\[3pt]
\end{tabular}}
\end{tabular}\hspace*{2em}\begin{tabular}[t]{@{}l@{}}\small $(\mathbb Z,-)$, computational:\\[-1pt]\setlength{\fboxsep}{0pt}\fcolorbox{black}{white}{\begin{tabular}{@{\hspace{5pt}}r@{\,:\ }l@{\ =\ }l@{\hspace{5pt}}}
\multicolumn{3}{@{\hspace{5pt}}l@{\hspace{5pt}}}{\rule{0pt}{1.3em}\textcolor{black!50}{{\fontfamily{pcr}\selectfont "0"}$\,\in \mathbb N$;\ \ $i\in \mathbb N\Rightarrow$ {\fontfamily{pcr}\selectfont "0ic"}$\,\in \mathbb N$}}\\
\multicolumn{3}{@{\hspace{5pt}}l@{\hspace{5pt}}}{\textcolor{black}{$|S_1|=\aleph_0$;\ \ $i\in \mathbb N\Rightarrow$ {\fontfamily{pcr}\selectfont "i0c"}$\,\in S_1$, {\fontfamily{pcr}\selectfont "0ic"}$\,\in S_1$}}\\[2pt]\hline
\multicolumn{2}{@{\hspace{5pt}}r@{\ =\ }}{\rule{0pt}{1.25em}\textcolor{black!50}{\texttt{+}}} & \textcolor{black!50}{{\fontfamily{pcr}\selectfont "iib0jcfjd"}}\\
\multicolumn{2}{@{\hspace{5pt}}r@{\ =\ }}{\textcolor{black!50}{\texttt{S}}} & \textcolor{black!50}{{\fontfamily{pcr}\selectfont "iajb+ibja+c"}}\\
\multicolumn{2}{@{\hspace{5pt}}r@{\ =\ }}{\textcolor{black!50}{\texttt{1}}} & \textcolor{black!50}{{\fontfamily{pcr}\selectfont "00c"}}\\
\multicolumn{2}{@{\hspace{5pt}}r@{\ =\ }}{\textcolor{black!50}{\texttt{$<$}}} & \textcolor{black!50}{{\fontfamily{pcr}\selectfont "jiibjbf1d0d"}}\\
\multicolumn{2}{@{\hspace{5pt}}r@{\ =\ }}{\textcolor{black!50}{\texttt{$-$}}} & \textcolor{black!50}{{\fontfamily{pcr}\selectfont "jibjbfid"}}\\
\multicolumn{2}{@{\hspace{5pt}}r@{\ =\ }}{\textcolor{black!50}{\texttt{R}}} & \textcolor{black!50}{{\fontfamily{pcr}\selectfont "iaib$<$0ibia$-$ciaib$-$0cd"}}\\
$-$ & $S_1\times S_1\to S_1$ & {\fontfamily{pcr}\selectfont "ijSR"}\\[3pt]
\end{tabular}}
\end{tabular}}\par
\end{minipage}}

\caption{Equivalence by composition. Top row: three presentations of $C_3$, with the product alone, with product, inverse and identity ($a^{-1}=a\circ a$, $e=a\circ a\circ a$) and with division alone ($a\circ b=a/((b/b)/b)$). Middle rows: five presentations of the two-element Boolean algebra, each of whose functions is a composition of the others'. Bottom row: computational presentations of the integers with zero, addition and subtraction, and of the integers under subtraction alone ($0=x-x$, $x+y=x-((y-y)-y)$).}
\label{ClosureEquivFig}
\end{figure}

Model theory historically tends to call a presentation a ``structure", sometimes identified up to isomorphism.
As described in \Sec{TerminologySubsec}, we let ``presentation" refer to a specific realization, and a ``mathematical structure" to its equivalence class under isomorphism.
However, it is interesting to further broaden the equivalence class definition, to also identify presentations that {\it generate each other via function composition}, because that is closer to the identification that many mathematicians make when they say that two presentations are ``the same thing, just described differently''.
Consider, for example, the first two presentations in the top row of \fig{ClosureEquivFig}.
The second one describes the cyclic group $C_3$, tabulating not only its binary (multiplication) function 
$f_1(g_1,g_2)=g_1 g_2$, but also its inverse function  $f_2(g)=g^{-1}$ and identity $f_3=e$.
Is the first presentation, listing only the multiplication table $f_1$, not also describing the group $C_3$?
Claiming that it does not would arguably go against standard mathematics terminology, since when finite groups are defined in the mathematics literature
and on Wikipedia, they are normally described by merely presenting this multiplication table 
(known as their Cayley table).
Moreover, the inverse and identity can be uniquely {\it defined} from the multiplication table alone, by composing the function $f_1$ with itself:
\begin{eqnarray}
f_2(g)&=&f_1(g,g)=g^2=g^{-1},\\
f_3&=&f_1(f_1(g,g),g)=g^3=e.
\end{eqnarray}
Indeed, for any finite group of order $n$, $g^{-1}=g^{n-1}$ and $e=g^n$.
In other words, $f_1$ generates $f_2$ and $f_3$ via composition.

\def\fnand{f_|}

The rest of \fig{ClosureEquivFig} shows more such examples.
The five shown presentations of Boolean functions on $\{0,1\}$ are all equivalent under composition, 
because they generate each other's generator functions.
For example, the binary function $\fnand$ (NAND) generates all of the others:
\beqa{ShefferEquivalenceEq}
f_\neg(x)&=&\fnand(x,x),\\ 
f_\land(x,y)&=&\fnand(\fnand(x,y),\fnand(x,y)),\nonumber\\ 
f_\lor(x,y)&=&\fnand(\fnand(x,x),\fnand(y,y)),\nonumber\\ 
f_\Rightarrow(x,y)&=&\fnand(x,\fnand(y,y)),\nonumber\\ 
f_\oplus(x,y)&=&\fnand(\fnand(x,\fnand(x,y)),\fnand(y,\fnand(x,y))),\nonumber\\
f_\downarrow(x,y)&=&f_\neg(f_\lor(x,y)).\nonumber
\eeqa
They in fact have the {\it same compositional closure}:
they all generate {\it all} Boolean functions of any number of variables.
Working mathematicians typically refer to this simply as Boolean algebra, regardless of which presentation was used to specify it, which further supports including compositional closure in the equivalence definition.

The final example in \fig{ClosureEquivFig} shows two different presentations of integers: 
$(\mathbb{Z},0, +, -)$ and $(\mathbb{Z},-)$, respectively.
They are compositionally equivalent because $f_0 = f_-(x,x)=x-x=0$ and
$f_+(x,y)=x+y=x-((y-y)-y)$.

In model theory and universal algebra, the compositional closure is called the {\it clone},
and equivalence under compositional closure is called ``term equivalence''.
We will also refer to a clone as a {\it closed mathematical structure}.
So a closed mathematical structure consists of abstract sets of elements and 
functions (the clone) \cite{Hodges1993,Post1941,Hall1958,Cohn1965}, closed under composition. We can think of it as all the functions on sets that are defined (via composition) by the presentation.

In our atlas, we ultimately choose to define mathematical structures up to isomorphism rather than
up to clone isomorphism, because  the latter sometimes lumps together two objects that mathematicians conventionally view as different. 
For example, there are two different groups of order 260 that have the same compositional closure
\cite{KearnesSzendrei2004} (see Appendix~\ref{CountingAppendix}):
with our definitions, these two groups are different mathematical structures, but define the same
closed mathematical structure.

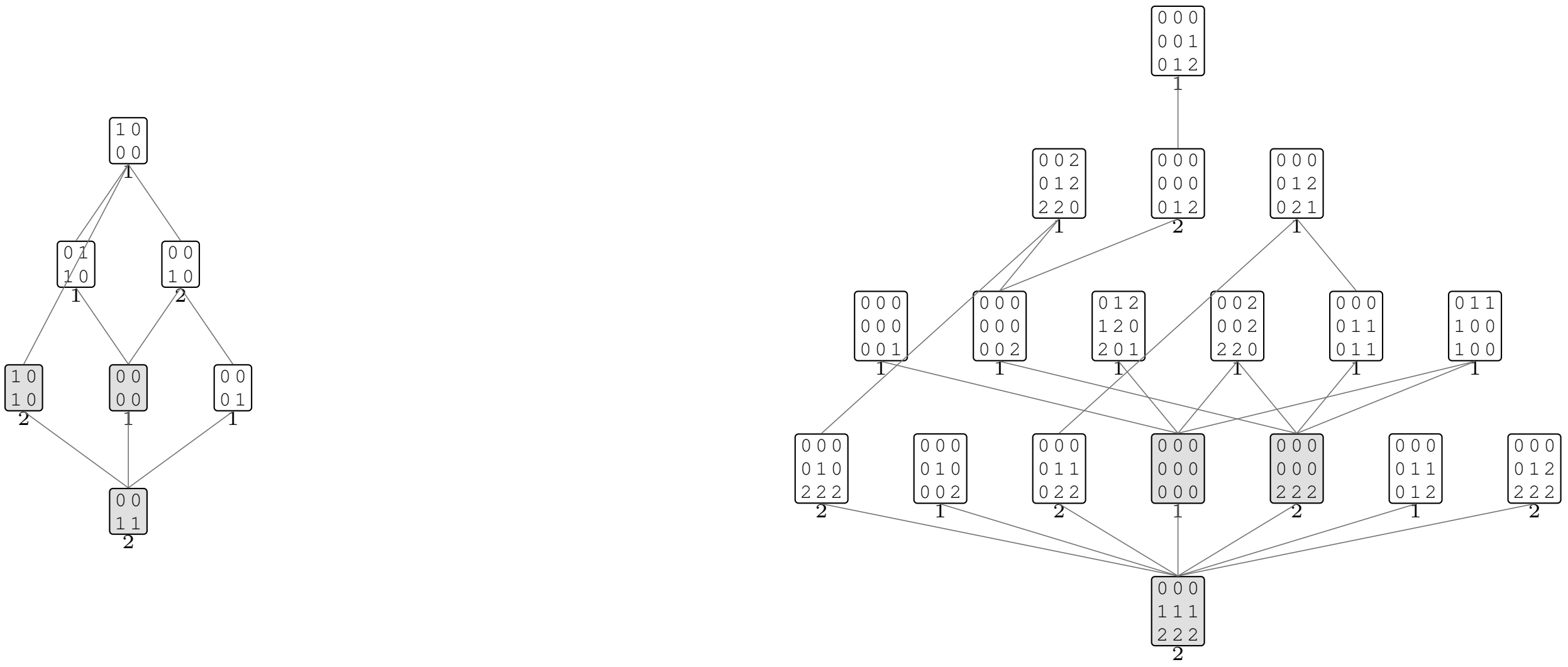
\begin{figure}[H]
\centering
\begin{tabular}[c]{@{}c@{}}\adjustbox{max width=0.3\linewidth}{\begin{tikzpicture}[x=1cm,y=1cm,every node/.style={inner sep=1.5pt}]
\node[draw,fill=black!12,font=\tiny\ttfamily,align=center,rounded corners=1pt] (n0) at (0.000,0.000) {0\,0\\1\,1};
\node[font=\tiny,inner sep=0pt,anchor=north] at (n0.south) {2};
\node[draw,fill=black!12,font=\tiny\ttfamily,align=center,rounded corners=1pt] (n4) at (-1.100,1.300) {1\,0\\1\,0};
\node[font=\tiny,inner sep=0pt,anchor=north] at (n4.south) {2};
\node[draw,fill=black!12,font=\tiny\ttfamily,align=center,rounded corners=1pt] (n1) at (0.000,1.300) {0\,0\\0\,0};
\node[font=\tiny,inner sep=0pt,anchor=north] at (n1.south) {1};
\node[draw,fill=white,font=\tiny\ttfamily,align=center,rounded corners=1pt] (n2) at (1.100,1.300) {0\,0\\0\,1};
\node[font=\tiny,inner sep=0pt,anchor=north] at (n2.south) {1};
\node[draw,fill=white,font=\tiny\ttfamily,align=center,rounded corners=1pt] (n3) at (-0.550,2.600) {0\,1\\1\,0};
\node[font=\tiny,inner sep=0pt,anchor=north] at (n3.south) {1};
\node[draw,fill=white,font=\tiny\ttfamily,align=center,rounded corners=1pt] (n5) at (0.550,2.600) {0\,0\\1\,0};
\node[font=\tiny,inner sep=0pt,anchor=north] at (n5.south) {2};
\node[draw,fill=white,font=\tiny\ttfamily,align=center,rounded corners=1pt] (n6) at (0.000,3.900) {1\,0\\0\,0};
\node[font=\tiny,inner sep=0pt,anchor=north] at (n6.south) {1};
\draw[black!55,line width=0.3pt] (n0.north) -- (n1.south);
\draw[black!55,line width=0.3pt] (n0.north) -- (n2.south);
\draw[black!55,line width=0.3pt] (n0.north) -- (n4.south);
\draw[black!55,line width=0.3pt] (n1.north) -- (n3.south);
\draw[black!55,line width=0.3pt] (n1.north) -- (n5.south);
\draw[black!55,line width=0.3pt] (n2.north) -- (n5.south);
\draw[black!55,line width=0.3pt] (n3.north) -- (n6.south);
\draw[black!55,line width=0.3pt] (n4.north) -- (n6.south);
\draw[black!55,line width=0.3pt] (n5.north) -- (n6.south);
\end{tikzpicture}}\end{tabular}\hfill
\begin{tabular}[c]{@{}c@{}}\adjustbox{max width=0.62\linewidth}{\begin{tikzpicture}[x=1cm,y=1cm,every node/.style={inner sep=1.5pt}]
\node[draw,fill=black!12,font=\tiny\ttfamily,align=center,rounded corners=1pt] (n0) at (0.000,0.000) {0\,0\,0\\1\,1\,1\\2\,2\,2};
\node[font=\tiny,inner sep=0pt,anchor=north] at (n0.south) {2};
\node[draw,fill=white,font=\tiny\ttfamily,align=center,rounded corners=1pt] (n5) at (-3.750,1.500) {0\,0\,0\\0\,1\,0\\2\,2\,2};
\node[font=\tiny,inner sep=0pt,anchor=north] at (n5.south) {2};
\node[draw,fill=white,font=\tiny\ttfamily,align=center,rounded corners=1pt] (n2) at (-2.500,1.500) {0\,0\,0\\0\,1\,0\\0\,0\,2};
\node[font=\tiny,inner sep=0pt,anchor=north] at (n2.south) {1};
\node[draw,fill=white,font=\tiny\ttfamily,align=center,rounded corners=1pt] (n6) at (-1.250,1.500) {0\,0\,0\\0\,1\,1\\0\,2\,2};
\node[font=\tiny,inner sep=0pt,anchor=north] at (n6.south) {2};
\node[draw,fill=black!12,font=\tiny\ttfamily,align=center,rounded corners=1pt] (n1) at (0.000,1.500) {0\,0\,0\\0\,0\,0\\0\,0\,0};
\node[font=\tiny,inner sep=0pt,anchor=north] at (n1.south) {1};
\node[draw,fill=black!12,font=\tiny\ttfamily,align=center,rounded corners=1pt] (n4) at (1.250,1.500) {0\,0\,0\\0\,0\,0\\2\,2\,2};
\node[font=\tiny,inner sep=0pt,anchor=north] at (n4.south) {2};
\node[draw,fill=white,font=\tiny\ttfamily,align=center,rounded corners=1pt] (n3) at (2.500,1.500) {0\,0\,0\\0\,1\,1\\0\,1\,2};
\node[font=\tiny,inner sep=0pt,anchor=north] at (n3.south) {1};
\node[draw,fill=white,font=\tiny\ttfamily,align=center,rounded corners=1pt] (n7) at (3.750,1.500) {0\,0\,0\\0\,1\,2\\2\,2\,2};
\node[font=\tiny,inner sep=0pt,anchor=north] at (n7.south) {2};
\node[draw,fill=white,font=\tiny\ttfamily,align=center,rounded corners=1pt] (n10) at (-3.125,3.000) {0\,0\,0\\0\,0\,0\\0\,0\,1};
\node[font=\tiny,inner sep=0pt,anchor=north] at (n10.south) {1};
\node[draw,fill=white,font=\tiny\ttfamily,align=center,rounded corners=1pt] (n8) at (-1.875,3.000) {0\,0\,0\\0\,0\,0\\0\,0\,2};
\node[font=\tiny,inner sep=0pt,anchor=north] at (n8.south) {1};
\node[draw,fill=white,font=\tiny\ttfamily,align=center,rounded corners=1pt] (n17) at (-0.625,3.000) {0\,1\,2\\1\,2\,0\\2\,0\,1};
\node[font=\tiny,inner sep=0pt,anchor=north] at (n17.south) {1};
\node[draw,fill=white,font=\tiny\ttfamily,align=center,rounded corners=1pt] (n11) at (0.625,3.000) {0\,0\,2\\0\,0\,2\\2\,2\,0};
\node[font=\tiny,inner sep=0pt,anchor=north] at (n11.south) {1};
\node[draw,fill=white,font=\tiny\ttfamily,align=center,rounded corners=1pt] (n9) at (1.875,3.000) {0\,0\,0\\0\,1\,1\\0\,1\,1};
\node[font=\tiny,inner sep=0pt,anchor=north] at (n9.south) {1};
\node[draw,fill=white,font=\tiny\ttfamily,align=center,rounded corners=1pt] (n12) at (3.125,3.000) {0\,1\,1\\1\,0\,0\\1\,0\,0};
\node[font=\tiny,inner sep=0pt,anchor=north] at (n12.south) {1};
\node[draw,fill=white,font=\tiny\ttfamily,align=center,rounded corners=1pt] (n16) at (-1.250,4.500) {0\,0\,2\\0\,1\,2\\2\,2\,0};
\node[font=\tiny,inner sep=0pt,anchor=north] at (n16.south) {1};
\node[draw,fill=white,font=\tiny\ttfamily,align=center,rounded corners=1pt] (n13) at (0.000,4.500) {0\,0\,0\\0\,0\,0\\0\,1\,2};
\node[font=\tiny,inner sep=0pt,anchor=north] at (n13.south) {2};
\node[draw,fill=white,font=\tiny\ttfamily,align=center,rounded corners=1pt] (n15) at (1.250,4.500) {0\,0\,0\\0\,1\,2\\0\,2\,1};
\node[font=\tiny,inner sep=0pt,anchor=north] at (n15.south) {1};
\node[draw,fill=white,font=\tiny\ttfamily,align=center,rounded corners=1pt] (n14) at (0.000,6.000) {0\,0\,0\\0\,0\,1\\0\,1\,2};
\node[font=\tiny,inner sep=0pt,anchor=north] at (n14.south) {1};
\draw[black!55,line width=0.3pt] (n0.north) -- (n1.south);
\draw[black!55,line width=0.3pt] (n0.north) -- (n2.south);
\draw[black!55,line width=0.3pt] (n0.north) -- (n3.south);
\draw[black!55,line width=0.3pt] (n0.north) -- (n4.south);
\draw[black!55,line width=0.3pt] (n0.north) -- (n5.south);
\draw[black!55,line width=0.3pt] (n0.north) -- (n6.south);
\draw[black!55,line width=0.3pt] (n0.north) -- (n7.south);
\draw[black!55,line width=0.3pt] (n1.north) -- (n10.south);
\draw[black!55,line width=0.3pt] (n1.north) -- (n11.south);
\draw[black!55,line width=0.3pt] (n1.north) -- (n12.south);
\draw[black!55,line width=0.3pt] (n1.north) -- (n17.south);
\draw[black!55,line width=0.3pt] (n4.north) -- (n8.south);
\draw[black!55,line width=0.3pt] (n4.north) -- (n9.south);
\draw[black!55,line width=0.3pt] (n4.north) -- (n11.south);
\draw[black!55,line width=0.3pt] (n4.north) -- (n12.south);
\draw[black!55,line width=0.3pt] (n5.north) -- (n16.south);
\draw[black!55,line width=0.3pt] (n6.north) -- (n15.south);
\draw[black!55,line width=0.3pt] (n8.north) -- (n13.south);
\draw[black!55,line width=0.3pt] (n8.north) -- (n16.south);
\draw[black!55,line width=0.3pt] (n9.north) -- (n15.south);
\draw[black!55,line width=0.3pt] (n13.north) -- (n14.south);
\end{tikzpicture}}\end{tabular}
\caption{Hasse diagram of closed mathematical structures generated by a 
single binary operation on 2 elements (left) 
and by a semigroup on three elements (right), ordered by inclusion.
Each generates its children. Numbers show the number of mathematical 
structures in each equivalence class.}
\label{CloneHasseFig}
\end{figure}

As illustrated in \fig{CloneHasseFig}, closed mathematical structures generated by binary functions are partially ordered by inclusion, 
where each generates its children. In other words, the clone of a child is a subset of the clone of a  parent. 
For example, for a single binary operation on 3 elements, there are 
$3^{3^2} = 19683$ different presentations, which form 3330 mathematical structures (magmas up to isomorphism).  
24 of them are also semigroups, but some generate each other, leaving only 18 clones. 15 of these are truly binary,
2 depend only on one variable and 1 is a constant. 
In this example, each class consists of a semigroup and its opposite (which may already be isomorphic to it), since $x\circ y$ and $y\circ x$ generate each other. 
More generally, classes can be quite large.

\subsection{Counting mathematical structures}

\tab{StructureCountTable} shows the number of mathematical structures with $n$ elements
that are models of various theories; most of these counts are listed in OEIS \cite{OEIS}.
Monounary algebras are simply a single unary function $f\colon S\to S$.
Finite distributive lattices are the same thing as finite frames.
The rows are seen to scale quite differently:  
the group rows are relatively flat and arithmetic in nature, 
the semigroup rows grow roughly like $n^{cn^2}$
 and the magma row is $n^{n^2}/n!$ up to a factor tending to $1$.

\tab{ClosedStructureCountTable} compares the numbers of mathematical structures and closed mathematical structures with $n$ elements, computed
as described in Appendix~\ref{CountingAppendix}.
The numbers of mathematical structures (the ``str'' columns) are known from the literature (cited in \tab{StructureCountTable}).
The new entries are the numbers of closed mathematical structures (the ``cl'' columns). Some of these might not be in the existing literature, 
for example the $411$ clones generated by a single binary operation on three elements.

The last column of the table shows $n_{\rm diff}$, defined as the smallest $n$ for which $\mathrm{cl}<\mathrm{str}$, that is, for which two non-isomorphic models of the theory generate each other,
as detailed in Appendix~\ref{CountingAppendix}. A ``--'' means that this never happens for any $n$.

\begin{table}[H]\centering
\footnotesize\setlength{\tabcolsep}{3.4pt}
\begin{tabular}{l|rrrrrrrr|l}\toprule
$n$ & 1 & 2 & 3 & 4 & 5 & 6 & 7 & 8 & Source\\\midrule
Monounary algebras & 1 & 3 & 7 & 19 & 47 & 130 & 343 & 951 & A001372\\
Rooted trees & 1 & 1 & 2 & 4 & 9 & 20 & 48 & 115 & A000081\\\midrule
Digraphs & $2$ & $10$ & $104$ & $3044$ & $291\,968$ & $96\,928\,992$ & $1.1\cdot10^{11}$ & $4.6\cdot10^{14}$ & A000595\\
Graphs & $1$ & $2$ & $4$ & $11$ & $34$ & $156$ & $1044$ & $12\,346$ & A000088\\
Preorders & $1$ & $3$ & $9$ & $33$ & $139$ & $718$ & $4535$ & $35\,979$ & A001930\\
Equivalence relations & $1$ & $2$ & $3$ & $5$ & $7$ & $11$ & $15$ & $22$ & A000041\\
Total orders & $1$ & $1$ & $1$ & $1$ & $1$ & $1$ & $1$ & $1$ & one per $n$\\\midrule
Magmas & $1$ & $10$ & $3330$ & $178\,981\,952$ & $2.5\cdot10^{15}$ & $1.4\cdot10^{25}$ & $5.1\cdot10^{37}$ & $1.6\cdot10^{53}$ & A001329\\
Semigroups & $1$ & $5$ & $24$ & $188$ & $1915$ & $28\,634$ & $1\,627\,672$ & $3\,684\,030\,417$ & A027851\\
Bands & $1$ & $3$ & $10$ & $46$ & $251$ & $1682$ & $13\,213$ & $119\,826$ & A058112\\
Monoids & $1$ & $2$ & $7$ & $35$ & $228$ & $2237$ & $31\,559$ & $1\,668\,997$ & A058129\\
Commutative monoids & $1$ & $2$ & $5$ & $19$ & $78$ & $421$ & $2637$ & $20\,486$ & A058131\\
Groups & $1$ & $1$ & $1$ & $2$ & $1$ & $2$ & $1$ & $5$ & A000001\\
Abelian groups & $1$ & $1$ & $1$ & $2$ & $1$ & $1$ & $1$ & $3$ & A000688\\
Semilattices & $1$ & $1$ & $2$ & $5$ & $15$ & $53$ & $222$ & $1078$ & A006966$(n{+}1)$\\\midrule
Semirings & $1$ & $2$ & $6$ & $40$ & $295$ & $3246$ & $59\,314$ & -- & arXiv:2507.03709\\
Rings (with 1) & $1$ & $1$ & $1$ & $4$ & $1$ & $1$ & $1$ & $11$ & A037291\\
Commutative rings & $1$ & $1$ & $1$ & $4$ & $1$ & $1$ & $1$ & $10$ & A127707\\
Integral domains & $0$ & $1$ & $1$ & $1$ & $1$ & $0$ & $1$ & $1$ & prime powers\\
Division rings & $0$ & $1$ & $1$ & $1$ & $1$ & $0$ & $1$ & $1$ & prime powers\\
Fields & $0$ & $1$ & $1$ & $1$ & $1$ & $0$ & $1$ & $1$ & prime powers\\
Lattices & $1$ & $1$ & $1$ & $2$ & $5$ & $15$ & $53$ & $222$ & A006966\\
Distributive lattices & $1$ & $1$ & $1$ & $2$ & $3$ & $5$ & $8$ & $15$ & A006982\\
Boolean algebras & $1$ & $1$ & $0$ & $1$ & $0$ & $0$ & $0$ & $1$ & powers of $2$\\\bottomrule
\end{tabular}
\caption{Mathematical structures with $n$ elements}
\label{StructureCountTable}
\end{table}

\begin{table}[H]\centering
\footnotesize\setlength{\tabcolsep}{3.5pt}
\begin{tabular}{l|rr|rr|rr|rr|rr|rr|c}\toprule
 & \multicolumn{4}{c|}{$n=2$} & \multicolumn{4}{c|}{$n=3$} & \multicolumn{4}{c|}{$n=4$} & \\
Theory & str & cl & str$_{\rm tb}$ & cl$_{\rm tb}$ & str & cl & str$_{\rm tb}$ & cl$_{\rm tb}$ & str & cl & str$_{\rm tb}$ & cl$_{\rm tb}$ & $n_{\rm diff}$\\\midrule
Monounary algebras & $3$ & $3$ & & & $7$ & $7$ & & & $19$ & $19$ & &  & 6\\
Rooted trees & $1$ & $1$ & & & $2$ & $2$ & & & $4$ & $4$ & &  & --\\
\midrule
Magmas & $10$ & $7$ & $5$ & $4$ & $3330$ & $411$ & $3317$ & $404$ & $178\,981\,952$ & --\footnotemark & $178\,981\,915$ & -- & 2\\
Semigroups & $5$ & $4$ & $2$ & $2$ & $24$ & $18$ & $19$ & $15$ & $188$ & $126$ & $179$ & $121$ & 2\\
Bands & $3$ & $2$ & $1$ & $1$ & $10$ & $6$ & $8$ & $5$ & $46$ & $26$ & $44$ & $25$ & 2\\
Monoids & $2$ & $2$ & $2$ & $2$ & $7$ & $6$ & $7$ & $6$ & $35$ & $27$ & $35$ & $27$ & 3\\
Commutative monoids & $2$ & $2$ & $2$ & $2$ & $5$ & $5$ & $5$ & $5$ & $19$ & $19$ & $19$ & $19$ & --\\
Groups & $1$ & $1$ & $1$ & $1$ & $1$ & $1$ & $1$ & $1$ & $2$ & $2$ & $2$ & $2$ & 24--260\\
Abelian groups & $1$ & $1$ & $1$ & $1$ & $1$ & $1$ & $1$ & $1$ & $2$ & $2$ & $2$ & $2$ & --\\
Semilattices & $1$ & $1$ & $1$ & $1$ & $2$ & $2$ & $2$ & $2$ & $5$ & $5$ & $5$ & $5$ & --\\
\midrule
Semirings & $2$ & $2$ & $2$ & $2$ & $6$ & $6$ & $6$ & $6$ & $40$ & $38$ & $40$ & $38$ & 4\\
Rings (with 1) & $1$ & $1$ & $1$ & $1$ & $1$ & $1$ & $1$ & $1$ & $4$ & $4$ & $4$ & $4$ & 16\\
Commutative rings & $1$ & $1$ & $1$ & $1$ & $1$ & $1$ & $1$ & $1$ & $4$ & $4$ & $4$ & $4$ & 16\\
Integral domains & $1$ & $1$ & $1$ & $1$ & $1$ & $1$ & $1$ & $1$ & $1$ & $1$ & $1$ & $1$ & --\\
Division rings & $1$ & $1$ & $1$ & $1$ & $1$ & $1$ & $1$ & $1$ & $1$ & $1$ & $1$ & $1$ & --\\
Fields & $1$ & $1$ & $1$ & $1$ & $1$ & $1$ & $1$ & $1$ & $1$ & $1$ & $1$ & $1$ & --\\
Lattices & $1$ & $1$ & $1$ & $1$ & $1$ & $1$ & $1$ & $1$ & $2$ & $2$ & $2$ & $2$ & 5\\
Boolean algebras & $1$ & $1$ & $1$ & $1$ & $0$ & $0$ & $0$ & $0$ & $1$ & $1$ & $1$ & $1$ & --\\
\bottomrule\end{tabular}
\caption{Number of mathematical structures ``str'' (presentations up to isomorphism) and closed mathematical structures (clones) ``cl'' for small $n$; the second pair of columns restricts to ones whose binary operations are all truly binary by depending on both input variables, so it does not apply to the two unary rows. The last column gives the smallest $n$ for which cl $<$ str, with a dash if this never happens and a range if the value is not yet determined; see Appendix~\ref{CountingAppendix}.}
\label{ClosedStructureCountTable}
\end{table}
\footnotetext{The magma clone counts for $n=4$ are left open. There are $178\,981\,952$ magmas on four elements up to isomorphism ($178\,981\,915$ of them truly binary), and each needs a canonical form of its clone, whose binary part can hold up to $4^{16}\approx4.3\cdot10^9$ functions; with the closure algorithm used for the other rows a typical magma takes hours and tens of gigabytes, so the row would cost far more than $10^5$ CPU-years. Completing it needs a different method, such as Rosenberg's completeness criterion for the Sheffer operations and a relational invariant for the rest.}

\subsection{Determining compositional equivalence in practice}

There is an interesting analogy between mathematical theories and closed mathematical structures.
Both are specified by a list of ``base cases" (axioms and generating functions, respectively; the list is often finite) and rules for generating 
new ones by combining old ones (via deduction and composition, respectively). 
Although the sets of theorems and functions are generally infinite, equivalence between two theories/presentations 
can be proven by showing that
the base cases of one formal system/presentation can be generated from the axioms/functions of the other.
For two finite presentations there is a halting algorithm determining whether they generate each other; one never needs to explore the whole infinite clone, merely the generated functions up to the highest
arity $d$ occurring among the generators of the two presentations. Consider first a single set with $n$ elements. To check whether a generator of arity $k\le d$ of one presentation equals some composition of functions from the other presentation, 
we only need to consider compositions with $k$ input variables, and these compute at most $n^{n^k}\le n^{n^d}$ different functions, the number of all $k$-ary functions on a set with $n$ elements. So we start from the $k$ projections $(x_1,\dots,x_k)\mapsto x_i$ and iteratively compose the functions generated so far with the generators until no new function is created. A constant counts as generated if some generated function is constant with that value. For presentations with several sets the argument is the same, with each variable and each function carrying the sets of its inputs and of its output, and with compositions allowed only when these match; the number of functions with a given input and output profile is again finite. This is how the ``cl" numbers in \tab{ClosedStructureCountTable} were computed (described in greater detail in Appendix~\ref{CountingAppendix}).
For reasons that are not yet fully understood, most of the functions used in science and mathematics are 
unary and binary, so $d=2$ often suffices in practice.

\subsection{Relation to physics: What should be considered equivalent?}

Historically, mathematics has often been developed to model physical systems.
Considering a physical system as described by a mathematical structure 
(or optionally even being one, as controversially explored in \cite{TegmarkMathematicalUniversebook})
provides an interesting perspective both on infinite structures and 
on which presentations should be considered equivalent, as summarized in 
Table~\ref{PresentationEquivalenceLevelTable}.

\begin{table}[H]
\centering
\begin{tabular}{|l|p{6.3cm}|p{5.2cm}|}
\hline
{\bf Equivalence}	&{\bf Two presentations are equivalent if...}	&{\bf An observer who...}\\
\hline
Isomorphic		&they differ only in how elements are labeled	&cannot see element labels\\\hline
Compositional		&each one's functions are compositions of the other's	&can also compose measurements\\\hline
Computational		&the same holds once equality tests ``='' and branching ``?'' are allowed	&can also tell whether two measurement results agree and act on it\\\hline
Definitional		&each one's functions can be defined from the other's using $\forall$ and $\exists$	&can also search through all elements\\
\hline
\end{tabular}
\caption{Four successively broader notions of equivalence between presentations. Each row also allows everything in the rows above it. The last column describes an observer for whom presentations equivalent in that sense are indistinguishable.}
\label{PresentationEquivalenceLevelTable}
\end{table}

Consider an observer who inhabits a physical world described by a mathematical structure, and is able to perform certain measurements and reasoning steps.
The world is made up of elements (examples may include points, magnetic field vectors, \etc) and various relations that hold between them.
For example, in classical physics, three nearby points in space may satisfy the relation that they lie on a line, or one magnetic field vector may be 2 times another.
Which relations or other functions would the observer consider well-defined and inherent in the physical system?
The answer depends on what she is able to do, and \tab{PresentationEquivalenceLevelTable} lists four successively more capable observers.
The more she can do, the more functions she can obtain from the ones that nature hands her, and the more presentations therefore become equivalent to her.
Let us go through the four rows in turn.

\paragraph{Isomorphism}
There is not a shred of evidence that notation matters in physics.
In our physical world, nature never seems to reveal specific coordinate systems or other entity labels to us.
For example, we do not physically observe a point in space being labeled ${\bf r}$, or our universe having any labeled $x$-axis.
This motivates describing the physical system as a mathematical structure (an equivalence class of isomorphic presentations), since  those unobservable labels are
the only difference between the presentations in the equivalence class.

\paragraph{Compositional equivalence}
If the observer is able to measure certain quantities (\eg, distances between points) and use them as inputs determining
subsequent experimental measurements, then she is able to {\it compose} the functions that nature gives her access to.
This motivates describing the physical system using a closed mathematical structure.

\paragraph{Computational equivalence}
Suppose that the observer can also tell whether two measurement results agree, and let the answer decide what she does next. 
Then she can also define functions by cases.
This amounts to adding two functions to the presentation for each of its sets $S$:
\begin{enumerate}
\item the equality test $=\colon S\times S\to\{0,1\}$, which returns 1 if its two inputs are the same element and 0 otherwise, and 
\item the ``if statement" $?\colon\{0,1\}\times S\times S\to S$, which returns its second input if the first is 1 and its third input otherwise
(analogously to MLISP's {\texttt d}-command).
\end{enumerate}

\paragraph{Definitional equivalence}
Finally, if the observer can also quantify over a full set of elements, then even more presentations become equivalent.
This is what model theory calls {\it definitional equivalence}.
As an example, consider the integers with addition alone, $(\Z,+)$, and the integers with addition and zero, $(\Z,+,0)$. 
There is no procedure for producing zero by merely adding different integers together (there would be if $-$ were also defined, as $x-x$ for any $x$).
But an observer able to search through all integers can identify and point to $0$ as the unique element $z$ satisfying $x+z=x$ for all $x$.

For a presentation with a single finite set, this fourth level has a simple description in terms of symmetries, \ie, relabelings of the elements that leave all functions of the presentation unchanged.
If some symmetry exchanges two elements, then no property whatsoever can distinguish them. Conversely, on a finite set, 
anything that all symmetries preserve can be defined using quantifiers \cite{Krasner1938,Hodges1993}: 
one says ``there is a relabeling of the elements that respects all the lookup tables and carries one of the listed cases to this one''.
Two presentations on the same $n$ elements are therefore equivalent in this sense exactly when they have the same symmetries, 
and two arbitrary ones exactly when their symmetry groups are conjugate subgroups of the permutation group $S_n$.
For example, a total order on $n$ elements has no symmetries, since its smallest element, its second smallest element and so on can each be singled out.
It is therefore equivalent to a set of $n$ elements that each have their own name (a constant), and indeed to every other presentation without symmetries.
Conversely, given a presentation where each element is labeled by a constant function, the order relation $\le$ can be defined via nested if-statements.
At this level, the number of different structures with $n$ elements thus drops to the number of conjugacy classes of subgroups of $S_n$, which is merely 
$1$, $2$, $4$, $11$ and $19$ for $n=1,\dots,5$.

In the physical universe we inhabit, intelligent observers can arguably be approximated as ``computational", with the caveat that their ability to measure and compute is limited.
In contrast, fully ``definitional" observers appear nonexistent because various forms of locality deny us access to more than a small fraction of what physically exists.

\subsection{Infinity in physics and mathematics}

How large is the set of mathematical structures? 
For starters, the number of finite mathematical structures is infinite but countable, since any presentation tableau
can be described in a finite number of bits of information and can therefore be encoded as a natural number whose binary representation
encodes the presentation bit-string. 
The same clearly holds for infinite {\it computable} structures, whose functions are defined by computer programs as in our presentation tableaus.
But what about structures whose relations cannot all be defined by halting computations? 
To sharpen this question, consider these four candidate mathematical structures, which each include
exactly one function on a single set (the set of natural numbers for the first three structures and the set of real numbers for the fourth one):
\begin{enumerate}
\item $P(n)$:  Given a natural number $n$, $P(n)$ denotes the smallest prime greater than $n$.
\item $T(n)$: Given a natural number $n$, $T(n)$ denotes the smallest twin prime greater than $n$
(a twin prime is a prime $p$ for which $p+2$ is also prime).
\item $H(m,n)$: Given two natural numbers $m$ and $n$, $H(m,n)$ equals $0$ if the $m$th computer program on an enumerated list of all  computer programs will keep running forever when fed the bits of $n$ as input, and $H(m,n)$ equals $1$ if the program instead halts after a finite number of steps.
\item $D(x)$: Given a real number $x$, $D(x)$ denotes $2x$.
\end{enumerate}
Which of these four should qualify as a valid mathematical structure, and which get disqualified for being insufficiently well-defined?
The first function, $P(n)$, is a piece of cake: it is easy to write a program that starts checking whether the numbers following $n$ are prime and stops as soon as it finds one, and we are guaranteed that this program will halt after a finite number of steps because Euclid proved that there are infinitely many primes. So $P(n)$ is an example of what we call a {\it computable function}.

The second function, $T(n)$, is trickier: it is again easy to write a program that checks each number following $n$ to see whether it is a twin prime, but if you plug in a number $n$ greater than $2996863034895\times 2^{1290000}-1$ (the largest twin prime known as of writing this),
there is no guarantee that the program will ever stop and deliver an answer, because despite the best efforts of our most talented mathematicians, we still do not know whether there are infinitely many twin primes. So for now, we do not know whether 
$T(n)$ is defined for all $n$, and it is debatable whether a mathematical structure containing such a sloppily specified function qualifies as well-defined.

The third function, $H(m,n)$, is even more nefarious: 
Alonzo Church and Alan Turing established
that there is {\it no} program that can compute $H(m,n)$ for arbitrary input numbers $m$ and $n$ in a finite number of steps, so $H(m,n)$ is an example of what we call an {\it uncomputable function}.
In other words, there is no program that can determine which other programs will eventually halt.
Of course, any given program will or will not halt, but the catch is, just as with the twin primes, that you might have to wait forever to find out.
The Church--Turing discovery of uncomputable functions is closely related to the discovery by Kurt G\"odel that some statements of arithmetic are undecidable, meaning that they can neither be proved nor disproved in a finite number of steps in a given formal system.

The fourth function, $D(x)$, is even worse: whereas $H(m,n)$ is computable for some but not all inputs, $D(x)$ is uncomputable for almost all inputs (for all real numbers except a subset of Lebesgue measure zero).
The reason is that a generic real number has infinitely many decimals that go on without any discernible pattern, and therefore requires an infinite
number of bits of information to describe: the problem is not merely performing an infinite number of computational steps on a finite input as for the $H$-example, but that of inputting and outputting an infinite amount of information.

So should a mathematical structure be considered well-defined even if it contains a function such as $H$ or $D$ that cannot be evaluated even on an arbitrarily powerful computer? 
I am not going to answer this question here, but find it interesting that closely related issues have been hotly debated among mathematicians without any reference to physics. 
According to the finitist school of mathematicians, which included Leopold Kronecker, 
Hermann Weyl and Reuben Goodstein,
a mathematical object does not exist unless it can be constructed from whole numbers in a finite number of steps. 
With this criterion, an existing mathematical structure can contain rational numbers and algebraic numbers, but not generic real numbers
or other generic members of continuous/uncountable sets. 

Turning now to physics, how well-defined do mathematical structures need to be to correspond to something in the physical world?
The above examples suggest a hierarchy of possible answers:
\begin{enumerate}
\item Finite structures
\item Computable structures (defined by halting computations)
\item Structures defined by computations that might not halt
\item Uncountable structures (involving {\eg} real numbers)
\end{enumerate}

Although physics textbooks are full of uncountable structures, it is striking that physics has thus far provided no solid experimental evidence that anything infinite (let alone uncountably infinite) exists physically. 
The contrast is particularly noteworthy for real numbers, where we cannot define generic ones
(or functions thereof) computationally.
Most physicists rarely bother distinguishing between a real number and 
a \texttt{float64} variable; indeed, computer languages including Fortran and Pascal even refer to the float datatype as 
``real".
This is probably because {\it the distinction never seems to matter}.
A \texttt{float64} variable represents a number of the form $x = \pm a\times 2^b$, 
with 52 bits for $a$ and 11 for $b$, and can approximate any number
between $10^{-308}$ and $10^{308}$ to about 15 significant digits.
In contrast, physicists never need such a large dynamic range in practice, since
the largest accessible distance (the diameter of our observable universe) is ``only" about $10^{62}$
times the smallest meaningful distance (the Planck length), and there are ``only"
about $10^{88}$ particles in our observable universe.
Indeed, the popular but controversial holographic principle 
\cite{BaoCarrollSingh2017} %
suggests that our universe 
can store only about $10^{123}$ bits of information.
No measurement in physics has reached a relative precision better than about one part in $10^{22}$ (length changes in gravitational wave detectors), for which \texttt{float128} still suffices. 
So when physicists casually talk about computing ``real numbers", usually they {\it de facto} mean not 
members of the uncountable set $\mathbb{R}$, but 
something that they can have their computer approximate to any reasonable precision they desire. 

Whereas many early physicists assumed that the physical laws they discovered were exact,
modern views tend to align more with Steven Weinberg's \cite{WeinbergEssay}
humbler stance that all of today's physics theories are probably mere effective theories, 
approximations of something deeper and more accurate. 
Some have even speculated that the deepest theory does not involve the continuum or any form of infinity
\cite{Gisin2021,DelSantoGisin2019,EllisMeissnerNicolai2018,TegmarkMathematicalUniversebook}.

Approximating something discrete as continuous is often convenient, greatly simplifying calculations. 
For example, consider the air in front of you. Keeping track of the positions and speeds of octillions of atoms would be hopelessly complicated. But if you ignore the fact that air is made of atoms and instead approximate it as a continuum, a smooth substance that has a density, pressure and velocity at each point, you find that this idealized air obeys a beautifully simple partial differential equation that explains many important phenomena: 
how sound waves propagate through air, how winds work, {\etc}
Yet despite all that convenience, air is not truly continuous. 
Many physicists guess that it may be the same way for space, time and all the other seemingly 
continuous building blocks of our physical world, even though fully workable non-continuous alternatives are still lacking. Humility aside, this guess is appealing because 
some of the thorniest open problems in modern physics rest on assumptions about infinity and the continuum, from the measure problem in eternal inflation \cite{Linde2015,Hebecker2020} to black hole and Big Bang singularities 
\cite{SenovillaGarfinkle2015,CrowtherDeHaro2022}.
On the theoretical physics side, it will be interesting to see whether discrete or finite theories are found that match the experimental successes of today's leading physics theories. On the experimental side, it will be interesting to see whether any observations can determine whether any assumed physical continuum satisfies the least-upper-bound property.

\section{Conclusions}\label{ConclusionsSec}

We have introduced MLISP, a tiny Turing-complete language, and used it to define a family tree of mathematical theories and associated models.
Emphasis was placed on consistently applying a standardized terminology throughout.

We used a consistent four-way ``map versus territory" terminology: formal systems and theories are descriptions; presentations and mathematical structures are what they describe. Theories and structures are both equivalence classes that strip away notation.
Every structure is defined by a presentation tableau, with lookup tables for finite sets and short halting programs for countably infinite ones. 
This standardizes the side of the family tree where the literature has converged least. 
For finite presentations on a set with $n$ elements, equivalence is decided by a halting algorithm that never explores more than the $n^{n^d}$ functions of the highest arity $d$ among the generators.
The definitions, key theorems, model relations and the correctness of MLISP are checked in Lean. 
Lean checks MLISP rather than defines the presentations, which keeps the trusted base small.

This work naturally raises many open questions. Here are a few examples:
\begin{itemize}
\item What is the lowest order (between 24 and 260) at which two different groups are compositionally equivalent?
\item How can the closed mathematical structures of various types and sizes be counted more efficiently? 
These integer sequences arguably deserve to be added to OEIS like their non-closed counterparts.
\item Is there generally a canonical normal form or ``simplest'' presentation defining a mathematical structure or a closed mathematical structure?
\item The partial tree presented is biased toward what appears in university courses, Wikipedia and Mathlib; is AI now capable of automatically extending it within, \eg, 
the $\tau$~Ceti project \cite{TauCeti}? 
\end{itemize}

As AI proposes and proves ever more mathematical theorems, compact human-readable maps of what the theorems are about become more valuable. 
This project is a small step in that direction, which will hopefully inspire others to continue further.

\bigskip
{\bf Acknowledgements:} The author would like to thank Meia Chita-Tegmark for encouragement with this long-running project. 
Codex and Claude Code assisted with plots, tables, Lean proofs, Appendix A and proofreading.
The mathematics tree was organized without AI, starting in 1996. The author created MLISP and the programs in \tab{tab:MLISPprograms} around 2018.

\clearpage
\bibliography{mathtree}

\begin{thebibliography}{10}

\bibitem{MarathonStatement}
{Current and former Caltech mathematicians}.
\newblock Open letter about the mathathon.
\newblock Proofs and Prompts,
  \url{https://proofsandprompts.com/2026/09/10/open-letter-about-the-mathathon/},
  September 2026.
\newblock Open letter of 10 September 2026 opposing the Caltech Mathathon, a
  hackathon on open research problems with LLMs; 771 signatories at
  publication.

\bibitem{FieldsMedalistStatement}
{Twenty-eight Fields Medalists}.
\newblock A severe misalignment of {AI} in mathematics.
\newblock Declaration, \url{https://mathandai.org/}, September 2026.
\newblock Signed by 28 Fields Medal recipients, including Pierre Deligne, Peter
  Scholze, Maryna Viazovska, Martin Hairer, C\'edric Villani, Manjul Bhargava
  and Terence Tao; published 11 September 2026 and open to further signatures.

\bibitem{Mathlib2020}
{The mathlib Community}.
\newblock The {L}ean mathematical library.
\newblock In {\em Proceedings of the 9th ACM SIGPLAN International Conference
  on Certified Programs and Proofs (CPP 2020)}, pages 367--381. ACM, 2020.

\bibitem{Tegmark9704009}
Max Tegmark.
\newblock Is ``the theory of everything'' merely the ultimate ensemble theory?
\newblock {\em Annals of Physics}, 270(1):1--51, 1998.
\newblock arXiv:gr-qc/9704009.

\bibitem{CarnapBachmann1936}
Rudolf Carnap and Friedrich Bachmann.
\newblock {\"U}ber {E}xtremalaxiome.
\newblock {\em Erkenntnis}, 6:166--188, 1936.

\bibitem{Schiemer2020}
Georg Schiemer.
\newblock Carnap's structuralist thesis.
\newblock In Erich~H. Reck and Georg Schiemer, editors, {\em The Prehistory of
  Mathematical Structuralism}, pages 383--420. Oxford University Press, Oxford,
  2020.
\newblock \url{https://doi.org/10.1093/oso/9780190641221.003.0015}.

\bibitem{Bourbaki1950}
Nicolas Bourbaki.
\newblock The architecture of mathematics.
\newblock {\em The American Mathematical Monthly}, 57(4):221--232, 1950.

\bibitem{Hilbert1918}
David Hilbert.
\newblock Axiomatisches {D}enken.
\newblock {\em Mathematische Annalen}, 78:405--415, 1918.

\bibitem{MacLane1986}
Saunders Mac~Lane.
\newblock {\em Mathematics: Form and Function}.
\newblock Springer, New York, 1986.

\bibitem{Corry1996}
Leo Corry.
\newblock {\em Modern Algebra and the Rise of Mathematical Structures},
  volume~17 of {\em Science Networks}.
\newblock Birkh\"auser, Basel, 1996.

\bibitem{Gowers2008}
Timothy Gowers, June Barrow-Green, and Imre Leader, editors.
\newblock {\em The Princeton Companion to Mathematics}.
\newblock Princeton University Press, Princeton, 2008.

\bibitem{OEIS}
{OEIS Foundation Inc.}
\newblock The {O}n-{L}ine {E}ncyclopedia of {I}nteger {S}equences, 2026.
\newblock \url{https://oeis.org}.

\bibitem{Lean4}
Leonardo de~Moura and Sebastian Ullrich.
\newblock The {L}ean 4 theorem prover and programming language.
\newblock In {\em Automated Deduction -- CADE 28}, volume 12699 of {\em Lecture
  Notes in Computer Science}, pages 625--635. Springer, 2021.

\bibitem{Rocq}
{The Rocq Development Team}.
\newblock The {R}ocq prover (formerly {C}oq), reference manual.
\newblock \url{https://rocq-prover.org/}, 2025.

\bibitem{Isabelle}
Tobias Nipkow, Lawrence~C. Paulson, and Markus Wenzel.
\newblock {\em Isabelle/{HOL}: A Proof Assistant for Higher-Order Logic},
  volume 2283 of {\em Lecture Notes in Computer Science}.
\newblock Springer, Berlin, 2002.

\bibitem{Bishop1967}
Errett Bishop.
\newblock {\em Foundations of Constructive Analysis}.
\newblock McGraw-Hill, New York, 1967.

\bibitem{LeanBug}
Leonardo de~Moura.
\newblock Postmortem for kernel soundness bug \#14576.
\newblock
  \url{https://leodemoura.github.io/blog/2026-8-1-postmortem-for-kernel-soundness-bug-14576/},
  August 2026.

\bibitem{CalkinWilf2000}
Neil Calkin and Herbert~S. Wilf.
\newblock Recounting the rationals.
\newblock {\em The American Mathematical Monthly}, 107(4):360--363, 2000.

\bibitem{BishopBridges1985}
Errett Bishop and Douglas Bridges.
\newblock {\em Constructive Analysis}, volume 279 of {\em Grundlehren der
  mathematischen Wissenschaften}.
\newblock Springer, Berlin, 1985.

\bibitem{Wildberger2005}
N.~J. Wildberger.
\newblock {\em Divine Proportions: Rational Trigonometry to Universal
  Geometry}.
\newblock Wild Egg, Sydney, 2005.

\bibitem{TegmarkMathematicalUniversebook}
Max Tegmark.
\newblock {\em Our Mathematical Universe: My Quest for the Ultimate Nature of
  Reality}.
\newblock Knopf, New York, 2014.

\bibitem{Nicod2017}
Jean Nicod.
\newblock A reduction in the number of the primitive propositions of logic.
\newblock {\em Proceedings of the Cambridge Philosophical Society}, 19:32--41,
  1917.

\bibitem{Hodges1993}
Wilfrid Hodges.
\newblock {\em Model Theory}, volume~42 of {\em Encyclopedia of Mathematics and
  its Applications}.
\newblock Cambridge University Press, Cambridge, 1993.

\bibitem{Post1941}
Emil~L. Post.
\newblock {\em The Two-Valued Iterative Systems of Mathematical Logic},
  volume~5 of {\em Annals of Mathematics Studies}.
\newblock Princeton University Press, Princeton, 1941.

\bibitem{Hall1958}
Philip Hall.
\newblock Unpublished lectures on universal algebra, 1958.
\newblock The term ``clone'' for a composition-closed set of operations
  containing the projections is attributed to Hall by Cohn \cite{Cohn1965}.

\bibitem{Cohn1965}
Paul~M. Cohn.
\newblock {\em Universal Algebra}.
\newblock Harper \& Row, New York, 1965.

\bibitem{KearnesSzendrei2004}
Keith~A. Kearnes and {\'A}gnes Szendrei.
\newblock Groups with identical subgroup lattices in all powers.
\newblock {\em Journal of Group Theory}, 7(3):385--402, 2004.

\bibitem{Krasner1938}
Marc Krasner.
\newblock Une g{\'e}n{\'e}ralisation de la notion de corps.
\newblock {\em Journal de Math{\'e}matiques Pures et Appliqu{\'e}es},
  17:367--385, 1938.

\bibitem{BaoCarrollSingh2017}
Ning Bao, Sean~M. Carroll, and Ashmeet Singh.
\newblock {The Hilbert Space of Quantum Gravity Is Locally Finite-Dimensional}.
\newblock {\em Int. J. Mod. Phys. D}, 26(12):1743013, 2017.
\newblock arXiv:1704.00066.

\bibitem{WeinbergEssay}
Steven Weinberg.
\newblock What is quantum field theory, and what did we think it is?
\newblock In Tian~Yu Cao, editor, {\em Conceptual Foundations of Quantum Field
  Theory}, pages 241--251. Cambridge University Press, Cambridge, 1999.
\newblock arXiv:hep-th/9702027.

\bibitem{Gisin2021}
Nicolas Gisin.
\newblock {Indeterminism in Physics, Classical Chaos and Bohmian Mechanics: Are
  Real Numbers Really Real?}
\newblock {\em Erkenntnis}, 86(6):1469--1481, 2021.
\newblock arXiv:1803.06824.

\bibitem{DelSantoGisin2019}
Flavio Del~Santo and Nicolas Gisin.
\newblock {Physics without determinism: Alternative interpretations of
  classical physics}.
\newblock {\em Phys. Rev. A}, 100(6):062107, 2019.
\newblock arXiv:1909.03697.

\bibitem{EllisMeissnerNicolai2018}
George F.~R. Ellis, Krzysztof~A. Meissner, and Hermann Nicolai.
\newblock {The physics of infinity}.
\newblock {\em Nature Phys.}, 14(8):770--772, 2018.

\bibitem{Linde2015}
Andrei Linde.
\newblock {A brief history of the multiverse}.
\newblock {\em Rept. Prog. Phys.}, 80(2):022001, 2017.
\newblock arXiv:1512.01203.

\bibitem{Hebecker2020}
Arthur Hebecker.
\newblock {\em {Naturalness, String Landscape and Multiverse: A Modern
  Introduction with Exercises}}, volume 979 of {\em Lecture Notes in Physics}.
\newblock Springer, March 2021.
\newblock arXiv:2008.10625.

\bibitem{SenovillaGarfinkle2015}
Jos{\'e} M.~M. Senovilla and David Garfinkle.
\newblock {The 1965 Penrose singularity theorem}.
\newblock {\em Class. Quant. Grav.}, 32(12):124008, 2015.
\newblock arXiv:1410.5226.

\bibitem{CrowtherDeHaro2022}
Karen Crowther and Sebastian De~Haro.
\newblock {Four Attitudes Towards Singularities in the Search for a Theory of
  Quantum Gravity}.
\newblock In Antonio Vassallo, editor, {\em The Foundations of Spacetime
  Physics: Philosophical Perspectives}, pages 223--250. Routledge, 2022.
\newblock arXiv:2112.08531.

\bibitem{TauCeti}
{Tau Ceti Project}.
\newblock {Tau Ceti}: an {AI}-built {Lean} library of formalized mathematics
  downstream of {Mathlib}.
\newblock \url{https://github.com/TauCetiProject/TauCeti}, 2026.

\bibitem{Jablonskij1958}
S.~V. Jablonskij.
\newblock Functional constructions in a $k$-valued logic.
\newblock {\em Trudy Mat. Inst. Steklov.}, 51:5--142, 1958.
\newblock In Russian.

\bibitem{Slupecki1939}
Jerzy S{\l}upecki.
\newblock Kryterium pe{\l}no{\'s}ci wielowarto{\'s}ciowych system{\'o}w logiki
  zda{\'n} ({A} criterion of fullness of many-valued systems of propositional
  logic).
\newblock {\em Comptes Rendus des S{\'e}ances de la Soci{\'e}t{\'e} des
  Sciences et des Lettres de Varsovie, Classe III}, 32:102--110, 1939.

\bibitem{Zadori1997}
L\'aszl\'o Z\'adori.
\newblock Categorical equivalence of finite groups.
\newblock {\em Bulletin of the Australian Mathematical Society},
  56(3):403--408, 1997.

\end{thebibliography}

\appendix

\clearpage
\section{Counting closed mathematical structures}
\label{CountingAppendix}

This appendix describes how the entries of \tab{ClosedStructureCountTable} were computed. 
The str columns reproduce numbers known from the literature (see \tab{StructureCountTable}); some of the cl entries may be new. The str and cl counts were obtained by exhaustive computer enumeration; Section~\ref{NdiffSubsec} explains the last column, which also rests on proofs and on searches at larger $n$. The Python scripts that produced the numbers are named below in parentheses.

\subsection{What is counted}

Fix a set $S=\{0,\dots,n-1\}$. A presentation of one of the theories in the table is a list of constants, unary functions $S\to S$ and binary functions $S\times S\to S$ satisfying the axioms of the theory.
\begin{itemize}[nosep]
\item {\bf str} is the number of presentations up to isomorphism, \ie, up to relabeling the elements of $S$ by a permutation.
\item {\bf cl} is the number of distinct clones that these presentations generate, again up to relabeling. Two presentations are counted once if, after some relabeling, every function of each is a composition of functions of the other.
\item The columns {\bf str$_{\rm tb}$} and {\bf cl$_{\rm tb}$} repeat the two counts for the presentations all of whose binary functions are {\it truly binary},
\ie, depend essentially on both arguments: $f(x,y)\ne f(x,z)$ for some $x,y,z$ and $f(x,y)\ne f(z,y)$ for some $x,y,z$.
\end{itemize}

\subsection{Step 1: enumerating the presentations}

For each $n$, we generate every labeled presentation of each theory and keep one representative per isomorphism class. The canonical form of a presentation is the lexicographically smallest of its $n!$ relabelings (the tables of all its functions written out as one tuple), and two presentations are isomorphic exactly when their canonical forms agree. The number of distinct canonical forms is str. The labeled presentations are generated as follows.
\begin{itemize}[nosep]
\item {\bf Monounary algebras:} all $n^n$ maps $f:S\to S$. {\bf Rooted trees:} those maps with exactly one fixed point that every element reaches under iteration of $f$.
\item {\bf Magmas:} all $n^{n^2}$ multiplication tables ($16$ for $n=2$ and $19\,683$ for $n=3$).
\item {\bf Semigroups:} all associative tables, found by backtracking: the table is filled in one entry at a time and a branch is abandoned as soon as an instance of $(xy)z=x(yz)$ with all entries known fails. This gives $8$, $113$ and $3492$ labeled semigroups for $n=2,3,4$ (\texttt{semigroups.py}).
\item {\bf Bands, semilattices, monoids, commutative monoids, groups, Abelian groups:} these are selected from the list of semigroups by testing the additional axioms (idempotency; idempotency and commutativity; existence of an identity; and so on). A monoid is stored with its identity as a constant, and a group with its identity and its inverse function, so that these count as generators in Step 2.
\item {\bf Semirings:} for every labeled commutative monoid $(S,+,0)$ from the list above and every choice of the element $1$, all multiplication tables are generated by backtracking subject to associativity, $1$ being an identity, $0$ being absorbing and distributivity over $+$ (\texttt{semirings.py}). {\bf Rings} are the semirings in which every element has an additive inverse; {\bf commutative rings, integral domains, division rings} and {\bf fields} are selected from the rings by testing their axioms.
\item {\bf Lattices:} for every semilattice $(S,\land)$ from the list above, with its order $x\le y\Leftrightarrow x\land y=x$, the join $x\lor y$ is computed as the least upper bound of $x$ and $y$; the semilattice is kept, as the pair $(\land,\lor)$, if all joins exist. {\bf Boolean algebras} are the lattices that are distributive and in which every element has a complement; they are stored with $0$, $1$ and the complement as additional generators.
\end{itemize}

\subsection{Step 2: computing the clone of a presentation}

Every function in these presentations has at most two arguments. In that case two presentations on $S$ generate the same clone if and only if they generate the same set of {\it binary} functions: each generator of one is then among the binary functions generated by the other (a unary function or a constant being regarded as a binary function that ignores one or both of its arguments), and compositions of generated functions are generated. It therefore suffices to compute, for each presentation, the set $B$ of all binary functions that it generates.

A binary function on $S$ is an $n\times n$ table, so $B$ is a set of tables. We compute it as a closure (\texttt{clones.py}):
\begin{enumerate}[nosep]
\item Start with the table $X$ of the function $(x,y)\mapsto x$, the table $Y$ of $(x,y)\mapsto y$ and one constant table for each constant of the presentation.
\item Apply every generator entrywise to the tables found so far: a unary generator $u$ turns a table $F$ into the table with entries $u(F_{ij})$, and a binary generator $g$ turns a pair of tables $F,G$ into the table with entries $g(F_{ij},G_{ij})$. Add every table not seen before.
\item Repeat step 2 until no new table appears.
\end{enumerate}
This halts because there are only $n^{n^2}$ tables, and it yields exactly $B$: every table produced is the table of a composition of generators applied to $x$ and $y$, and every such composition is produced, by induction on its length.

Three shortcuts avoid the closure in the cases where $B$ is large:
\begin{itemize}[nosep]
\item {\bf All functions, $n=3$:} A single binary operation on three elements generates all functions if and only if it belongs to none of the $18$ maximal clones on three elements, which were classified by Jablonskij \cite{Jablonskij1958}. We test each of the $19\,683$ operations for membership in these $18$ clones once (\texttt{sheffer3.py}); $3774$ of them belong to none and are assigned the clone of all functions without running the closure.
\item {\bf All functions, $n\ge 3$:} By S{\l}upecki's criterion \cite{Slupecki1939}, a presentation generates all functions if it generates all $n^n$ unary functions and has a generator that is surjective and truly binary. The unary functions generated are computed by the same closure applied to tables of unary functions, which is cheap.
\item {\bf Semirings and their descendants, $n=4$:} By distributivity, every term of a semiring is a sum of products of variables and constants. We therefore first close $\{x,y,0,1\}$ under multiplication and then close the result under addition; this gives the same set $B$ as the general closure but takes far fewer steps (\texttt{ringlike\_clones.py}). A ring, division ring or field generates the same clone as its semiring, because $-x$ is a sum of copies of $x$ and, in a field with $q$ elements, the inverse (with the convention $0^{-1}=0$) is $x^{-1}=x^{q-2}$ for $q>2$ and $x^{-1}=x$ for $q=2$. For $n=2$ and $3$ we verified that this method and the general closure produce identical sets.
\end{itemize}

\subsection{Step 3: counting clones up to relabeling}

A permutation $\sigma$ of $S$ maps a table $F$ to the table $F^\sigma$ with $F^\sigma(x,y)=\sigma(F(\sigma^{-1}x,\sigma^{-1}y))$, and maps the set $B$ to $B^\sigma=\{F^\sigma: F\in B\}$. The canonical form of $B$ is the lexicographically smallest of the $n!$ sorted lists $B^\sigma$. Two presentations generate the same closed mathematical structure exactly when the canonical forms of their sets $B$ coincide, so cl is the number of distinct canonical forms occurring in a row. For cl$_{\rm tb}$, only the presentations all of whose binary generators are truly binary are included.

\subsection{Checks}

The str columns reproduce the published sequences listed in \tab{StructureCountTable} for every row where one exists. 
An operation and its opposite $x\circ' y=y\circ x$ always generate each other, so the classes of presentations generating the same clone are unions of classes under isomorphism or anti-isomorphism. For semigroups, the cl entries $4,18,126$ for $n=2,3,4$ equal the published numbers of semigroups up to isomorphism or anti-isomorphism (OEIS sequence A001423 \cite{OEIS}); for bands ($2,6,26$) and monoids ($2,6,27$) they equal the corresponding numbers, which we computed ourselves. Since the counts agree, there are no further identifications: for $n\le 4$ the only distinct associative operations generating each other are an operation and its opposite. 
For magmas with $n=3$, the $3330$ isomorphism classes give $411$ clones, of which the clone of all functions accounts for $629$ classes. 
The magma entries for $n=4$ were not computed, for the reason given in the footnote to the table.

\subsection{When compositional closure makes a difference}\label{NdiffSubsec}

In many rows of \tab{ClosedStructureCountTable}, cl equals str for every $n$ shown. The last column of the table records the smallest $n$, denoted $n_{\rm diff}$, for which cl $<$ str, \ie, for which two non-isomorphic models of the theory generate each other. Throughout, only finite models are considered. For some theories this never happens, and for others the equality is a coincidence of small $n$. In each case the argument is the same: list the binary functions that a model generates, and ask which of them could be the operation of another model of the same theory.

\paragraph{Theories where it never makes a difference}
\begin{itemize}[nosep]
\item {\bf Commutative monoids} $(M,+,0)$: The binary functions generated are $(x,y)\mapsto ax+by$, where $x$ and $y$ range over the elements of $M$ and $a$, $b$ are natural numbers; here $ax$ denotes $x$ added to itself $a$ times. Suppose that $x\circ y=ax+by$ is another commutative monoid operation, with identity $e$. The only constant that the monoid generates is $0$, so $e=0$. Then $x\circ 0=x$ gives $ax=x$ for all $x$, and $0\circ y=y$ gives $by=y$ for all $y$, so $x\circ y=x+y$. A commutative monoid thus generates no commutative monoid other than itself.
\item {\bf Abelian groups:} The same argument applies, with $a$ and $b$ integers.
\item {\bf Semilattices:} The only binary functions generated by $\land$ are $x$, $y$ and $x\land y$. For $n\ge2$ only the last is commutative, and for $n=1$ there is only one semilattice.
\item {\bf Rooted trees:} A unary function $f$ generates exactly its iterates $x, f, f^2,\ldots$, so $f$ and $g$ generate each other only if each is an iterate of the other. Unless $f=g$, this forces $f=f^m$ for some $m>1$. A rooted tree with $f=f^m$ is a star, in which every node points to the root, and all iterates of a star are equal to it.
\item {\bf Integral domains, division rings, fields and Boolean algebras:} For every finite $n$ there is at most one of each with $n$ elements, so there is nothing to identify.
\end{itemize}

\paragraph{Theories where it makes a difference from the start}
A non-commutative operation $x\circ y$ generates its opposite $y\circ x$. This accounts for all identifications in the semigroup, band and monoid rows for $n\le 4$: the first $n$ with a model that is not isomorphic to its opposite is $n=2$ for semigroups and bands and $n=3$ for monoids. Magmas ($n_{\rm diff}=2$) and semirings ($n_{\rm diff}=4$) also have identifications of other kinds.

\paragraph{Theories where the difference appears only later}
\begin{itemize}[nosep]
\item {\bf Lattices} ($n_{\rm diff}=5$): The binary functions generated by $\land$ and $\lor$ are $x$, $y$, $x\land y$ and $x\lor y$, so the only other lattice that a lattice generates is its dual, obtained by exchanging $\land$ and $\lor$. Hence cl counts lattices up to duality. Every lattice with at most four elements is isomorphic to its dual; of the five lattices with five elements, two are each other's duals, giving four clones.
\item {\bf Monounary algebras} ($n_{\rm diff}=6$): If two non-isomorphic functions $f$ and $g$ are iterates of each other, then $f=f^m$ for some $m>1$, so $f$ permutes its image: it consists of cycles with further elements (leaves) attached directly to cycle elements, and $g=f^k$ runs through the same cycles in a different order. Usually $g$ is then isomorphic to $f$. The smallest exception has six elements: a cycle $0\to1\to2\to0$ with two leaves attached to $0$ and one leaf attached to $1$. Here $f^2$ runs through the cycle backwards and $(f^2)^2=f^4=f$, so $f$ and $f^2$ generate each other, but they are not isomorphic, because the numbers of leaves met along the direction of the cycle are $(2,1,0)$ for $f$ and $(1,2,0)$ for $f^2$. An enumeration of all $n^n$ functions confirms that this is the first exception: cl $=$ str $=1,3,7,19,47$ for $n\le5$, whereas str $=130$ and cl $=129$ for $n=6$ (\texttt{monounary\_clones.py}).
\item {\bf Rings} ($n_{\rm diff}=16$): A ring generates its opposite ring, with multiplication $x\cdot' y=yx$, so cl $<$ str as soon as some ring is not isomorphic to its opposite. The ring of triples $(a,b,c)$ with $a\in\Z_4$ and $b,c\in\Z_2$, with componentwise addition and the product $$(a,b,c)(a',b',c')=(aa',\ \bar a b'+bc',\ cc'),$$ where $\bar a$ denotes $a$ reduced modulo $2$ (the upper triangular matrices $\left(\begin{smallmatrix}a&b\\0&c\end{smallmatrix}\right)$), has $16$ elements and is such a ring. For an element $x$, let $r(x)$ be the number of $y$ with $xy=0$ and $l(x)$ the number of $y$ with $yx=0$. Over the $16$ elements, $r$ takes the values $1,2,4,8,16$ respectively $4,4,3,4,1$ times, whereas $l$ takes them $4,2,6,3,1$ times; an isomorphism with the opposite ring would exchange the two lists (\texttt{ring\_invariants.py}). For $n\le 15$ nothing is identified. For $5\le n\le15$, the only $n$ with more than one ring with identity are $8$ (eleven rings), $9$ (four) and $12$ (four), and for each of these $n$ the rings are distinguished by quantities that two rings generating each other must share: the number of constants generated, the number of unary functions generated, the number of subrings and the numbers of ideals and of minimal ideals (\texttt{ring\_invariants.py}).
\item {\bf Commutative rings} ($n_{\rm diff}=16$): A commutative ring is its own opposite, so the mechanism above is not available, and the computation just described shows that nothing is identified for $n\le15$. For $n=16$, however, the two rings
$$R_1=\Z_4[g]/(g^2-2)\quad\hbox{and}\quad R_2=\Z_4[g]/(g^2-2g-2)$$
generate each other. They have the same elements $u+vg$ with $u,v\in\Z_4$, the same addition and the same $0$ and $1$, and differ only in the product $g\cdot g$, which is $2$ in $R_1$ and $2+2g$ in $R_2$. In each of the two rings, the product of the other ring is
$$x\otimes y=xy+(x^2-x^4)(y+y^4),$$
which we verified by evaluating both sides on all $256$ pairs $(x,y)$ in each ring (\texttt{comm\_rings16\_\allowbreak pairs.py}). The rings are not isomorphic: $R_1$ contains a square root of $2$, whereas $R_2$ is the ring $\Z_4[i]$ of Gaussian integers modulo $4$ (with $i=g-1$, so that $i^2=-1$), in which $(u+vi)^2=u^2-v^2+2uvi$ and $u^2-v^2$ is never $2$ modulo $4$. We found this pair by enumerating all $37$ commutative rings with identity with $16$ elements (\texttt{comm\_rings16.py}) and comparing the quantities listed above; these distinguish all of the rings except this pair and the pair $\Z_4[g]/(g^2)$, $\Z_4[g]/(g^2-2g)$, for which we do not know whether they generate each other.
\item {\bf Groups} ($24\le n_{\rm diff}\le 260$): Two non-isomorphic groups with $260$ elements that generate each other are known: the smallest member of a family constructed by Kearnes and Szendrei \cite[Example 2.14]{KearnesSzendrei2004}. Let $G_1=\Z_4\ltimes(\Z_5\times\Z_{13})$, where the generator of $\Z_4$ acts on $\Z_5$ as multiplication by $2$ and on $\Z_{13}$ as multiplication by $5$, and let $G_2$ be the same group with $5$ replaced by its inverse $8$. Then $x\circ y=xy\,[y,x]^{40}$, with $[y,x]=y^{-1}x^{-1}yx$, is a group operation on the set $G_1$ that is isomorphic to $G_2$ and not to $G_1$, and the product of $G_1$ is obtained back from $\circ$ by the same formula; we verified these statements by computer (\texttt{termeq\_counterexample.py}). This is equivalence of the generated functions; the earlier result of Z\'adori \cite{Zadori1997} concerns the weaker categorical equivalence of finite groups. For a lower bound, suppose that the groups $G$ and $G'$ on the same set generate each other. The powers of an element are the same in both, so each element has the same order in $G$ and in $G'$; and an Abelian group generates no group other than itself, so $G'$ is non-Abelian if $G$ is. For $n\le23$ with $n\ne16$, any two non-isomorphic non-Abelian groups with $n$ elements differ in the number of elements of some order, and for $n=16$ we verified by computing, for each of the $14$ groups, all group operations that it generates, that each of them is isomorphic to the original group (\texttt{termeq\_groups.py}). Hence $n_{\rm diff}\ge24$.
\end{itemize}

\end{document}